\documentclass[%
 reprint,
 superscriptaddress,
 showpacs,
 preprintnumbers,
 amsmath,
 amssymb,
 aps,
 prx,
]{revtex4-2}

\usepackage{array,booktabs}
\usepackage{scrextend}
\usepackage{hyperref}
\usepackage{lipsum}
\usepackage{graphicx}
\usepackage{amsmath,amssymb,amsfonts,dsfont}
\usepackage{color}
\usepackage[normalem]{ulem} 
\usepackage{multirow}
\usepackage{array}
\usepackage{upgreek}
\usepackage{mathrsfs}
\usepackage{capt-of}
\usepackage{siunitx}
\usepackage[hints,insection]{minitoc}
\usepackage{multirow}
\usepackage{float} 
\usepackage{threeparttable,array} 
\usepackage{placeins}
\usepackage{soul} 
\setstcolor{red}
\usepackage{xcolor}

\newcommand{\ketbare}[1]{$|\mathrm{#1}\rangle$}
\newcommand{\ketidle}[1]{$|\mathrm{#1}\rangle$}

\begin{document}

\title{Realization of high-fidelity CZ and $ZZ$-free iSWAP gates with a tunable coupler
}

\def\RLEaffil{Research Laboratory of Electronics, Massachusetts Institute of Technology, Cambridge, MA 02139, USA}
\def\LLaffil{MIT Lincoln Laboratory, Lexington, MA 02421, USA}
\def\Physaffil{Department of Physics, Massachusetts Institute of Technology, Cambridge, MA 02139, USA}
\def\EECSaffil{Department of Electrical Engineering and Computer Science, Massachusetts Institute of Technology, Cambridge, MA 02139, USA}

\author{Youngkyu Sung}
\email{youngkyu@mit.edu}
\affiliation{\RLEaffil}
\affiliation{\EECSaffil}
\author{Leon Ding}
\affiliation{\RLEaffil}
\affiliation{\Physaffil}
\author{Jochen Braum\"uller}
\affiliation{\RLEaffil}
\author{Antti Veps\"al\"ainen}
\affiliation{\RLEaffil}
\author{\mbox{Bharath Kannan}}
\affiliation{\RLEaffil}
\affiliation{\EECSaffil}
\author{\mbox{Morten Kjaergaard}}
\affiliation{\RLEaffil}
\author{Ami Greene}
\affiliation{\RLEaffil}
\affiliation{\EECSaffil}
\author{Gabriel O. Samach}
\affiliation{\RLEaffil}
\affiliation{\EECSaffil}
\affiliation{\LLaffil}
\author{\mbox{Chris McNally}}
\affiliation{\RLEaffil}
\affiliation{\EECSaffil}
\author{David Kim}
\affiliation{\LLaffil}
\author{Alexander Melville}
\affiliation{\LLaffil}
\author{\mbox{Bethany M. Niedzielski}}
\affiliation{\LLaffil}
\author{Mollie E. Schwartz}
\affiliation{\LLaffil}
\author{\mbox{Jonilyn L. Yoder}}
\affiliation{\LLaffil}
\author{Terry P. Orlando}
\affiliation{\RLEaffil}
\author{Simon Gustavsson}
\affiliation{\RLEaffil}
\author{\mbox{William D. Oliver}}
\email{william.oliver@mit.edu}
\affiliation{\RLEaffil}
\affiliation{\EECSaffil}
\affiliation{\Physaffil}
\affiliation{\LLaffil}

\date{\today}
\begin{abstract} 
High-fidelity two-qubit gates at scale are a key requirement to realize the full promise of quantum computation and simulation.
The advent and use of coupler elements to tunably control two-qubit interactions has improved operational fidelity in many-qubit systems by reducing parasitic coupling and frequency crowding issues.
Nonetheless, two-qubit gate errors still limit the capability of near-term quantum applications.
The reason, in part, is the existing framework for tunable couplers based on the dispersive approximation does not fully incorporate three-body multi-level dynamics, which is essential for addressing coherent leakage to the coupler and parasitic longitudinal ($ZZ$) interactions during two-qubit gates. 
Here, we present a systematic approach that goes beyond the dispersive approximation to exploit the engineered level structure of the coupler and optimize its control.
Using this approach, we experimentally demonstrate CZ and $ZZ$-free iSWAP gates with two-qubit interaction fidelities of $99.76 \pm 0.07$\% and $99.87 \pm 0.23$\%, respectively, which are close to their $T_1$ limits.
\end{abstract}

\maketitle
A key challenge for large-scale quantum computation and simulation is the extensible implementation of high-fidelity entangling gates~\cite{Krantz2019}. Over the past two decades, superconducting qubits have made great strides in gate fidelities and scalability~\cite{Kjaergaard2020}, heralding the era of noisy intermediate scale quantum (NISQ) systems~\cite{Preskill2018, Arute2019}. 
The introduction of tunable couplers, which dynamically control the qubit-qubit interaction,
is an architectural breakthrough that helps resolve many scalability issues such as frequency crowding and parasitic coupling between adjacent qubits, and enables fast, high-fidelity two qubit gates~\cite{Hime2006, Niskanen2007, Ploeg2007, Harris2007, Chen2014, Mckay2016, Weber2017, Neill2017, Yan2018, Kounalakis2018, Mundada2019, Arute2019, Li2020, Foxen2020, Collodo2020, Xu2020, Han2020}. 
Recently, two-qubit gates with bosonic qubits have also been demonstrated by using a driven transmon coupler~\cite{Rosenblum2018,Gao2019}.
Despite tremendous progress, however, the two-qubit gate error still remains a major bottleneck for realizing the full promise of NISQ hardware and ultimately building error-corrected logical qubits~\cite{Gambetta2017,Preskill2018}.

To further improve the fidelity of coupler-mediated entangling gates, a systematic approach for optimizing control and level-structure of the coupler is required. 
However, the existing theoretical framework based on the perturbative approach, which assumes a dispersive qubit-coupler interaction~\cite{Yan2018}, has several limitations. 
First, when performing fast two-qubit gates, the qubit-coupler coupling generally enters into the non- or weakly-dispersive regime. 
Therefore, the perturbative approach breaks down and coherent energy exchange between the qubit and coupler arises,
which is not captured within the existing framework.
In other words, theoretical treatments are simplified at the cost of overlooking coherent leakage to the coupler -- non-adiabatic error -- when performing fast two-qubit gates. 
Furthermore, the perturbative treatment of tunable couplers disregards the presence of higher levels of the coupler~\cite{Yan2018}.
This is a significant omission; the higher level of the coupler participates in the multi-level dynamics of two-qubit gates, and thereby, adds a considerable amount of residual two-qubit interactions.

In this paper, we engineer the control and level-structure of the coupler by going beyond the dispersive approximation in order to realize high-fidelity two-qubit gates. We implement both longitudinal (CZ) and transversal (iSWAP) two-qubit gates; the availability of both type of gates generally reduces gate overheads of NISQ algorithms~\cite{Preskill2018,Kivlichan2018}. We propose an intuitive, yet systematic approach for optimizing control to suppress coherent leakage to the coupler. Via optimized control, we significantly reduce the non-adiabatic error of a \SI{60}{\nano\second}-long CZ gate, thereby demonstrating a two-qubit interaction
fidelity of $99.76\pm0.07\%$.
in interleaved randomized benchmarking. We also address a fundamental issue of the iSWAP gate when coupling two transmon qubits: parasitic $ZZ$ interaction due to their higher levels~\cite{O'Malley2015, Mckay2016, Barends2019, Foxen2020, Krinner2020}. 
We successfully suppress the residual $ZZ$ interaction of the iSWAP gate in a passive manner, by exploiting the engineered coupler level structure and demonstrate 
a two-qubit interaction
fidelity of $99.87\pm0.23\%$ with a \SI{30}{\nano\second} gate duration. 

\begin{figure}[h!]
    \begin{center}
    \includegraphics[width=\columnwidth]{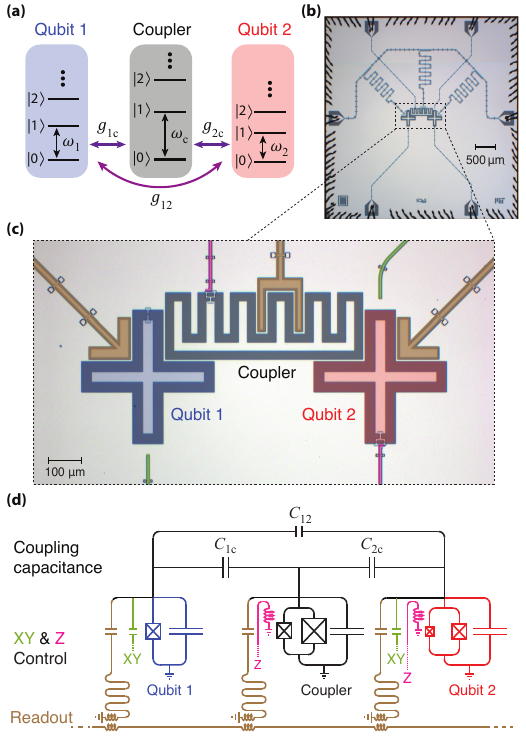}
    \end{center}
    \caption{
     \textbf{(a)} Schematic diagram of a pairwise interacting three-body system. Each constituent body has anharmonic multiple energy level structure.
     \textbf{(b-c)} Experimental realization of a three-body system in superconducting circuits.
     \textbf{(d)} Circuit schematic. In \textbf{(c)} false colors (blue, red, pink, green, and brown) are used to indicate the corresponding circuit components in \textbf{(d)}.
    }
    \label{fig:Fig1}
\end{figure}

We shall consider a pairwise interacting three-body quantum system, in which each constituent body is a multi-level anharmonic oscillator~(Fig.~\ref{fig:Fig1}a). Quantum bits are encoded in the first two levels of the leftmost and rightmost anharmonic oscillators with resonant frequencies $\omega_1$ and $\omega_2$, respectively. 
The middle anharmonic oscillator serves as the coupler.
These two distant qubits and the coupler are coupled through exchange-type interactions with coupling strengths $g_{\mathrm{1c}}$, $g_{\mathrm{2c}}$, and $g_{\mathrm{12}}$. 
We assume the qubit-coupler interactions to be much stronger than the direct qubit-qubit interaction $g_{1\mathrm{c}} = g_{2\mathrm{c}} \gg g_{\mathrm{12}}$.
This is the case for our device, and is a practical parameter regime for tunable couplers, in general~\cite{Yan2018}.
We approximate the qubits and the coupler as Duffing oscillators, a common model for anharmonic multi-level qubit systems such as the transmon~\cite{Koch2007} and the C-shunt flux qubit~\cite{Yan2016}. Thus, the system Hamiltonian can be written as follows ($\hbar\equiv 1$),
\begin{align}
    \label{eq:H_lab}
    H = &\sum_{i} \left( \omega_{i} b_{i}^{\dagger}b_{i} + \frac{\eta_{i}}{2} b_{i}^{\dagger}b_{i}^{\dagger}b_{i}b_{i}\right) 
    +   \sum_{i<j} g_{ij}(b_{i}-b_{i}^{\dagger})(b_{j}-b_{j}^{\dagger}),
\end{align}
where $b^{\dagger}_{i}$ and $b_{i}$ ($i,j\in \{1, 2, \mathrm{c}\}$) are, respectively, the raising and lowering operators defined in the eigenbasis of the corresponding oscillators. The level anharmonicity of each oscillator is denoted by $\eta_{i}$. 
As shown in Ref.~\cite{Yan2018}, the destructive interference between the coupler-mediated and direct qubit-qubit couplings enables the resulting net qubit-qubit coupling to be turned on and off by adjusting the coupler frequency $\omega_{\mathrm{c}}$.

We realize this pairwise interacting three-body system in a circuit quantum electrodynamics  setup~\cite{Blais2004,Wallraff2004} using three capacitively coupled transmons (Figs.~\ref{fig:Fig1}b-d)~\cite{Koch2007,Barends2013}.
The transmon coupler at the center mediates interaction between the two distant transmon qubits. 
While the resonant frequency $\omega_1/2\pi$ of qubit 1 (QB1) is fixed at 4.16 GHz, the frequencies of qubit 2 (QB2) and the coupler (CPLR) are tunable ($\omega_{2}/2\pi=$ 3.7--4.7 GHz and $\omega_{\mathrm{c}}/2\pi=$ 3.7--6.7 GHz) by modulating the external magnetic flux threading through their asymmetric SQUID loops~\cite{Hutchings2017}. 
More details about the device are provided in Appendix~\ref{suppsec:device_setup}.
Coupler-mediated two-qubit gates are implemented by dynamically tuning $\omega_2$ and $\omega_{\mathrm{c}}$. 
Both qubits have microwave control lines to drive single qubit X- and Y-rotation gates. 
Both the qubits and the coupler are dispersively coupled to coplanar waveguide resonators for their state readout. 
We discriminate between the ground, first- and second-excited states, such that we can distinguish 27 different states of the system (see Appendix~\ref{suppsec:state_readout} for details).

\begin{figure}[h!]
    \begin{center}
    \includegraphics[width=8.9cm]{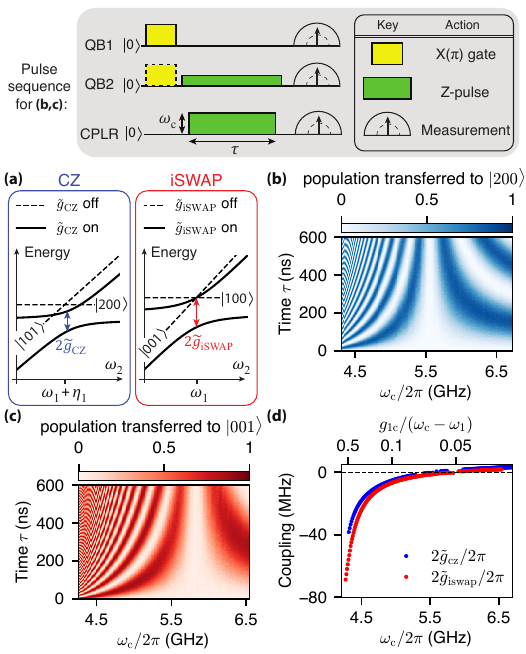}
    \end{center}
    \caption{
    \textbf{The tunable coupling for the CZ and the iSWAP gates.}
    \textbf{(a)} Illustrations of level crossings relevant to the CZ and the iSWAP gates. The energy splittings (2$\tilde{g}_{\mathrm{CZ}}$ and 2$\tilde{g}_{\mathrm{iSWAP}}$) are tunable by adjusting $\omega_{\mathrm{c}}$. See main text for details.
     \textbf{(b-c)} Experimental data for the energy exchange between \ketidle{200} and \ketidle{101}, and \ketidle{100} and \ketidle{001} as function of the coupler frequency $\omega_c$, respectively.
     The pulse sequences are illustrated at the top.
    \textbf{(d)} By fitting the oscillations with sinusoidal curves, we extract the swap rates $|2\tilde{g}_{\mathrm{CZ}}|/2\pi$ and $|2\tilde{g}_{\mathrm{iSWAP}}|/2\pi$  (circles). The top $x$-axis shows the corresponding perturbation parameter $g_{1\mathrm{c}}/(\omega_{\mathrm{c}}-\omega_{1})$ at each $\omega_{\mathrm{c}}$. 
    }
    \label{fig:Fig2}
\end{figure}

We use the notation \ketidle{QB1, CPLR, QB2} to represent the eigenstates of the system (Eq.~\eqref{eq:H_lab}) in the idling configuration where CPLR is placed at the frequency such that the effective QB1-QB2 coupling is nearly zero (dashed lines in Fig.~\ref{fig:Fig2}a). 
Note that these states approximate the diabatic (bare) states, \textit{i.e.}, the eigenstates of the uncoupled system, because QB1 and QB2 are effectively decoupled and both are far-detuned from CPLR ($g_{i\mathrm{c}}/(\omega_{\mathrm{c}}-\omega_{i}) < 1/20$, $i\in\{1,2\}$).
To implement CZ and iSWAP gates, we use non-adiabatic transitions between \ketidle{101} and \ketidle{200}, and \ketidle{100} and \ketidle{001}, respectively~\cite{Strauch2003, DiCarlo2009, Yamamoto2010, Barends2019}. 
The non-adiabatic transitions are regulated by adjusting $\omega_{\mathrm{c}}$, which effectively tunes the coupling strengths between \ketidle{101} and \ketidle{200} ($2\tilde{g}_{\mathrm{CZ}}$), or between \ketidle{100} and \ketidle{001} ($2\tilde{g}_{\mathrm{iSWAP}}$). 
For example, biasing $\omega_{\mathrm{c}}$ closer to $\omega_{1}$ and $\omega_{2}$ leads to opening of the avoided crossings ($|\tilde{g}_{\mathrm{CZ}}|>0$, $|\tilde{g}_{\mathrm{iSWAP}}|>0$) and downward level shifts induced by qubit-coupler interactions (solid curves in Fig.~\ref{fig:Fig2}a). 
The CZ gate is performed by suddenly bringing the states \ketbare{101} and \ketbare{200} into resonance at their ``bare'' energy degeneracy point, which projects these ``bare'' states onto the dressed states formed by the coupling $\tilde{g}_{\mathrm{CZ}}$ and results in Larmor precession within the dressed-state basis.
We let them complete a single period of an oscillation, such that \ketidle{101} picks up an overall phase $e^{i\pi}$.
To implement the iSWAP gate, we put \ketidle{100} and \ketidle{001} on resonance and let them complete half an oscillation, so that the two states are fully swapped.

We first demonstrate the tunability of the effective QB1-QB2 coupling strengths $\tilde{g}_{\mathrm{CZ}}$ and $\tilde{g}_{\mathrm{iSWAP}}$ by measuring the energy exchange between \ketidle{101} and \ketidle{200}, and \ketidle{100} and \ketidle{001}, respectively, as a function of CPLR frequency $\omega_{\mathrm{c}}$. 
To measure the energy exchange between \ketidle{101} and \ketidle{200}, we first prepare \ketidle{101} by applying $\pi$ pulses to both QB1 and QB2 at the idling configuration. Next, we rapidly adjust QB2 frequency $\omega_2$ so that \ketidle{101} and \ketidle{200} are on resonance and then turn on $\tilde{g}_{\mathrm{CZ}}$ by shifting $\omega_{\mathrm{c}}$. We wait a variable delay time $\tau$ and measure the state population of \ketidle{200}. We repeat these measurements with varying $\omega_{\mathrm{c}}$ (Fig.~\ref{fig:Fig2}b). %
In the similar manner, to measure $g_{\mathrm{iSWAP}}$, we prepare \ketidle{100} and measure the state population transferred to \ketidle{001} as a function of $\tau$ and $\omega_{\mathrm{c}}$ (Fig.~\ref{fig:Fig2}c).

In Fig.~\ref{fig:Fig2}d, we plot the effective coupling strengths $2\tilde{g}_{\mathrm{CZ}}/2\pi$ and $2\tilde{g}_{\mathrm{iSWAP}}/2\pi$ as a function of CPLR frequency $\omega_{\mathrm{c}}$ by fitting the excitation exchange oscillations. 
To implement fast two-qubit gates (we use a 60 ns-long CZ gate and a 30 ns-long iSWAP gate), a strong coupling strength is required, which strongly hybridizes the CPLR with both QB1 and QB2 ($g_{i\mathrm{c}}/(\omega_{\mathrm{c}}-\omega_{i}) \approx 1/3$). However, dynamically entering and exiting such a non-dispersive regime easily leads to coherent leakage into the CPLR (non-adiabatic error). Hence, well-engineered control is required to avoid the coherent leakage when implementing fast two-qubit gates.

\begin{figure}[ht!]
    \centering
    \includegraphics[width=8.9cm]{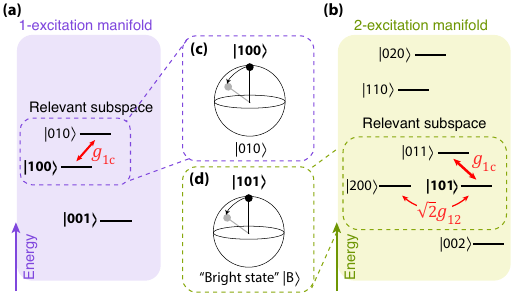}
    \caption{ 
    \textbf{(a, b)} Energy level diagrams of the single- and double-excitation manifolds. The dashed boxes indicate subspaces spanned by energy levels that are relevant to coherent leakage during the CZ gate. The red double-headed arrows denote exchange interactions between the energy levels.
    \textbf{(c)} Bloch-sphere representation of the relevant subspace in the single-excitation manifold.
    \textbf{(d)} Bloch-sphere representation of the two-level approximation for the relevant subspace in the double-excitation manifold. When $g_{\mathrm{1c}} \gg g_{12}$, and because the transition between \ketidle{200} and \ketidle{011} is dipole-forbidden, the state \ketbare{101} primarily interacts with a bright state $|\mathrm{B}\rangle \equiv \cos\Theta$\ketbare{011}$+\sin\Theta$\ketbare{200}, where  $\Theta\equiv\tan^{-1}(\sqrt{2}g_{12}/g_{1\mathrm{c}})$. 
    }
    \label{fig:Fig3}
\end{figure}

To implement an optimized control scheme, we propose a tractable model for analyzing the leakage dynamics. We first note that the energy levels of the system interact via excitation-preserving exchange within the rotating wave approximation, such that the dynamics can be analyzed in the two independent manifolds, one involving single excitation and one involving double excitations  (Figs.~\ref{fig:Fig3}a and \ref{fig:Fig3}b, respectively). 
In each manifold, we identify the subspaces spanned by the states which strongly interact with computational qubit states and cause leakage during the CZ gate (dashed boxes in Fig.~\ref{fig:Fig3}, see Appendix~\ref{suppsec:effective_H} for details).
For the sake of simplifying the leakage dynamics, we intentionally chose a small anharmonicity for the coupler to avoid strong hybridization of $|020\rangle$ with other states during CZ gates (see Appendix~\ref{suppsec:advantages_small_eta_c} for details).
Of these states, \ketbare{100} and \ketbare{101} are computational qubit states and all others are leakage states. 
In the double-excitation manifold, the transition between \ketbare{200} and \ketbare{011} is dipole-forbidden (requires a second-order process), and is therefore suppressed. 
This allows the description of the corresponding three-level dynamics to be further simplified by introduction of a partially hybridized basis:
a bright state $|\mathrm{B}\rangle \equiv \cos\Theta$\ketbare{011}$+\sin\Theta$\ketbare{200} and a dark state $|\mathrm{D}\rangle \equiv \cos\Theta$\ketbare{200}$-\sin\Theta$\ketbare{011}, where $\Theta\equiv\tan^{-1}(\sqrt{2}g_{12}/g_{1\mathrm{c}})$~\cite{Lambropoulos2007}.

\begin{figure*}[ht!]
    \begin{center}
    \includegraphics[width=18.3cm]{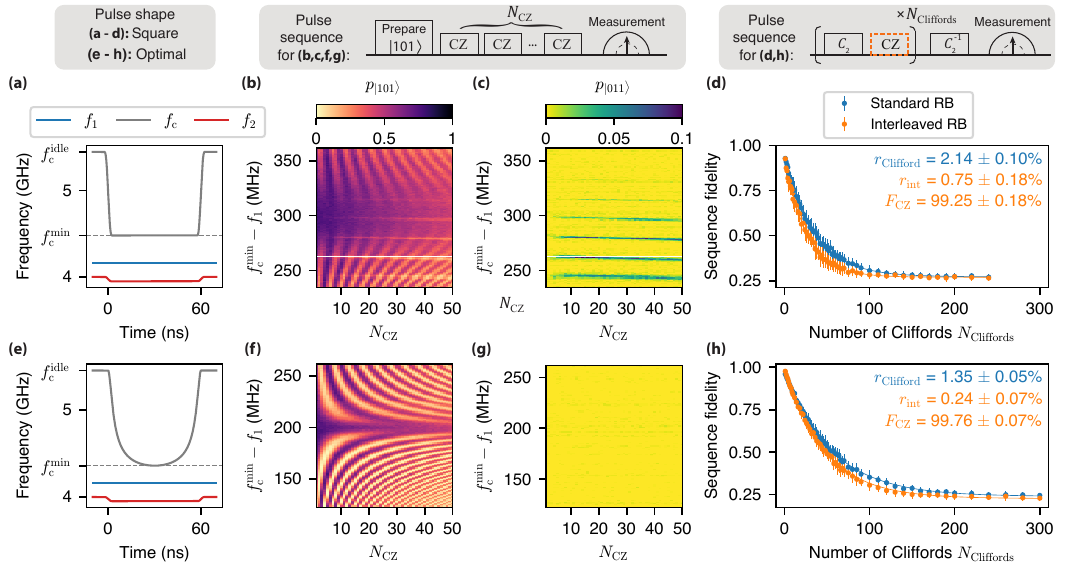}
    \end{center}
    \caption{
    \textbf{Suppressing leakage to the coupler by optimizing the coupler control.}
    \textbf{(a,e)} Square-shaped and Slepian-based optimal control waveforms for \SI{60}{\nano\second}-long CZ gates, respectively. \textbf{(b,f)} State population of \ketidle{101} after applying repeated CZ gates vs. the coupler pulse amplitude ($f_{\mathrm{c}}^{\mathrm{min}}-f_1$) and the number of CZ gates $N_{\mathrm{CZ}}$. \textbf{(c,g)} Leakage population to \ketidle{011} after applying repeated CZ gates. The square-shaped pulse shows periodic leakage to \ketidle{011}, which is suppressed down to the background noise limit by optimizing the pulse shape. \textbf{(d,h)} Interleaved randomized benchmarking (RB) results of the CZ gates. The pulse sequence is illustrated at the top; we apply $N_{\mathrm{Clifford}}$ random two-qubit Clifford gates ($C_2$) and the recovery Clifford gate $C_2^{-1}$, which makes the total sequence equal to the identity operation. Errors bars represent $\pm1$ standard deviations. We measure 30 random sequences for each sequence length $N_{\mathrm{Cliffords}}$. To ensure accurate uncertainties of the error rates ($r_{\mathrm{Clifford}}$ and $r_{\mathrm{int}}$), we perform a weighted least-squares fit using the inverse of variance as the weights.
    }
    \label{fig:Fig4}
\end{figure*}

If $g_{\mathrm{1c}}\gg g_{12}$, within this truncated three-level subspace, the computational state \ketbare{101} only interacts with the bright state $|\mathrm{B}\rangle$ and we can neglect the dark state $|\mathrm{D}\rangle$. 
Consequently, the leakage dynamics within the single- and double-excitation manifolds are described by the following effective two-level Hamiltonians $H_{\mathrm{1}}(t)$ and $H_{\mathrm{2}}(t)$, respectively.
\begin{align}
\label{eq:H_eff}
H_{\mathrm{1}}(t)&=  \stackrel{\mbox{\scalebox{0.7}{$|100\rangle$ \ \  \ketbare{010}\ \ }}}{%
    \begin{pmatrix}
    \omega_{1} & g_{\mathrm{1c}} \\
    g_{\mathrm{1c}} & \omega_{\mathrm{c}}(t) 
    \end{pmatrix}%
  },
  &H_{\mathrm{2}}(t) &= \stackrel{\mbox{\scalebox{0.7}{ \ketbare{101}  \qquad \ \ $|\mathrm{B}\rangle$   }}}{%
    \begin{pmatrix}
    \omega_{1}+\omega_{2} & {g}_{\mathrm{B}} \\
    {g}_{\mathrm{B}} & {\omega}_{\mathrm{B}}(t) 
    \end{pmatrix}%
  },      
\end{align}
where the coupling strength between \ketbare{101} and \ketbare{B} is given by ${g_{\mathrm{B}}=g_{\mathrm{1c}}\cos\Theta  +  \sqrt{2}g_{\mathrm{12}}\sin\Theta}$ and the energy of the \ketbare{B} is ${{\omega}_{\mathrm{B}}= \cos^2\Theta(\omega_{\mathrm{c}}(t)+\omega_2)+ \sin^2\Theta(\omega_1+\omega_2)}$. 
Such a mapping of the multi-level dynamics onto two-level systems is useful, because optimal control techniques are well-studied for two-level cases~\cite{Martinis2014}.
This technique of simplifying multi-level leakage dynamics using bright and dark states is also used to optimize the control pulse for our iSWAP gate (see Appendix~\ref{suppsec:effective_H}).
%

Since $g_{\mathrm{1c}}\gg g_{12}$ $( \Theta\approx0,\  |\mathrm{B}\rangle\approx|011\rangle, \ \mathrm{and} \  |\mathrm{D}\rangle\approx|200\rangle)$, the effective Hamiltonians $H_{\mathrm{1}}(t)$ and $H_{\mathrm{2}}(t)$ are equivalent up to offset energies.
This equivalence enables us to suppress leakage in both single- and double-excitation manifolds by optimizing a single control parameter $\omega_{\mathrm{c}}(t)$.
Note that although \ketbare{200} behaves as a dark state in this truncated subspace, it still interacts with \ketbare{101} via a second-order process through the intermediate state \ketbare{110} (outside the truncated subspace), which enables the CZ gate. 
Our goal here is to suppress fast, non-adiabatic transitions between $|101\rangle$ and $|\mathrm{B}\rangle\approx |011\rangle$ (a first-order interaction) that occur much faster than a slow swapping between $|101\rangle$ and $|\mathrm{D}\rangle\approx |200\rangle$ (a second-order interaction through the intermediate state $|110\rangle$). Therefore our two-level system model addresses only the predominant, leading-order leakage dynamics.
Developing a theoretical framework for addressing additional leakage dynamics, such as leakage into \ketbare{110}, will be the subject of future work.
Following Ref.~\cite{Martinis2014}, we take the Slepian-based approach to devise an optimal control waveform $\omega_{c}(t)$ that targets adiabatic evolution within the effective two-level systems. 
In Appendix~\ref{suppsec:slepian_approach}, we present numerical simulation results that validate the suppression of leakage to CPLR when using the optimized pulse shape for both CZ and iSWAP gates.

We experimentally assess the performance of an optimized control pulse for the CZ gate by comparing its performance to a simple square pulse (Fig~\ref{fig:Fig4}). 
First, to characterize the leakage into CPLR, we vary the control pulse amplitude and measure the leakage of the CZ gates into \ketidle{011} (Figs.~\ref{fig:Fig4}b-c and f-g).
The amplitude is parametrized by the minimum point of CPLR frequency $f_{\mathrm{c}}^{\mathrm{min}}$ (see Figs.~\ref{fig:Fig4}a and e). 
The chevron patterns of the \ketidle{101} population $p_{|{101}\rangle}$ represent coherent energy exchange between \ketidle{101} and \ketidle{200}. 
We predistort the pulses to eliminate non-idealities, such as transients in the control lines, to ensure the desired pulse shape is faithfully delivered to the device (see Appendix~\ref{suppsec:Z-pulse_transient_calibration} for details) and thereby achieve symmetric chevron patterns~\cite{Foxen2018, Rol2020, Sung2020_predistortion}. 
On top of the chevrons, we observe distinctive periodic resonances for the square pulse, which are due to the leakage to \ketidle{011}. 
We suppress this leakage via an optimized control pulse shape (Fig.~\ref{fig:Fig4}g). 
Although we only present measurements of the leakage population to \ketidle{011} in Fig.~\ref{fig:Fig4}, we have experimentally confirmed that the leakage to other states in the two-photon manifolds -- \ketidle{020}, \ketidle{110} and \ketidle{002} --  are negligible (see Appendix~\ref{suppsec:supp_exp_cz_leakage}), thereby validating our two-level system model in Eq.~\eqref{eq:H_eff}).
%

Next, we confirm the improvement due to optimal pulse shaping by comparing the gate errors of the CZ gates.
The tune-up procedures for the CZ gate are illustrated in Appendix~\ref{suppsec:cz_tune-up}.
The single-qubit XY gates are performed in the idling configuration where the static $ZZ$ interaction between QB1 and QB2 is eliminated (see Appendices~\ref{suppsec:static_ZZ} and \ref{suppsec:1qb_rb}).
In Figs.~\ref{fig:Fig4}d and \ref{fig:Fig4}h, we measure the fidelities of the CZ gates via interleaved randomized benchmarking (RB) ~\cite{Corcoles2013, Megesan2012, Barends2014}.
Due to a dynamic change of the qubit frequencies during a two-qubit gate, additional Z-rotations accrue during the gate. 
Such Z-rotations need to be undone by applying compensating Z-rotations.
Therefore, the error rate of an interleaved two-qubit gate consists of two factors: the error rate of the native two-qubit interaction (which contains unwanted Z-rotations due to the change of the qubit frequencies) and the error rate of the compensation Z-rotations. %
Throughout this paper, we focus on the quality of the native two-qubit interaction.
In the case of CZ gates, we correct the additional Z-rotations by applying virtual Z gates that are essentially error-free~\cite{Mckay2017}. Therefore, the error rate $r_{\mathrm{int}}$ of the interleaved CZ gate is equivalent to the two-qubit interaction  error rate $r_{\mathrm{CZ}}\equiv1-F_{\mathrm{CZ}}$ of the native CZ gate.
The CZ gate with optimal pulse-shaping shows a higher two-qubit interaction fidelity $F_{\mathrm{CZ}}=1-r_{\mathrm{int}}=99.76 \pm 0.07\%$, which amounts to a $70\%$ error reduction compared to the square-shaped control pulse.

Based on the average gate composition for the two-qubit Cliffords~\cite{Barends2014}, we have estimated the two-qubit Clifford error rates $r_{\mathrm{Clifford,est}}$ using the following formula: $r_{\mathrm{Clifford,est}}=8.25\times r_{\mathrm{1qb}}+1.5\times r_{\mathrm{CZ}}$. The estimated Clifford error rates $r_{\mathrm{Clifford,est}}$ for the square and optimal pulses are $1.79\pm0.29\%$ and $1.02\pm0.11\%$, respectively. Differences between $r_{\mathrm{Clifford,est}}$ and $r_{\mathrm{Clifford}}$ are possibly due to residual distortion of the two-qubit gate pulses, which may additionally degrade the quality of subsequent single-qubit gates. By comparing Figs.~\ref{fig:Fig4}(d) and (h), we find that change in $r_{\mathrm{Clifford}}$ ($0.79\%$) is very close to $1.5\times$the change in $r_{\mathrm{int}}$ ($0.77\%$), conforming to the theory.

By solving a Lindblad master equation, we find that the $T_1$ limit for a \SI{60}{\nano\second}-long CZ gate is approximately 99.85\% (see Appendix~\ref{suppsec:T1_contribution_to_gate_errors}). 
We also simulate the contribution of $1/f^{\alpha}$ flux noise (which predominantly limits the pure dephasing times $T_2^*$ of QB2 and CPLR at the idling configuration) to the gate error rate and find it to be an order of magnitude smaller than the $T_1$-contribution (see Appendices~\ref{suppsec:flux_noise_in_the_device} and \ref{suppsec:Flux_noise_contribution_to_gate_errors} for details).
The gap between the measured fidelity $F_{\mathrm{CZ}}$ and its coherence limit implies additional coherent, leakage errors due to imperfect control. 
We find the leakage rate of the CZ gate with an optimal pulse shape is $0.06 \pm 0.07\%$ possibly due to residual pulse distortion of Z pulses (see Appendix~\ref{suppsec:cz_residual_leakage} for details).

\begin{figure}
    \begin{center}
    \includegraphics[width=8.9cm]{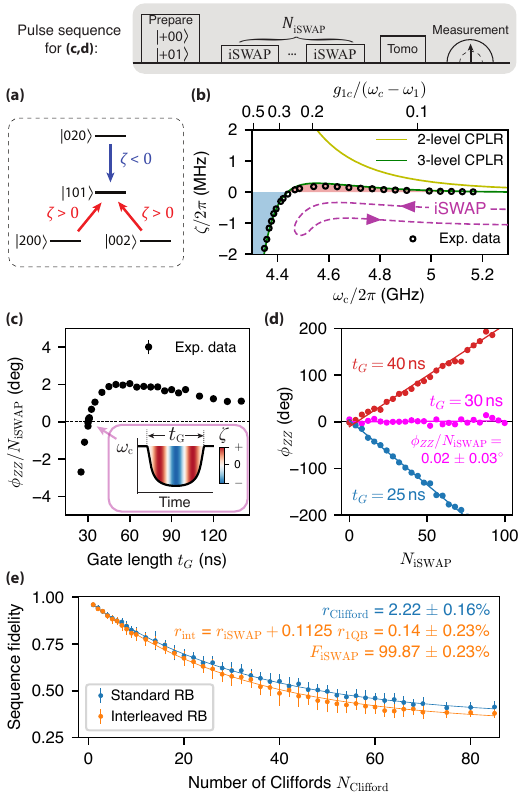}
    \end{center}
    \caption{
    \textbf{Cancelling out residual $ZZ$ interaction of the iSWAP by exploiting the engineered coupler level structure.} \textbf{(a)} The residual $ZZ$ interaction during iSWAP is originated from the level repulsion between $|101\rangle$ and the second excited level of the qubits (red arrows). This level repulsion is counteracted by utilizing the level repulsion from the 2nd-excited state of the coupler (blue arrow).
    \textbf{(b)} Residual $ZZ$ strength $\zeta$ 
    as a function of the coupler frequency $\omega_{\mathrm{c}}$, when the two qubits are on resonance. The top $x$-axis shows the corresponding perturbation parameter $g_{1\mathrm{c}}/(\omega_{\mathrm{c}}-\omega_{1})$. The solid curves correspond to numerical simulation assuming either $\eta_{\mathrm{c}}=\infty$ (yellow) or $\eta_{\mathrm{c}}/2\pi=-\SI{90}{\mega\hertz}$ (green). \textbf{(c)} $ZZ$ angle of the iSWAP gate $\phi_{{ZZ}}/N_{\mathrm{iSWAP}}$ as a function of the gate length $t_{\mathrm{G}}$. We cancel out the $ZZ$ angle by exploiting the tunability of $\zeta$ from positive to negative values. Inset shows the dynamic change of $\zeta$ during the excursion of $\omega_{\mathrm{c}}$ for a 30-ns long iSWAP gate. 
    Each data point is obtained by fitting the accumulated $ZZ$ angle $\phi_{{ZZ}}$ of $N_{\mathrm{iSWAP}}$-times repeated iSWAP gates with a linear function as shown in \textbf{(d)}. 
    \textbf{(e)} The results of interleaved randomized benchmarking (RB) for the $ZZ$-free \SI{30}{\nano\second}-long iSWAP gate (see main text for details). We measure 30 random sequences for each sequence length ($N_{\mathrm{Cliffords}}$). Errors bars represent $\pm1$ standard deviations. The error rates ($r_{\mathrm{Clifford}}$ and $r_{\mathrm{int}}$) and their uncertainties are extracted by performing a weighted least-squares fit using the inverse of variance as the weights. 
    }
    \label{fig:Fig5}
\end{figure}

Now, we move on to engineering the level structure of the coupler to suppress residual $ZZ$ interactions during the iSWAP gate.
The transmon qubit has a weak negative anharmonicity~\cite{Koch2007}. 
Therefore, the second excited levels of the transmons \ketbare{200} and \ketbare{002} are located near the computational qubit state \ketbare{101} when the two qubits are in resonance.
Interaction between these three energy levels leads to level repulsion (red arrows in Fig.~\ref{fig:Fig5}a). Due to the repulsion, the frequency of \ketidle{101} is shifted upward (note that \ketbare{200} and \ketbare{002} are located below \ketbare{101}), which results in a positive $ZZ$ interaction of strength $\zeta\equiv (E_{|{101}\rangle}-E_{|{001}\rangle})-(E_{|{100}\rangle}-E_{|{000}\rangle})$, where $E_{|m\rangle}$ denotes the eigenenergy of $|m\rangle$. 
Such residual $ZZ$ interactions have generally either been accomodated~\cite{Barends2019} or actively corrected by applying a partial CZ gate~\cite{Foxen2020} within the transmon qubit architecture. Recently, an approach to suppress the ZZ interactions by using qubits with opposing signs for their anharmonicity (e.g., a transmon qubit and a C-shunt flux qubit) has been proposed and demonstrated ~\cite{Zhao2020,Ku2020}. In this work, we utilize the higher level of the coupler \ketbare{020} to counteract the level repulsion while only using transmon qubits. Note that \ketbare{020} is located above \ketbare{101}, thereby providing a means to cancel the unwanted $ZZ$ term (blue arrow in Fig.~\ref{fig:Fig5}a).

In Fig.~\ref{fig:Fig5}b, we measure the residual $ZZ$ strength $\zeta$ as a function of $\omega_{\mathrm{c}}$, when QB1 and QB2 are in resonance. To measure $\zeta$, we perform a cross-Ramsey type experiment, which measures the conditional phase accumulation $\phi_{{ZZ}}$ of QB1 while initializing QB2 in either its ground or excited state. We measure $\phi_{{ZZ}}$ at full periods of the swap oscillation, where the net amount of excitation exchange is zero. Dividing $\phi_{{ZZ}}$ by the swap period ($2\pi/\tilde{g}_{\mathrm{iSWAP}}$), we extract $\zeta/2\pi$. The experimental data show good agreement with numerical simulation (green curve in Fig.~\ref{fig:Fig5}b, see Appendix~\ref{suppsec:numerical_simulation} for details about the simulation). We also compare the experimental data with simulated $\zeta$ for a 2-level CPLR (yellow curve in Fig.~\ref{fig:Fig5}b). Owing to the presence of the higher level of CPLR, $\zeta$ is significantly suppressed. We also note that levels beyond the 2nd-excited level of CPLR have little impact on the dynamics, since they are outside the relevant manifolds. This result clearly indicates that using a well-engineered multi-level coupler can significantly reduce a residual $ZZ$ error of the iSWAP gate, thereby further enhancing the fidelity.

When performing the iSWAP gate, its residual $ZZ$ angle $\phi_{{ZZ}}$ is accumulated by a dynamic change of $\zeta$ during the excursion of CPLR frequency $\omega_{\mathrm{c}}$. 
If the negative and positive portions of $\zeta$ during the gate are equal, the overall $ZZ$ phase is completely cancelled out. 
We measure the residual $ZZ$ angle $\phi_{{ZZ}}$ of the iSWAP gate by adjusting the pulse length in sync with the pulse amplitude such that the excitation is always fully swapped (Fig.~\ref{fig:Fig5}c).
We optimize the iSWAP pulse shape in the same manner to suppress coherent leakage to CPLR (see Appendices~\ref{suppsec:effective_H} and \ref{suppsec:slepian_approach} for details). 
Therefore, we simultaneously address both coherent leakage to CPLR and residual $ZZ$ interaction by optimizing the pulse shape and duration.
Owing to the cancellation induced by the higher level of CPLR, the iSWAP gate with a \SI{30}{\nano\second} duration features negligible residual $ZZ$ ($\phi_{{ZZ}}/N_{\mathrm{iSWAP}}=0.02\pm 0.03 \SI{}{\degree}$), which we refer to as the $ZZ$-free iSWAP gate (Fig.~\ref{fig:Fig5}d).
Note that the duration of a $ZZ$-free iSWAP gate depends on the coupler’s anharmonicity $\eta_{\mathrm{c}}$. Here, we engineer the coupler’s level structure such that its anharmonicity is relatively small ($\eta_{\mathrm{c}}/2\pi=-\SI{90}{\mega\hertz}$), in order to implement a faster $ZZ$-free iSWAP gate than what would be possible with larger $\eta_{\mathrm{c}}$ (see Appendix~\ref{suppsec:advantages_small_eta_c} for details). 
%

We measure the two-qubit interaction fidelity of the  $ZZ$-free iSWAP gate by performing  interleaved randomized benchmarking (RB) in Fig.~\ref{fig:Fig5}e.
The tune-up procedures for the iSWAP gate are described in Appendix~\ref{suppsec:cz_tune-up}.
Unlike the CZ gate, when performing single qubit gates, we bias QB1 and QB2 in resonance to synchronize their XY axes in the Bloch sphere (see Appendix~\ref{suppsec:synchronization_xy_axes} for details). 
This is facilitated by the tunable coupler, which switches off the effective transverse coupling between QB1 and QB2. Since they are put in resonance, the microwave crosstalk between the XY drive tones becomes critical. We cancel out this microwave crosstalk by applying active cancellation tone for each of the drive lines (see Appendix~\ref{suppsec:mw_crosstalk_calibration} for details). We find that using a long microwave pulse is desirable for better active cancellation. Hence, we apply \SI{70}{\nano\second}-long microwave pulses when implementing X and Y single-qubit gates, even though they show lower average gate fidelities (QB1 = 99.92\%, QB2 = 99.81\%) than the \SI{30}{\nano\second}-long pulses used in the CZ gate benchmarking experiments (QB1 = 99.94\%, QB2 = 99.90\%).
See Appendix~\ref{suppsec:1qb_rb} for single-qubit Clifford randomized benchmarking data. 

Unlike the CZ gate, we implement actual (not virtual) Z gates using XY gates (see Appendix~\ref{suppsec:euler_z}) to cancel out Z-rotations that are accompanied by the iSWAP gate.
As a consequence, when performing the interleaved RB, the iSWAP-interleaved sequence acquires 0.1125 additional XY gates per Clifford on average (see Appendix~\ref{suppsec:iswap_rb} for details). We extract the two-qubit interaction fidelity $F_{\mathrm{iSWAP}}$ by subtracting the contribution of single-qubit gate error $0.1125 \times r_{\mathrm{1QB}}\approx 0.015\%$ from the error rate $r_{\mathrm{int}}$ of the interleaved gate: $F_{\mathrm{iSWAP}} \equiv 1-r_{\mathrm{iSWAP}} = 1 - (r_{\mathrm{int}} - 0.1125\times r_{\mathrm{1QB}})$.
Owing to the ZZ-cancellation and a short gate length, the measured iSWAP gate exhibits high two-qubit interaction fidelity $F_{\mathrm{iSWAP}}=99.87 \pm 0.23 \%$.
Based on the average gate composition for the two-qubit Cliffords (see Appendix~\ref{suppsec:iswap_rb} for details), we have estimated the two-qubit Clifford error rates $r_{\mathrm{Clifford,est}}$ using the following formula: $r_{\mathrm{Clifford,est}})=10.9375\times r_{\mathrm{1qb}}+1.5\times r_{\mathrm{iSWAP}}$. 
The estimated Clifford error rate is $1.73\pm0.37\%$.
The difference between $r_{\mathrm{Clifford,est}}$ and $r_{\mathrm{Clifford}}$ could be due to residual distortion of the two-qubit gate pulses.
The measured two-qubit interaction fidelity is close to the $T_1$ limit of 99.91\% obtained by solving the Lindblad master equation (see Appendix~\ref{suppsec:T1_contribution_to_gate_errors}).
We confirm that error contribution of $1/f^{\alpha}$ flux noise is relatively insignificant (see Appendices~\ref{suppsec:flux_noise_in_the_device} and \ref{suppsec:Flux_noise_contribution_to_gate_errors} for details).
Again, note that this two-qubit interaction fidelity only quantifies the quality of the native iSWAP gate, which contains additional unwanted single-qubit Z rotations.
In practice, such additional Z-rotations can be compensated by compiling the correctional Z-rotations into adjacent single-qubit gates (see Appendix~\ref{suppsec:iswap_rb} for example).

We note a large uncertainty in an estimate for the two-qubit interaction fidelity $F_{\mathrm{iSWAP}}$. The large uncertainty is mainly due to relatively low single-qubit gate fidelities ($\approx99.86\%$), which degrades the fidelity of reference Clifford sequences~\cite{Epstein2014}. 
These low single-qubit gate fidelities arise from biasing the qubits on resonance to avoid phase swapping, which necessitates microwave crosstalk cancellation. 
Further research should be undertaken to improve the single-qubit gate fidelities in this architecture. 
One alternative is to bias the qubits off-resonantly and correct for the accumulated single qubit phases in software~\cite{Mi2020}. 
Exploring the viability of this approach will be the subjects of our future work.
In addition, applying iterative randomized benchmarking~\cite{Sheldon2016,Ficheux2020} would be useful to better characterize the contributions of systematic coherent errors versus incoherent noise to the two-qubit gates.

Looking forward, our work provides a path towards building quantum information processors that are capable of running near-term quantum applications and ultimately achieving fault tolerant quantum computation.
Our optimal control approaches to suppressing coherent leakage of multi-qubit gates is of particular importance, because leakage error is especially detrimental to the implementation of quantum error correcting codes ~\cite{Aliferis2005,Fowler2013, Suchara2015, Barends2014, Kelly2015,Andersen2020}.
Additionally, the high-fidelity $ZZ$-free iSWAP gate (more generally, $XY$ entangling gates without residual $ZZ$) is beneficial for NISQ applications, since it enables efficient circuit compilation and improves the accuracy of NISQ algorithms such as quantum simulation of molecular eigenstates~\cite{Barkoutsos2018,Ganzhorn2019,Arute2020}, quantum approximate optimization algorithms (QAOA) for solving high-dimensional graph problems~\cite{Abrams2020,Harrigan2021}, and quantum simulation of many-body condensed matter systems (e.g., the 2D $XY$ model)~\cite{Salathe2015,Mi2020,Google2020_Accurate,Wang2001}  

While the residual $ZZ$ of an iSWAP gate can be cancelled by applying an additional CPHASE gate, this inevitably increases the circuit depth, which degrades the performance of (NISQ) algorithms. Alternatively, one can implement error mitigation techniques to alleviate the detrimental effect of residual $ZZ$ on algorithms~\cite{Arute2020}, but this may also introduce overhead, such as additional measurements and classical post-processing, depending on the mitigation protocols. Notably, all these efforts to reduce the impact of residual $ZZ$ of $XY$ entangling gates can be simply avoided by using our $ZZ$ cancellation approach. 

Taken together, the principles and demonstrations shown in this work will help resolve major challenges in the implementation of quantum computing hardware for NISQ-era applications.
%

\section*{Acknowledgement}
It is a pleasure to thank A. Bengtsson, A. Di Paolo, P. Krantz, T. Menke, K. P. O'Brien, B. Royer, and F. Yan for insightful discussion; M. Pulido and C. Watanabe for generous assistance; J. Wang for the optical micrograph of the device. This research was funded in part by the U.S. Army Research Office Grant W911NF-18-1-0411 and the Assistant Secretary of Defense for Research \& Engineering under Air Force Contract No. FA8721-05-C-0002. Y.S. gratefully acknowledges support from the Korea Foundation for Advanced Studies. L.D. gratefully acknowledges support from the IBM PhD Fellowship. B.K. gratefully acknowledges support from the National Defense Science and Engineering Graduate Fellowship program. The views and conclusions contained herein are those of the authors and should not be interpreted as necessarily representing the official policies or endorsements, either expressed or implied, of the U.S. Government.

\section*{Data availability}
The data that support the findings of this study may be made available from the corresponding authors upon request and with the permission of the US Government sponsors who funded the work.

\section*{Author contribution}
Y.S. and L.D. performed the experiments and analyzed the data. Y.S. and L.D. developed the theoretical framework with constructive feedback from A.V., B.K., and W.D.O.. Y.S. carried out numerical simulations. Y.S. and J.B. designed the device and D.K., A.M., B.M.N., and J.L.Y. fabricated it. J.B., B.K., M.K., A.G., G.O.S., C.M., and M.E.S. assisted with the experimental setup. T.P.O, S.G., and W.D.O. supervised the project. All authors contributed to the discussion of the results and the manuscript. 

%
\appendix

\section{Measurement setup}
\label{suppsec:measurement_setup}
The experiments were performed in a BlueFors XLD-600 dilution refrigerator with a base temperature of $\SI{10}{\milli\kelvin}$. We magnetically shielded the device with a superconducting can surrounded by a Cryoperm-10 cylinder. All attenuators in the cryogenic samples are made by XMA and installed to remove excess thermal photons from higher-temperature stages. We apply microwave readout tones to measure the transmission of the device. We pump the Josephson travelling wave parametric amplifier (JTWPA)~\cite{Macklin2015} using an RF source (Berkeley Nucleonics Corp Model 845), in order to pre-amplify the readout signal at base temperature. A microwave isolator placed after the sample allows for the signal to pass through to the JTWPA without being attenuated, while removing all the reflected noise of the JTWPA and dumping it in a $\SI{50}{\ohm}$ termination. Two microwave isolators are placed after the JTWPA to prevent noise from higher-temperature stages to the JTWPA and the sample. We amplify the signal by using a high-electron mobility transistor (HEMT) amplifier, which is thermally connected to the $\SI{4}{\kelvin}$ stage. The output line is further amplified outside of the cryostat with an amplifier (MITEQ AMF-5D-00101200-23-10P) with a quoted noise figure of 2.3 dB, and a preamplifier (Stanford Research SR445A). 

\begin{figure}[ht!]
    \centering
    \includegraphics[width=8.9cm]{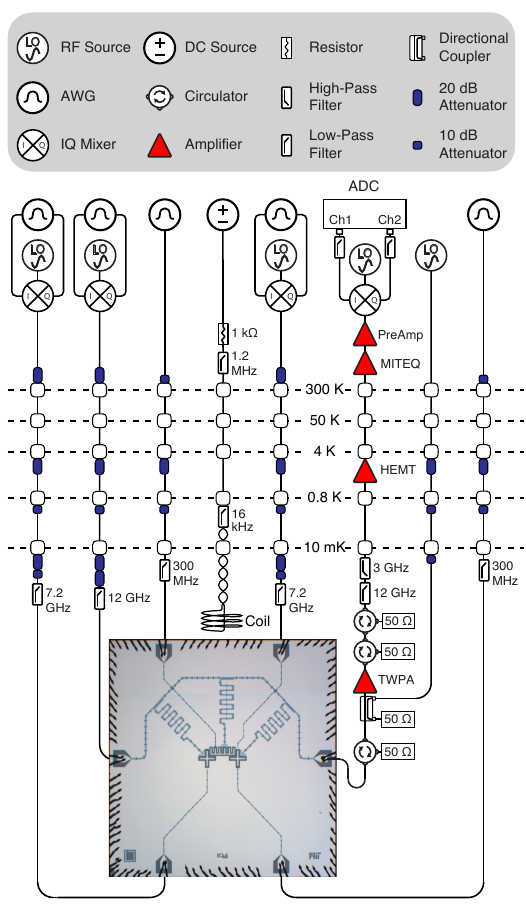}
    \caption{ 
    A schematic diagram of the experimental setup.
    }
    \label{suppfig:measurement_setup}
\end{figure}

Outside of the cryostat, we have all of the control electronics which allow us to apply signals used for the XY and Z controls of the qubits and the coupler. Pulse envelopes of XY control signals and readout signals are programmed in Labber software and then uploaded to arbitrary waveform generators (AWG Keysight M3202A). Subsequently, the pulses generated by AWGs are mixed with coherent tones from RF sources (Rohde and Schwarz SGS100A). Z control signals are generated by AWGs (AWG Keysight M3202A). We also apply magnetic flux globally through a coil installed in the device package as an additional DC flux bias source (Yokogawa GS200). All components for generating signals are frequency-locked by a common SRS rubidium frequency standard (10 MHz). A detailed schematic is given in Fig.~\ref{suppfig:measurement_setup}. 
\newpage

\section{Device setup}
\label{suppsec:device_setup}

\begin{figure}[H]
    \centering
    \includegraphics[width=8.9cm]{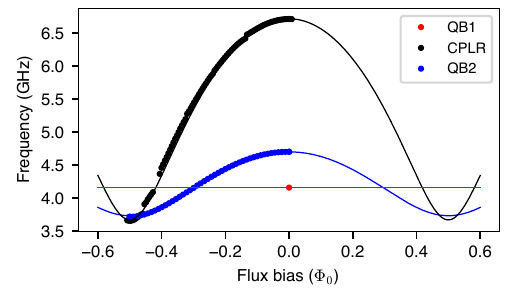}
    \caption{ 
    $|0\rangle\rightarrow|1\rangle$ transition frequencies of the qubits (red and blue) and the coupler (black). Circles correspond to experimental data. Solid curves correspond to simulations based on the fitted circuit parameters: QB1 ($E_{\mathrm{J}}/h=\SI{12.2}{\giga\hertz}$, $E_{\mathrm{c}}/h=\SI{0.195}{\giga\hertz}$), CPLR ($E_{\mathrm{J}}^{1}/h=\SI{46}{\giga\hertz}$, $E_{\mathrm{J}}^{2}/h=\SI{25}{\giga\hertz}$, $E_{\mathrm{c}}/h=\SI{0.085}{\giga\hertz}$), and QB2 ($E_{\mathrm{J}}^{1}/h=\SI{13}{\giga\hertz}$, $E_{\mathrm{J}}^{2}/h=\SI{2.8}{\giga\hertz}$, $E_{\mathrm{c}}/h=\SI{0.19}{\giga\hertz}$), where $E_{\mathrm{J}}$ and $E_{\mathrm{c}}$ denote the corresponding Josephson energy and the charging energy, respectively~\cite{Koch2007}.
    }
    \label{suppfig:qubit_spectroscopy}
\end{figure}

The device parameters are summarized in Table~\ref{supptable:device_params}. The $|0\rangle \rightarrow |1\rangle$ transition frequencies of the qubits and the coupler as a function of flux bias are shown in Fig.~\ref{suppfig:qubit_spectroscopy}. We note that the maximum frequencies have decreased 1--2 MHz in each cool-down, due to device aging. 

The coupling strengths between the qubit and coupler $g_{\mathrm{1c}}/2\pi,g_{\mathrm{2c}}/2\pi$ are approximately \SI{72}{\mega\hertz}. Note that further increasing $g_{\mathrm{1c}}$, $g_{\mathrm{2c}}$ would enable faster gates with fewer non-adiabatic effects. However, solely increasing $g_{\mathrm{1c}}$, $g_{\mathrm{2c}}$ would result in an increase of the idling coupler frequency $\omega_{\mathrm{c,idle}}$ at which the net qubit-qubit coupling is nearly zero. There are practical considerations that place an upper limit on $\omega_{\mathrm{c,idle}}$. For example, other modes in the system such as readout resonators or/and spurious package modes should not be within the operating frequency range of the coupler.  Alternatively, one could also increase the direct qubit-qubit coupling $g_{12}$ to compensate for the increased $g_{\mathrm{1c}}$, $g_{\mathrm{2c}}$ such that $\omega_{\mathrm{c,idle}}$ remains at a lower frequency and is within the ideal operating range (no other modes fall within this range). However, increasing $g_{12}$ can be potentially problematic when scaling up due to strong parasitic $ZZ$ coupling between next-nearest neighboring qubits. Given these constraints, we have chosen the coupling strengths that enable fast two-qubit gates (30--\SI{60}{\nano\second}) via our optimized control techniques and avoid unwanted resonances between a coupler and other modes on the chip.

\begin{figure}[H]
    \centering
    \includegraphics[width=8.9cm]{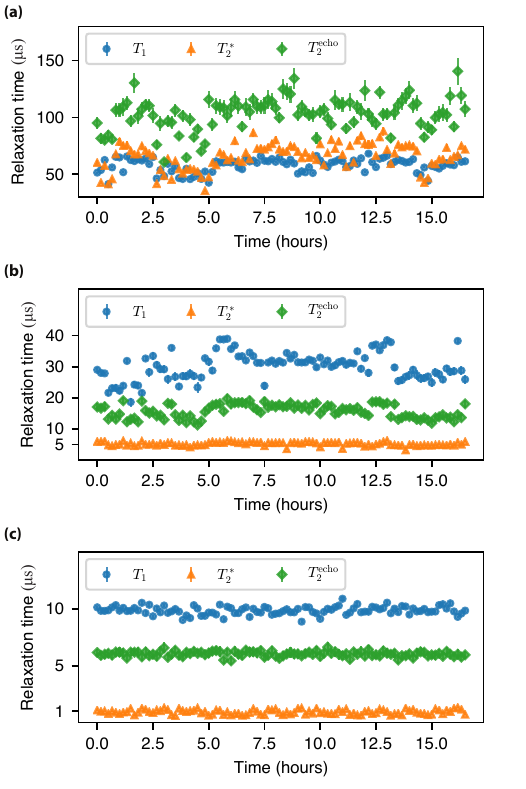}
    \caption{ 
    Coherence times of QB1 (\textbf{a}), QB2 (\textbf{b}), and CPLR as a function of time.
    }
    \label{suppfig:qubit_coherence}
\end{figure}

We measure coherence times of QB1, QB2, and CPLR at the idling configuration ($\omega_{1}/2\pi=\SI{4.16}{\giga\hertz}$, $\omega_{2}/2\pi=\SI{4.00}{\giga\hertz}$, $\omega_{\mathrm{c}}/2\pi=\SI{5.45}{\giga\hertz}$) for 16.5 hours (Fig.~\ref{suppfig:qubit_coherence}). At the idling configuration, we bias QB2 away from its flux-insensitive point (commonly referred to as a ``sweet spot'') in order to avoid two-level-systems (TLSs)~\cite{Klimov2018, Muller2019} during two-qubit gate operations. 

\begin{figure}[H]
    \centering
    \includegraphics[width=8.9cm]{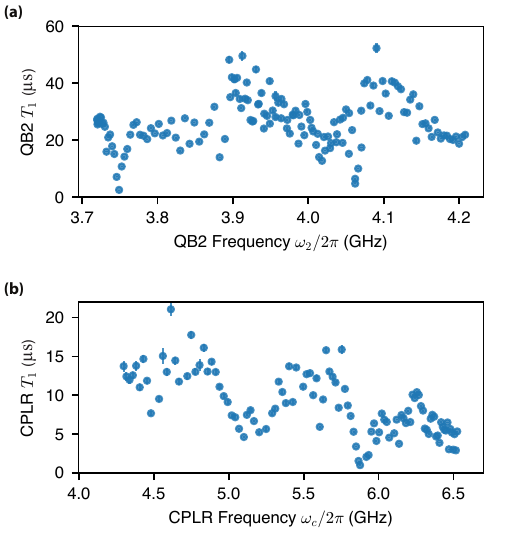}
    \caption{ 
    \textbf{(a)} Energy relaxation times of QB2 as a function of its frequency $\omega_2$. Its $T_1$ drops at $\omega_2/2\pi=$ \SI{3.75}{\giga\hertz} and \SI{4.07}{\giga\hertz} due to TLSs. \textbf{(b)} Energy relaxation times of CPLR as a function of its frequency $\omega_{\mathrm{c}}$. Its $T_1$ drops at $\omega_2/2\pi=$ \SI{5.1}{\giga\hertz} and \SI{5.9}{\giga\hertz} due to TLSs. 
    }
    \label{suppfig:qubit_coherence_vs_freq}
\end{figure}

\begin{figure}[H]
    \centering
    \includegraphics[width=8.9cm]{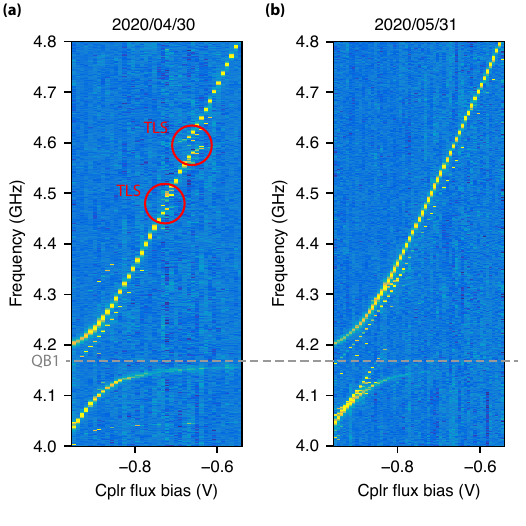}
    \caption{ 
\textbf{(a)} Experimental data of coupler spectroscopy measurement when TLSs appeared in the operating frequency range of CPLR. \textbf{(b)} We displaced the TLSs from the operating frequency range of the coupler by thermal cycling to room temperature.
    }
    \label{suppfig:tls_in_the_coupler}
\end{figure}

We also measure $T_1$ of QB2 and CPLR as functions of their frequencies $\omega_{\mathrm{2}}$ and $\omega_{\mathrm{c}}$ (Fig.~\ref{suppfig:qubit_coherence_vs_freq}). We find TLSs in both QB2 and coupler, but they are located out of the operating frequency ranges, so that they negligibly affect the performance of two-qubit gates. However we note that the TLS landscape varied between cool-downs, occasionally causing TLSs to jump into the operating range for CPLR. We observed degradation of the two qubit gate fidelities (below 99\%), when TLSs are strongly coupled to the coupler in its operating frequency range (Fig.~\ref{suppfig:tls_in_the_coupler}a).

In these experiments, we have used relatively large area Josephson junctions (1--$\SI{3}{\micro\meter}\times \SI{200}{\micro\meter}$). This is to: (1) achieve a much higher coupler frequency than the qubit frequencies while using only a single e-beam layer, and (2) use asymmetric SQUIDs, which are advantageous for their lower flux sensitivity. When only using a single e-beam layer and, therefore, a single critical current density, the need for both large and small $E_{\mathrm{J}}$ results in certain junctions being necessarily large in area.
These large junctions ultimately lead to an increase in the number of TLSs~\cite{Martinis2005, Oliver2013}. To mitigate this issue, one can employ multiple e-beam layers with different critical current densities, such that both large and small $E_{\mathrm{J}}$ values can be fabricated using only small-area Josepshon junctions  ($\approx\SI{200}{\nano\meter}\times\SI{200}{\nano\meter}$). These smaller junctions reduce the probability of strongly coupled TLSs appearing in the operating regime. This approach will be implemented in future work.

\vfill
\begin{table}[H]
\begin{ruledtabular}
\begin{threeparttable}
\begin{tabular}{c | c c c} 
\hspace*{4cm} & QB1 & CPLR & QB2 \\\hline  \\[-0.3cm]
$\omega/2\pi$\tnote{a}~ (GHz)&  4.16 & 5.45 & 4.00 \\ 
$\eta/2\pi$\tnote{b}~ (MHz)&  -220 & -90 & -210 \\ 
$g_{\mathrm{1c}}/2\pi$\tnote{c}~ (MHz)& \multicolumn{2}{c}{72.5} & \\ 
$g_{\mathrm{2c}}/2\pi$\tnote{c}~ (MHz)& & \multicolumn{2}{c}{71.5} \\ 
$g_{\mathrm{12}}/2\pi$\tnote{c}~ (MHz)& \multicolumn{3}{c}{5.0} \\ 
$T_1$\tnote{d}~ (\SI{}{\micro\second}) & 60 & 10 & 30  \\ 
$T_2^*$\tnote{d}~ (\SI{}{\micro\second})& 66 & 1  & 5   \\ 
$T_2^{\mathrm{echo}}$\tnote{d}~ (\SI{}{\micro\second})& 103 & 6 & 16\\ 
 $\omega_{\mathrm{r}}/2\pi$\tnote{e}~ (GHz) & 7.12 & 7.17 & 7.07\\ 
 $\kappa_{\mathrm{r}}/2\pi$\tnote{f}~ (MHz) & 0.5 & 0.5 & 0.5\\ 
 $\chi_{\mathrm{r}}^{(0,1)}/2\pi$\tnote{g}~ (kHz) & 170 & 392 & 140 \\ 
 $\chi_{\mathrm{r}}^{(1,2)}/2\pi$\tnote{h}~ (kHz) & 182 & 313 & 141\\ 
\end{tabular}

\footnotesize
\begin{tablenotes}
\item[a] $|0\rangle \rightarrow |1\rangle$ transition frequencies at the idling configuration.
\item[b] Anharmonicities at the idling configuration.
\item[c] Pairwise coupling strengths at $\omega_{1}/2\pi=\omega_{2}/2\pi=\omega_{\mathrm{c}}/2\pi= \SI{4.16}{\giga\hertz}$. Note that $g_{\mathrm{1c}}$, $g_{\mathrm{2c}}$, and $g_{\mathrm{12}}$ depend on $\omega_{1}$, $\omega_{2}$, and $\omega_{\mathrm{c}}$~\cite{Yan2018}.
\item[d] Energy decay time ($T_1$), Ramsey decay time ($T_2^*$), and spin-echo decay time ($T_2^{\mathrm{echo}}$) measured at the idling configuration.
\item[e] Readout resonator frequency.
\item[f] Readout resonator linewidth.
\item[g] Effective dispersive shifts for the $|0\rangle \rightarrow |1\rangle$ transition due to the interaction with the readout cavity mode.
\item[h] Effective dispersive shifts for the $|1\rangle \rightarrow |2\rangle$ transition due to the interaction with the readout cavity mode.
\end{tablenotes}
\caption{Device parameters.} 
\label{supptable:device_params}
\end{threeparttable}
\end{ruledtabular}
\end{table}

\section{$1/f^{\alpha}$ flux noise in the device}
\label{suppsec:flux_noise_in_the_device}
We characterize $1/f^{\alpha}$ flux noise which predominantly limits the dephasing times $T_2^{*}$, $T_2^{\mathrm{echo}}$ of QB2 and CPLR at the idling configuration. Following Ref~\cite{Yoshihara2006}, we estimate the power spectral densities $S_{\Phi}^{\mathrm{echo}}(\omega) = \frac{A^{\mathrm{echo}}_{\mathrm{\Phi}}}{\omega}$ of $1/f$ flux noise from spin-echo measurement data. The flux noise amplitude $\sqrt{A^{\mathrm{echo}}_{\Phi}}$ is calculated from the following equation, $\sqrt{A^{\mathrm{echo}}_{\Phi}} = \frac{\Gamma_{\phi}^{\mathrm{echo}}}{\sqrt{\ln{2}}} \left(\frac{\partial\omega}{\partial\Phi}\right)^{-1}$,
where $\Gamma_{\phi}^{\mathrm{echo}}$ is a Gaussian pure dephasing rate and $\frac{1}{2\pi}\frac{\partial\omega}{\partial\Phi}$ is a flux sensitivity of the $|0\rangle\rightarrow |1\rangle$ transition frequency. To compute $\Gamma_{\phi}^{\mathrm{echo}}$ , we perform a fit to the decay curve $f(t)\propto \exp{(-t/(2T_1^{\mathrm{exp}})-(\Gamma_{\phi}^{\mathrm{echo}}t)^2}$), where $T_1^{\mathrm{exp}}$ is the energy relaxation time measured in a preceding $T_1$ experiment. 
Table~\ref{supptab:spin_echo_flux_noise} presents the power spectral densities at 1Hz for QB2 and CPLR. We find that these values are comparable with the numbers in the literature~\cite{Yoshihara2006, Braumuller2020}. 

\begin{table}[H]
\begin{ruledtabular}
         \begin{tabular}{c | c c c}
             & $\Gamma_{\phi}^{\mathrm{echo}}$ ($10^{3}\mathrm{s}^{-1}$) & $\frac{1}{2\pi}\frac{\partial\omega}{\partial\Phi}$ $\mathrm{(GHz/\Phi_0})$ & $S_{\Phi}^{\mathrm{echo}}(f=1\mathrm{Hz})$ \\
            \hline \\[-0.3cm]
            QB2  & $55\pm9$ & $2.92$ & $(\SI{1.4}{\micro\Phi_0})^2/ \mathrm{Hz}$ \\
            CPLR & $118\pm4$ & $8.73$ & $(\SI{1.0}{\micro\Phi_0})^2/ \mathrm{Hz}$  \\
        \end{tabular}
    \caption{Power spectral densities $S_{\Phi}^{\mathrm{echo}} (f)$ at 1Hz of $1/f$ flux noise estimated from echo experiments.}
    \label{supptab:spin_echo_flux_noise}
\end{ruledtabular}
\end{table}

\begin{figure}[H]
    \centering
    \includegraphics[width=8.9cm]{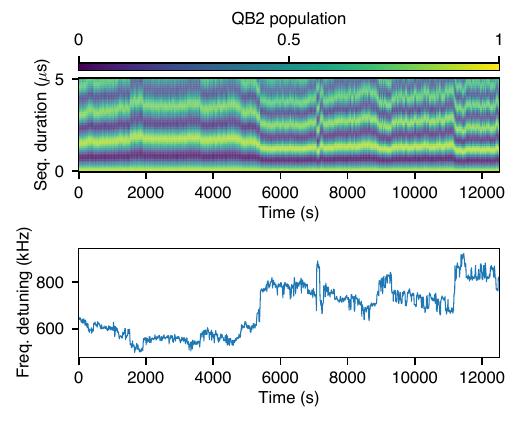}
    \caption{ 
 Fluctuation in the QB2 frequency as a function of time, measured via repeated Ramsey experiments. We compute the PSD of the frequency fluctuation by using the ‘scipy.signal.welch’ function from the open-source Python library SciPy.}
    \label{suppfig:repeated_ramsey_experiment_QB2}
\end{figure}

\begin{figure}[H]
    \centering
    \includegraphics[width=8.9cm]{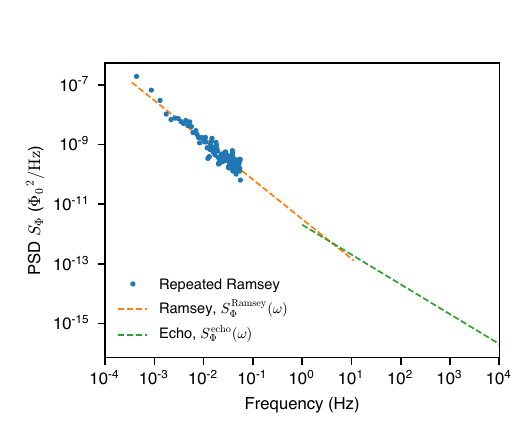}
    \caption{ 
 The estimated flux noise PSD affecting QB2 from repeated Ramsey measurements (blue circles and the corresponding fit: orange dashed curve) and spin-echo experiments (green dashed curve).}
    \label{suppfig:flux_noise_PSD_QB2}
\end{figure}

We compute the noise power spectral density at long time-scales ($5\times10^{-4}$--$10^{-1}\SI{}{\hertz}$) from repeated Ramsey measurements~\cite{Sank2012,Yan2012} (see Figs.~\ref{suppfig:repeated_ramsey_experiment_QB2} and ~\ref{suppfig:flux_noise_PSD_QB2}). The estimated power spectral density shows a $1/f^{\alpha}$ dependence and is fitted to the following equation (an orange dashed curve in Fig.~\ref{suppfig:flux_noise_PSD_QB2}): $S_{\Phi}^{\mathrm{Ramsey}}(\omega)=\frac{A_{\Phi}^{\mathrm{Ramsey}}}{\omega^{\alpha}}$, where the exponent $\alpha\approx 1.33$ and $A_{\Phi}^{\mathrm{Ramsey}}\approx (\SI{6}{\micro \Phi_0})^2\times(\SI{}{Hz})^{(\alpha-1)}$. We find that the corresponding power spectral density at 1Hz 
$S_{\Phi}^{\mathrm{Ramsey}}(\omega/2\pi=\SI{1}{\hertz})\approx (\SI{1.77}{\micro\Phi_0})^2/\mathrm{Hz}$.

\section{State readout}
\label{suppsec:state_readout}
We probe quantum states of QB1, QB2, and CPLR via the dispersive readout scheme~\cite{Blais2004}. We drive the readout resonators by applying a square-shaped \SI{3}{\micro\second}-long microwave pulse. We discriminate the three states -- ground state \ketbare{0}, the first excited state \ketbare{1}, and the second excited state \ketbare{2} -- for QB1, QB2, and CPLR. 
The first excited state $|1\rangle$ is prepared by applying a \SI{250}{\nano\second}-long $\pi$ pulse that drives the $|0\rangle \rightarrow |1\rangle$ transition. The second excited state $|2\rangle$ is prepared by applying two consecutive \SI{250}{\nano\second}-long $\pi$ pulses, which drive the $|0\rangle \rightarrow |1\rangle$ and $|1\rangle \rightarrow |2\rangle$ transitions respectively.

\begin{figure}[htbp]
    \centering
    \includegraphics[width=8.9cm]{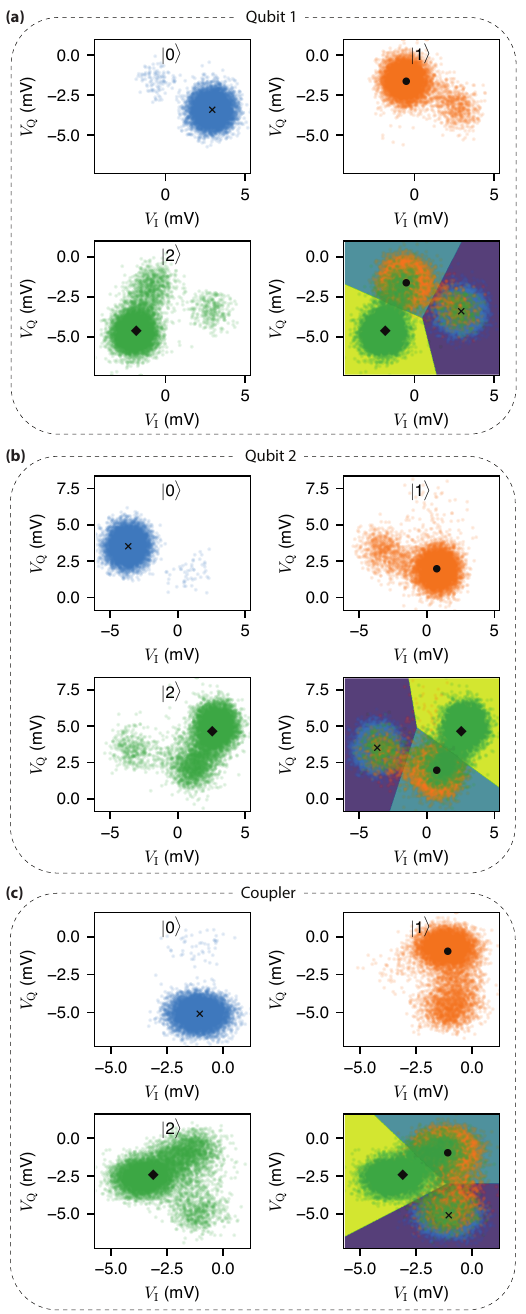}
    \caption{ 
    Single-shot measurements in the I-Q plane for the qubits (\textbf{a}-\textbf{b}) and the coupler \textbf{(c)}. For the same color points, we repetitively prepare at the corresponding state and measure the I-Q outcomes (the number of repetitions: 10,000). Black markers denote the median points.
    }
    \label{suppfig:state_readout}
\end{figure}

For state discrimination, we use a linear support vector machine, which finds hyper-planes separating the I-Q data into three parts which corresponding to \ketbare{0}, \ketbare{1} and \ketbare{2}, respectively (Fig.~\ref{suppfig:state_readout})~\cite{Magesan2015}. We characterize the readout performance by computing assignment probability matrices. See Table~\ref{supptab:readout_assignment_prob} for the assignment probability matrices. Note that the assignment probabilities include state preparation errors.

\begin{table}[H]
\begin{ruledtabular}
        \begin{tabular}{ll|lll}
            \multicolumn{2}{c|}{Qubit 1} & \multicolumn{3}{c}{Prepared state, $n$}\\
           \multicolumn{2}{c|}{$P_{\mathrm{1}}(m|n)$ } & \multicolumn{1}{l}{\ketbare{0}}  & \multicolumn{1}{l}{\ketbare{1}}  & \multicolumn{1}{l}{$|2\rangle$}  \\
            \hline \multirow{3}{*}{Assigned state, $m$} & \multicolumn{1}{l|}{\ketbare{0}} & \multicolumn{1}{l}{0.9885} & \multicolumn{1}{l}{0.0673} & \multicolumn{1}{l}{0.0304}  \\
                                 & \multicolumn{1}{l|}{\ketbare{1}} & \multicolumn{1}{l}{0.0115} & \multicolumn{1}{l}{0.9266} & \multicolumn{1}{l}{0.0907}\\
                                 & \multicolumn{1}{l|}{\ketbare{2}} & \multicolumn{1}{l}{0} & \multicolumn{1}{l}{0.0061} & \multicolumn{1}{l}{0.8789}\\[0.08cm]
            \hline \hline\\[-0.3cm]
            \multicolumn{2}{c|}{Qubit 2}& \multicolumn{3}{c}{Prepared state, $n$}\\
            \multicolumn{2}{c|}{$P_{\mathrm{2}}(m|n)$}& \multicolumn{1}{l}{\ketbare{0}}  & \multicolumn{1}{l}{\ketbare{1}}  & \multicolumn{1}{l}{\ketbare{2}}  \\
            \hline
            \multirow{3}{*}{Assigned state, $m$} & \multicolumn{1}{l|}{\ketbare{0}} & \multicolumn{1}{l}{0.9946} & \multicolumn{1}{l}{0.0772} & \multicolumn{1}{l}{0.0385}  \\
                                 & \multicolumn{1}{l|}{\ketbare{1}} & \multicolumn{1}{l}{0.0053} & \multicolumn{1}{l}{0.905} & \multicolumn{1}{l}{0.1734}\\
                                 & \multicolumn{1}{l|}{\ketbare{2}} & \multicolumn{1}{l}{0.0001} & \multicolumn{1}{l}{0.0178} & \multicolumn{1}{l}{0.7881}\\[0.08cm]
            \hline \hline \\[-0.3cm]
            \multicolumn{2}{c|}{Coupler}& \multicolumn{3}{c}{Prepared state, $n$}\\
            \multicolumn{2}{c|}{$P_{\mathrm{c}}(m|n)$}& \multicolumn{1}{l}{\ketbare{0}}  & \multicolumn{1}{l}{\ketbare{1}}  & \multicolumn{1}{l}{\ketbare{2}}  \\
            \hline
            \multirow{3}{*}{Assigned state, $m$} & \multicolumn{1}{l|}{\ketbare{0}} & \multicolumn{1}{l}{0.9915} & \multicolumn{1}{l}{0.1918} & \multicolumn{1}{l}{0.0741}  \\
                                 & \multicolumn{1}{l|}{\ketbare{1}} & \multicolumn{1}{l}{0.0052} & \multicolumn{1}{l}{0.7796} & \multicolumn{1}{l}{0.1891}\\
                                 & \multicolumn{1}{l|}{\ketbare{2}} & \multicolumn{1}{l}{0.0033} & \multicolumn{1}{l}{0.0286} & \multicolumn{1}{l}{0.7368}\\
        \end{tabular}
    \caption{Assignment probability matrices $P(m|n)_i$ ($i\in\{1,2,\mathrm{c}\}$) for the state readout of QB1, QB2, and CPLR.}
    \label{supptab:readout_assignment_prob}
\end{ruledtabular}
\end{table}

\newpage
\section{Static $ZZ$ interaction in the dispersive limit}
In this section, we present experimental data and perturbative calculations of the static $ZZ$ interaction $\zeta$ as a function of CPLR's frequency $\omega_{\mathrm{c}}$. For the perturbative analysis, we assume that QB1, QB2, and CPLR are dispersively coupled to each other $g_{ij}/|\omega_i-\omega_j| \ll 1$ ($i,j\in \{1,2,\mathrm{c} \}$, $i<j$).

\begin{figure}[H]
    \centering
    \includegraphics[width=8.9cm]{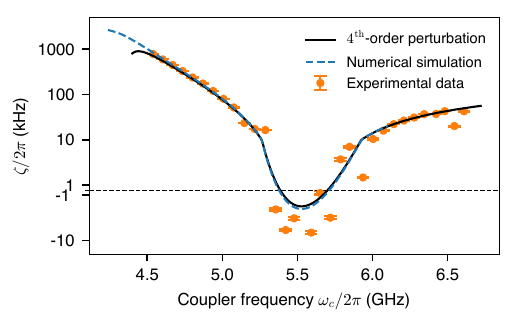}
    \caption{ 
    Static $ZZ$ interaction strength $\zeta$ as a function of the coupler frequency $\omega_{\mathrm{c}}$. QB1 and QB2 are biased at $\omega_1/2\pi=\SI{4.16}{\giga\hertz}$ and $\omega_2/2\pi=\SI{4.00}{\giga\hertz}$. The solid black curve corresponds to $\zeta$ obtained by the perturbation theory up to the fourth order without rotating wave approximation (Eq.~\eqref{suppeq:zeta_perturbative_analysis}).
    The blue dashed curve corresponds to $\zeta$ obtained by numerically diagonalizing the system Hamiltonian  (Eq.~\eqref{eq:H_lab}, see Appendix~\ref{suppsec:numerical_simulation} for the parameters used).}
    \label{suppfig:zz_vs_wc_idle}
\end{figure}

Fig.~\ref{suppfig:zz_vs_wc_idle} shows experimental data of $ZZ$ interaction strength $\zeta$ as a function of $\omega_{\mathrm{c}}$. 
In this measurement, we bias the frequencies of QB1 and QB2 at 4.16 GHz and 4.00 GHz, respectively. 
We measure $\zeta$ via a cross-Ramsey type experiment which measures the QB1 frequency while initializing QB2 in either its ground or excited state.
The static $ZZ$ interaction is nearly eliminated when $\omega_{\mathrm{c}}/2\pi=\SI{5.45}{\giga\hertz}$. At this bias point (the idling configuration), we perform single qubit gates in the CZ gate benchmarking experiments (Fig.~\ref{fig:Fig4} in the main text).

\label{suppsec:static_ZZ}

Following Ref.~\cite{Zhu2013}, we use perturbation theory to calculate theoretical values of $\zeta$ up to the fourth order according to the following formula:

\begin{align}
    \zeta = \left(E_{|101\rangle}-E_{|001\rangle}\right)-\left(E_{|100\rangle}-E_{|000\rangle}\right), 
\end{align}
where $E_{|m\rangle}$ denotes the eigenenergy of the eigenstate $|m\rangle \in \{|000\rangle, |100\rangle, |001\rangle, |101\rangle\}$.
Specifically, we calculate the $ZZ$ contributions of the $n$-th order perturbations $\zeta^{(n)}$ by computing the $n$-th order corrections $E^{(n)}_{|m\rangle}$ to the eigenenergies of $|m\rangle$ ($n \in \{ 2,3,4 \}$, $|m\rangle \in \{ |101\rangle, |001\rangle, |100\rangle, |000\rangle$) as follows:
\begin{align}
    \zeta^{(n)} = \left(E_{|101\rangle}^{(n)}-E_{|001\rangle}^{(n)}\right)-\left(E_{|100\rangle}^{(n)}-E_{|000\rangle}^{(n)}\right)
\end{align}
The $ZZ$ contributions from the $n$-th order perturbation terms can be split into the rapid counter-rotating-wave terms $\zeta^{(n)}_{\mathrm{CRW}}$ and the slow rotating-wave terms $\zeta^{(n)}_{\mathrm{RW}}$. 
In general, the rapid oscillating terms are neglected by applying the rotating wave approximation~\cite{Li2020}.
However, in our case, we note that the fast-oscillating terms considerably contribute to the static $ZZ$ interaction.

The total $ZZ$ contribution up to the fourth order perturbation is given as
\begin{align}
    \label{suppeq:zeta_perturbative_analysis}
    \zeta = & \zeta^{(2)}_{\mathrm{RW}} + \zeta^{(3)}_{\mathrm{RW}} + \zeta^{(4)}_{\mathrm{RW}} + \zeta^{(2)}_{\mathrm{CRW}} + \zeta^{(3)}_{\mathrm{CRW}} \nonumber \\ &+ \zeta^{(4)}_{\mathrm{CRW}}.
\end{align}
For brevity's sake, we introduce the following notations.
\begin{align}
    \Delta_{ij}\equiv\omega_i-\omega_j, \\ \Sigma_{ij}\equiv\omega_i+\omega_j,
\end{align} where $i,j\in \{1,2,\mathrm{c}\}$.
The $ZZ$ contributions from the $n$-th order perturbation terms are calculated as follows.
\begin{flalign}\label{suppeq:zeta_2_RWA}
    &\zeta^{(2)}_{\mathrm{RW}} = g_{12}^2 \left(\frac{2}{\Delta_{12}-\eta_2} + \frac{2}{\Delta_{21}-\eta_1}\right). &
\end{flalign}

\begin{flalign}\label{suppeq:zeta_3_RWA}
    \zeta^{(3)}_{\mathrm{RW}} = & g_{\mathrm{1c}}g_{\mathrm{2c}}g_{12} \Bigg(\frac{4}{(\Delta_{12}-\eta_2)\Delta_{\mathrm{1c}}} + \frac{4}{(\Delta_{21}-\eta_1)\Delta_{\mathrm{2c}}} & \nonumber \\
    & +\frac{2}{\Delta_{1\mathrm{c}}\Delta_{2\mathrm{c}}} - \frac{2}{\Delta_{1\mathrm{2}}\Delta_{1\mathrm{c}}}   -\frac{2}{\Delta_{2\mathrm{1}}\Delta_{2\mathrm{c}}}\Bigg). &
\end{flalign}
For the fourth-order slow rotating-wave terms, we omit smaller contributing terms containing $g_{12}$. 
\begin{flalign}\label{suppeq:zeta_4_RWA}
    \zeta^{(4)}_{\mathrm{RW}} = & g_{\mathrm{1c}}^2 g_{\mathrm{2c}}^2 \Bigg( 2 \left( \frac{1}{\Delta_{\mathrm{1c}}}+\frac{1}{\Delta_{\mathrm{2c}}} \right)^2 \frac{1}{\Delta_{\mathrm{1c}}+\Delta_{\mathrm{2c}} - \eta_{\mathrm{c}}} & \nonumber \\
    & + \frac{2}{\Delta_{\mathrm{2c}}^2(\Delta_{\mathrm{21}}-\eta_1)} + \frac{2}{\Delta_{\mathrm{1c}}^2(\Delta_{\mathrm{12}}-\eta_2)} & \nonumber \\
    & - \left( \frac{1}{\Delta_{\mathrm{2c}}} + \frac{1}{\Delta_{12}} \right) \frac{1}{\Delta_{\mathrm{1c}}^2} - \left( \frac{1}{\Delta_{\mathrm{1c}}} + \frac{1}{\Delta_{21}} \right) \frac{1}{\Delta_{\mathrm{2c}}^2} \Bigg). &
\end{flalign}
\begin{flalign}\label{suppeq:zeta_2_cRW}
    & \zeta^{(2)}_{\mathrm{CRW}} = g_{12}^2 \left(\frac{-4}{\Sigma_{12}+\eta_1+\eta_2} + \frac{2}{\Sigma_{12}+\eta_1} + \frac{2}{\Sigma_{12}+\eta_2} \right). &
\end{flalign}
Since $\zeta^{(3)}_{\mathrm{CRW}}$ and $\zeta^{(4)}_{\mathrm{CRW}}$ are expressed by a large number of terms, for the sake of clarity, we instead write the corresponding eigenenergy corrections $E^{(n)}_{|m\rangle}$ as follows. Namely $\zeta^{(3)}_{\mathrm{CRW}}$ is given by
\begingroup
\allowdisplaybreaks
\begin{flalign}\label{suppeq:zeta_3_cRW}
    &\zeta^{(3)}_{\mathrm{CRW}} = \left(E^{(3)}_{|101\rangle}-E^{(3)}_{|001\rangle}\right)-\left(E^{(3)}_{|100\rangle}-E^{(3)}_{|000\rangle}\right), &
   \end{flalign}
\endgroup 
where the eigenenergy corrections $E^{(n)}_{|m\rangle}$ are given as
\begingroup
\allowdisplaybreaks
\begin{flalign}\label{suppeq:E_101^3_cRW}
    E^{(3)}_{|101\rangle}=& g_{1\mathrm{c}} g_{2\mathrm{c}} g_{12} \Bigg( 
  \frac{8}{(\Sigma_{\mathrm{1c}} + \eta_1)(\Sigma_{\mathrm{12}} + \eta_1 + \eta_2)} & \nonumber \\ 
    & + \frac{8}{(\Sigma_{\mathrm{1c}} + \eta_1)(\Sigma_{\mathrm{2c}} + \eta_2)} & \nonumber \\
      & + \frac{8}{(\Sigma_{\mathrm{2c}} + \eta_2)(\Sigma_{\mathrm{12}} + \eta_1 + \eta_2)} & \nonumber \\
  & - \frac{4}{\Delta_{\mathrm{2c}}(\Sigma_{\mathrm{1c}} + \eta_1)} & \nonumber \\  
     & - \frac{4}{(\Delta_{12} - \eta_2)(\Sigma_{\mathrm{2c}} + \eta_2)}  & \nonumber \\  
  & - \frac{4}{(\Delta_{\mathrm{21}} - \eta_1)(\Sigma_{\mathrm{1c}} + \eta_1)} &\nonumber \\
  & - \frac{4}{\Delta_{\mathrm{1c}}(\Sigma_{\mathrm{2c}} + \eta_2)} + \frac{2}{\Delta_{\mathrm{1c}} \Sigma_{\mathrm{12}}} + \frac{2}{\Delta_{\mathrm{2c}}\Sigma_{\mathrm{12}}}
\Bigg), &
\end{flalign}
\endgroup
\begingroup
\allowdisplaybreaks
\begin{flalign}\label{suppeq:E_100^3_cRW}
  E_{|100\rangle}^{(3)} = &  g_{\mathrm{1c}} g_{\mathrm{2c}} g_{12} \Bigg(\frac{4}{(\Sigma_{\mathrm{1c}} + \eta_1)(\Sigma_{\mathrm{12}} + \eta_1)} & \nonumber \\
  & + \frac{4}{(\Sigma_{\mathrm{12}} + \eta_1)\Sigma_{\mathrm{2c}}} + \frac{4}{(\Sigma_{\mathrm{1c}} + \eta_1)\Sigma_{\mathrm{2c}}} & \nonumber \\
  & - \frac{2}{\Delta_{\mathrm{1c}}\Sigma_{\mathrm{2c}} } - \frac{2}{\Delta_{12}\Sigma_{\mathrm{2c}}}  \Bigg), &
\end{flalign}
\endgroup
\begingroup
\allowdisplaybreaks
\begin{flalign}\label{suppeq:E_001^3_cRW}
  E_{|001\rangle}^{(3)} = &  g_{\mathrm{1c}} g_{\mathrm{2c}} g_{12}  \Bigg(\frac{4}{(\Sigma_{\mathrm{2c}} + \eta_2)(\Sigma_{\mathrm{12}} + \eta_2)} & \nonumber \\
  & + \frac{4}{(\Sigma_{\mathrm{12}} + \eta_2)\Sigma_{\mathrm{1c}}} + \frac{4}{(\Sigma_{\mathrm{2c}} + \eta_2)\Sigma_{\mathrm{1c}}} & \nonumber \\
  & - \frac{2}{\Delta_{\mathrm{2c}}\Sigma_{\mathrm{1c}} } - \frac{2}{\Delta_{21}\Sigma_{\mathrm{1c}}}  \Bigg), &
\end{flalign}
\endgroup
\begingroup
\allowdisplaybreaks
\begin{flalign}\label{suppeq:E_000^3_cRW}
  E_{|000\rangle}^{(3)} = &  g_{\mathrm{1c}} g_{\mathrm{2c}} g_{12}  \Bigg(\frac{2}{\Sigma_{\mathrm{1c}}\Sigma_{\mathrm{12}}} + \frac{2}{\Sigma_{\mathrm{1c}}\Sigma_{\mathrm{2c}}} + \frac{2}{\Sigma_{\mathrm{2c}}\Sigma_{\mathrm{12}}} \Bigg).&
\end{flalign}
\endgroup
%
For the fourth-order fast oscillating terms, we omit smaller contributing terms. Specifically, terms of order $\mathcal{O}(g_{ic}^4 / \Delta_{ic}^2 \Sigma_{ic})$ and $\mathcal{O}(g_{ic}^4 / \Delta_{ic} \Sigma_{ic}^2)$ are calculated, whereas terms of order $\mathcal{O}(g_{ic}^4 / \Sigma_{ic}^3)$ ($i\in \{1,2 \}$) and $\mathcal{O}(g_{12})$ are neglected. Therefore $ZZ$ contribution $\zeta^{(4)}_{\mathrm{CRW}}$ from the fourth-order fast oscillating terms is given as 
\begin{flalign}
&\zeta^{(4)}_{\mathrm{CRW}} = \left(E_{|101\rangle}^{(4)} -  E_{|001\rangle}^{(4)}\right) - \left(E_{|100\rangle}^{(4)} - E_{|000\rangle}^{(4)}\right), &
\end{flalign}
where the eigenenergy corrections $E^{(n)}_{|m\rangle}$ for the predominantly contributing terms are given as follows.
\\
\paragraph{ Order $\mathcal{O}(g_{\mathrm{1c}}^2 g_{\mathrm{2c}}^2 / \Delta_{i\mathrm{c}}^2 \Sigma_{i\mathrm{c}})$, $i \in \{ 1,2 \}$\textrm{:}} 
\begingroup
\allowdisplaybreaks
\begin{flalign}\label{suppeq:E_101^4_cRW}
E_{|101\rangle}^{(4)} = & g_{\mathrm{1c}}^2 g_{\mathrm{2c}}^2 \Bigg( - \frac{4}{\Delta_{\mathrm{2c}}^2(\Delta_{12} + \eta_{1} + 2 \omega_c + \eta_{\mathrm{c}})} & \nonumber \\
& \ \quad \qquad -\frac{4}{\Delta_{\mathrm{1c}}^2(\Delta_{21} + \eta_{2} + 2 \omega_c + \eta_{\mathrm{c}})} + \frac{1}{\Delta_{\mathrm{1c}}^2 \Sigma_{12}} & \nonumber \\
& \ \quad \qquad + \frac{1}{\Delta_{\mathrm{2c}}^2 \Sigma_{12}} - \frac{4}{\Delta_{\mathrm{2c}}(\Delta_{21} - \eta_{1})(\Sigma_{\mathrm{1c}} + \eta_{1})} & \nonumber \\
& \ \quad \qquad- \frac{4}{\Delta_{\mathrm{1c}}(\Delta_{12} - \eta_{2})(\Sigma_{\mathrm{2c}} + \eta_{2})} & \nonumber \\
& \ \quad \qquad- \frac{4}{\Delta_{\mathrm{2c}}(2 \omega_c + \eta_{\mathrm{c}})\Delta_{\mathrm{1c}}} + \frac{2}{\Delta_{\mathrm{1c}}\Delta_{\mathrm{2c}}\Sigma_{12}} & \nonumber \\
& \ \quad \qquad + \frac{2}{\Delta_{\mathrm{2c}}^2(\Sigma_{\mathrm{1c}} + \eta_{1})} + \frac{2}{\Delta_{\mathrm{1c}}^2(\Sigma_{\mathrm{2c}} + \eta_{2})} \Bigg) & \nonumber \\
& - \frac{2 g_{\mathrm{2c}}^4}{\Delta_{\mathrm{2c}}^2(2\omega_c + \eta_{\mathrm{c}})} - \frac{2 g_{\mathrm{1c}}^4}{\Delta_{\mathrm{1c}}^2(2\omega_c + \eta_{\mathrm{c}})} & \nonumber \\
& + \frac{2 g_{\mathrm{1c}}^4}{\Delta_{\mathrm{1c}}^2(\Sigma_{\mathrm{1c}} + \eta_{1})} + \frac{2 g_{\mathrm{2c}}^4}{\Delta_{\mathrm{2c}}^2(\Sigma_{\mathrm{2c}} + \eta_{2})} & 
\end{flalign}
\endgroup
\begingroup
\allowdisplaybreaks
\begin{flalign}\label{suppeq:E_100^4_cRW}
E_{|100\rangle}^{(4)} = & g_{\mathrm{1c}}^2 g_{\mathrm{2c}}^2 \Bigg(-\frac{2}{\Delta_{\mathrm{1c}}^2(\Delta_{21} + 2 \omega_c + \eta_{\mathrm{c}})} & \nonumber \\
& \ \quad \qquad - \frac{2}{\Delta_{\mathrm{1c}} \Delta_{12} \Sigma_{\mathrm{2c}}} + \frac{1}{\Delta_{\mathrm{1c}}^2\Sigma_{\mathrm{2c}} } \Bigg) &\nonumber \\
& + \frac{2g_{\mathrm{1c}}^4}{\Delta_{\mathrm{1c}}^2} \Bigg( -\frac{1}{2\omega_c + \eta_{\mathrm{c}}} + \frac{1}{\Sigma_{\mathrm{1c}} + \eta_{1}} \Bigg) & 
\end{flalign}
\endgroup
\begingroup
\allowdisplaybreaks
\begin{flalign}\label{suppeq:E_001^4_cRW}
E_{|001\rangle}^{(4)} = & g_{\mathrm{1c}}^2 g_{\mathrm{2c}}^2\Bigg(-\frac{2}{\Delta_{\mathrm{2c}}^2(\Delta_{12} + 2 \omega_c + \eta_{\mathrm{c}})} & \nonumber \\
& \ \quad \qquad - \frac{2}{\Delta_{\mathrm{2c}} \Delta_{21} \Sigma_{\mathrm{1c}}} + \frac{1}{\Delta_{\mathrm{2c}}^2 \Sigma_{\mathrm{1c}}} \Bigg) & \nonumber \\ & +\frac{2g_{\mathrm{2c}}^4}{\Delta_{\mathrm{2c}}^2} \Bigg( -\frac{1}{2\omega_c + \eta_{\mathrm{c}}} + \frac{1}{\Sigma_{\mathrm{2c}} + \eta_{2}} \Bigg) &
\end{flalign}
\endgroup
\begin{flalign}
&E_{|000\rangle}^{(4)} = 0&
\end{flalign}

\paragraph{Order $\mathcal{O}(g_{\mathrm{1c}}^2 g_{\mathrm{2c}}^2 / \Delta_{i\mathrm{c}} \Sigma_{i\mathrm{c}}^2)$, $i\in \{ 1,2 \}$:} 

\begingroup
\allowdisplaybreaks
\begin{flalign}
E_{|101\rangle}^{(4)} = & g_{\mathrm{1c}}^2 g_{\mathrm{2c}}^2 \Bigg( \frac{8}{\Delta_{\mathrm{2c}}(\Delta_{12} + \eta_{1} + 2 \omega_c + \eta_{\mathrm{c}})(\Sigma_{\mathrm{1c}} + \eta_{1})} & \nonumber \\
& \ \quad \qquad + \frac{8}{\Delta_{\mathrm{1c}}(\Delta_{21} + \eta_{2} + 2 \omega_c + \eta_{\mathrm{c}})(\Sigma_{\mathrm{2c}} + \eta_{2})} & \nonumber \\
& \ \quad \qquad + \frac{8}{\Delta_{\mathrm{2c}}(2 \omega_c + \eta_{\mathrm{c}})(\Sigma_{\mathrm{1c}} + \eta_{1})} &\nonumber \\ 
& \ \quad \qquad + \frac{8}{\Delta_{\mathrm{1c}}(2 \omega_c + \eta_{\mathrm{c}})(\Sigma_{\mathrm{2c}} + \eta_{2})} & \nonumber \\
& \ \quad \qquad + \frac{2}{(\Delta_{12} - \eta_{2})(\Sigma_{\mathrm{2c}} + \eta_{2})^2} & \nonumber \\
& \ \quad \qquad + \frac{2}{(\Delta_{21} - \eta_{1})(\Sigma_{\mathrm{1c}} + \eta_{1})^2}& \nonumber \\
& \ \quad \qquad - \frac{2}{\Delta_{2\mathrm{c}}(\Sigma_{\mathrm{1c}} + \eta_{1})^2}  - \frac{2}{\Delta_{1\mathrm{c}}(\Sigma_{\mathrm{2c}} + \eta_{2})^2} \Bigg) & \nonumber \\
& + \frac{8 g_{\mathrm{2c}}^4}{\Delta_{\mathrm{2c}}(2 \omega_c + \eta_{\mathrm{c}})(\Sigma_{\mathrm{2c}} + \eta_{2})} & \nonumber \\
& + \frac{8 g_{\mathrm{1c}}^4}{\Delta_{\mathrm{1c}}(2 \omega_c + \eta_{\mathrm{c}})(\Sigma_{\mathrm{1c}} + \eta_{1})} & \nonumber \\
& - \frac{2 g_{\mathrm{1c}}^4}{\Delta_{\mathrm{1c}}(\Sigma_{\mathrm{1c}} + \eta_{1})^2} - \frac{2 g_{\mathrm{2c}}^4}{\Delta_{\mathrm{2c}}(\Sigma_{\mathrm{2c}} + \eta_{2})^2} &
\end{flalign}
\endgroup
\begingroup
\allowdisplaybreaks
\begin{flalign}
E_{|100\rangle}^{(4)} = & g_{\mathrm{1c}}^2 g_{\mathrm{2c}}^2 \Bigg( \frac{4}{\Delta_{\mathrm{1c}}\Sigma_{\mathrm{2c}}(\Delta_{21} + 2 \omega_c + \eta_{\mathrm{c}})}  & \nonumber \\
&  \ \quad \qquad + \frac{4}{\Delta_{\mathrm{1c}}\Sigma_{\mathrm{2c}}(2 \omega_c + \eta_{\mathrm{c}})} + \frac{1}{\Delta_{12} \Sigma_{\mathrm{2c}}^2 } & \nonumber \\
&  \ \quad \qquad - \frac{1}{\Delta_{\mathrm{1c}} \Sigma_{\mathrm{2c}}^2} \Bigg) & \nonumber \\
& + g_{\mathrm{1c}}^4 \Bigg( \frac{8}{\Delta_{\mathrm{1c}}(2\omega_c + \eta_{\mathrm{c}})(\Sigma_{\mathrm{1c}} + \eta_{1})} &\nonumber \\
&  \ \quad \qquad  - \frac{2}{\Delta_{\mathrm{1c}}(\Sigma_{\mathrm{1c}} + \eta_{1})^2} \Bigg) &
\end{flalign}
\endgroup

\begingroup
\allowdisplaybreaks
\begin{flalign}
E_{|001\rangle}^{(4)} = & g_{\mathrm{1c}}^2 g_{\mathrm{2c}}^2 \Bigg( \frac{4}{\Delta_{\mathrm{2c}}\Sigma_{\mathrm{1c}}(\Delta_{12} + 2 \omega_c + \eta_{\mathrm{c}})} & \nonumber \\
&  \ \quad \qquad + \frac{4}{\Delta_{\mathrm{2c}}\Sigma_{\mathrm{1c}}(2 \omega_c + \eta_{\mathrm{c}})} + \frac{1}{\Delta_{21} \Sigma_{\mathrm{1c}}^2} & \nonumber \\
&  \ \quad \qquad - \frac{1}{\Delta_{\mathrm{2c}} \Sigma_{\mathrm{1c}}^2} \Bigg) & \nonumber \\
& + g_{\mathrm{2c}}^4 \Bigg( \frac{8}{\Delta_{\mathrm{2c}}(2\omega_c + \eta_{\mathrm{c}})(\Sigma_{\mathrm{2c}} + \eta_{2})} & \nonumber \\
&  \ \quad \qquad - \frac{2}{\Delta_{\mathrm{2c}}(\Sigma_{\mathrm{2c}} + \eta_{2})^2} \Bigg)&
\end{flalign}
\endgroup
\begin{flalign}
& E^{(4)}_{|000\rangle} = 0 &
\end{flalign}

In Fig.~\ref{suppfig:zz_vs_wc_idle}, we plot the theoretical values of $\zeta$ obtained by the calculation of Eq.~\eqref{suppeq:zeta_perturbative_analysis}. The theoretical calculation (solid black curve) shows good agreement with both experimental data (orange circles) and a numerical simulation (blue dashed curve). 

\newpage 
\section{Z-pulse transient calibration}
\label{suppsec:Z-pulse_transient_calibration}
The shape of the Z control (flux control) pulses are distorted as they pass through various electric components. This pulse distortion can be analyzed in the frequency domain by measuring the step response. 
In general, the qubit is employed as a sensor to characterize the step response of the flux control line~\cite{Barends2014,Foxen2018,Rol2020}. 
Specifically, we measure a Ramsey-type experiment, which measures the dynamic frequency change of the qubit as a response to the flux change. 

The step response can be fitted by multiple exponential time constants $\tau_k$ and settling amplitudes $a_k$ ($k = 1, 2, \cdots$) as follows.
\begin{align}
    V_{\mathrm{out, step}}(t) = V_{\mathrm{in, step}}(t)\times (1+\sum_{k}a_k e^{-(t/\tau_k)}),
\end{align}
where $V_{\mathrm{in,step}}(t)$ corresponds to a step function generated by AWG and $V_{\mathrm{out,step}}(t)$ corresponds to the response of the qubit to the step function. Note that we express the qubit response $V_{\mathrm{out,step}}(t)$ in the unit of AWG voltage and calculate the relative amplitude change $V_{\mathrm{out, step}}(t)/V_{\mathrm{in, step}}(t)$.

\begin{figure}[H]
    \centering
    \includegraphics[width=8.9cm]{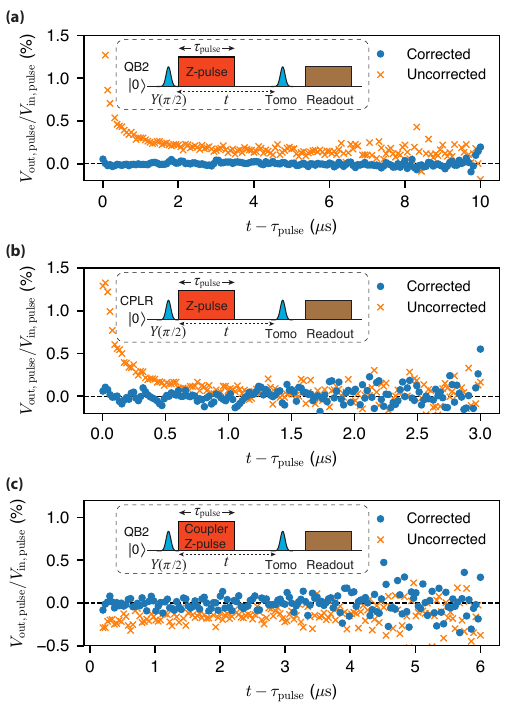}
    \caption{ 
    \textbf{(a)} Measurement of a turn-off transient of a \SI{5}{\micro\second}-long  QB2-Z pulse ($\tau_{\mathrm{pulse}}=\SI{5}{\micro\second}$) without predistortion (orange crosses) and with predistortion (blue circles).
    \textbf{(b)} Measurement of a turn-off transient for a \SI{1}{\micro\second}-long CPLR-Z pulse ($\tau_{\mathrm{pulse}}=\SI{1}{\micro\second}$). 
    \textbf{(c)} 
    Measurement of a turn-off transient of flux-crosstalk from the CPLR's flux line to QB2's SQUID. A \SI{3}{\micro\second}-long CPLR-Z pulse ($\tau_{\mathrm{pulse}}=\SI{3}{\micro\second}$) is applied. The pulse sequences are illustrated in the insets.
    }
    \label{suppfig:flux_transient_calibration}
\end{figure}

To reliably characterize long-time scale transients of the Z control pulses, we use a new protocol, which utilizes the Han spin echo technique~\cite{Hahn1950} (detailed procedures will be described in a forthcoming manuscript~\cite{Sung2020_predistortion}). We measure the turn-off transients of a square-shaped pulse with fixed duration $\tau_{\mathrm{pulse}}$ and fit the response with the following equation: 
\begin{align}
    V_{\mathrm{out, pulse}}(t) = &V_{\mathrm{in, pulse}}(t) \nonumber \\ 
    &\times\sum_{k}a_k \left(e^{-(t-\tau_{\mathrm{pulse}})/\tau_k)}- e^{-(t/\tau_k)}\right),
\end{align}
where $V_{\mathrm{in, pulse}}(t)$ corresponds to a $\tau_{\mathrm{pulse}}$-long square-shaped Z-pulse generated by AWG and $V_{\mathrm{out, pulse}}(t)$ corresponds to the response of the qubit to the pulse.
Figs.~\ref{suppfig:flux_transient_calibration}a and b show the turn-off transients of the QB2 and CPLR Z-pulses, respectively. 
The pulse sequences are illustrated in the insets. 
We plot the relative amplitude change $V_{\mathrm{out, step}}(t)/V_{\mathrm{in, step}}(t)$ as a function of the time delay between the Z-pulse and the tomography pulse $(t-\tau_{\mathrm{pulse}}$). 
We fit the transients with a sum of multiple exponential curves and extract the exponential time constants $\tau_k$ and the corresponding settling amplitudes (Table.~\ref{supptab:transient_calibration}) Notably, we observed long-time transients ($\approx\SI{30}{\micro\second}$) in our experimental setup, which are critical to correct in order to achieve high-fidelity two-qubit gates. We also measure and correct transients in the flux crosstalk (Fig.~\ref{suppfig:flux_transient_calibration}c), possibly due to an additional pulse distortion that occurs during the transmission from the end of CPLR's flux line to the QB2's SQUID. 

\begin{table}[H]
\begin{ruledtabular}
         \begin{tabular}{c | c c c}
         
             & QB2 & CPLR & Crosstalk (CPLR $\rightarrow$ QB2)\\
            \hline \\[-0.3cm]
            $a_1$ (\%) & $-$0.179 & $-$0.220 & -0.1152 \\
            $\tau_1$ (ns) & 21.9 & 31.0 & 1152 \\
            $a_2$ (\%) & $-$1.024 & $-$0.459 & -1.758 \\
            $\tau_2$ (ns) & 50 & 32.4 & 29770 \\
            $a_3$ (\%) & $-$0.251 & $-$0.567 \\
            $\tau_3$ (ns) & 87 & 45.7 \\
            $a_4$ (\%) & $-$0.484 & $-$0.938 \\
            $\tau_4$ (ns) & 158 & 127 \\
            $a_5$ (\%) & $-$0.487 & $-$0.358 \\
            $\tau_5$ (ns) & 773 & 730 \\
            $a_6$ (\%) & $-$1.143 & $-$1.36 \\
            $\tau_6$ (ns) & 26440 & 30000 \\
        \end{tabular}
    \caption{Summary of the fitted flux-transient parameters.}
    \label{supptab:transient_calibration}
\end{ruledtabular}
\end{table}

\section{The effective Hamiltonians for leakage dynamics during the two qubit gates}
\label{suppsec:effective_H}

\begin{figure}[htbp!]
    \centering
    \includegraphics[width=8.9cm]{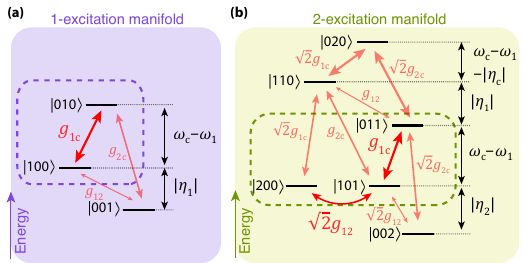}
    \caption{ 
    Energy level diagrams of the single-excitation manifold (\textbf{a}) and the double-excitation manifold (\textbf{b}) when performing the CZ gate. The dashed boxes indicate subspaces spanned by energy levels that are relevant to coherent leakage during the CZ gate. The red double-headed arrows denote exchange interactions between the energy levels.
    }
    \label{suppfig:effective_H_for_CZ}
\end{figure}

In this section, we derive the effective two-level Hamiltonians that describe the coherent leakage of CZ (Eq.~\eqref{eq:H_eff}) in the main text) and iSWAP gates. We first identify the states that strongly interact with the computational qubit states (\ketbare{000}, \ketbare{100}, \ketbare{001}, and \ketbare{101}) during the two-qubit gates and cause the coherent leakage. Subsequently, we truncate the system Hamiltonian (Eq.~\eqref{eq:H_lab}) into the relevant subspaces spanned by these leakage states and the associated computational qubit states.

We identify the leakage states for the CZ gate in the single- and double-excitation manifolds (Fig.~\ref{suppfig:effective_H_for_CZ}). Recall that, when performing the CZ gate, we bring \ketbare{101} in resonance with \ketbare{200} ($\omega_1+\eta_1=\omega_2$) and bias the coupler closer to the qubits to switch on the effective qubit-qubit coupling $\tilde{g}_{\mathrm{CZ}}$. Therefore, in the single excitation manifold, \ketbare{010} strongly interacts with $|100\rangle$, since \ketbare{010} (CPLR) is brought closer to $|100\rangle$ (QB1) in terms of energy. On the other hand, \ketbare{001} (QB2) is detuned from \ketbare{100} (QB1) by QB1's anharmonicity $\eta_1$, and thus \ketbare{001} is located farther from \ketbare{010} and is less hybridized with QB1 and CPLR.
Thus, we focus on the two-level dynamics between \ketbare{100} and \ketbare{010} and define the relevant subspace accordingly (a dashed purple box in Fig.~\ref{suppfig:effective_H_for_CZ}a).

Along the same line, in the double-excitation manifold, we identify the leakage states which strongly interact with the computational qubit state \ketbare{101} and cause the coherent leakage  during the CZ gate. We first rule \ketbare{020} out as a leakage state, since it couples to \ketbare{101} via a second-order process that is generally weaker than first-order interactions. Next, we rule out \ketbare{110} and \ketbare{002}, since they are relatively far-detuned from \ketbare{101} compared to \ketbare{011} and \ketbare{200}. Specifically, \ketbare{002} is detuned from \ketbare{101} by QB2's anharmonicity $\eta_2$, of which magnitude is much greater than the direct QB1-QB2 coupling strength $\sqrt{2}g_{12}$ ($|\eta_2|\gg\sqrt{2}g_{12}$). In addition, \ketbare{110} is located farther from \ketbare{101} than \ketbare{011} by QB1's anharmonicity $|\eta_1|$. 
After ruling these out as leakage states, we determine the relevant subspace as shown in Fig.~\ref{suppfig:effective_H_for_CZ}b (spanned by the states within the dashed green box). 

Next, we truncate the system Hamiltonian to the relevant subspaces  in both the single- and double-excitation manifolds and obtain the following effective Hamiltonians $H_{\mathrm{1}}^{\mathrm{CZ}}$ and $H_{\mathrm{2}}^{\mathrm{CZ}}$:
\begin{align}
\label{suppeq:H_cz_eff}
H_{\mathrm{1}}^{\mathrm{CZ}}&=  \stackrel{\mbox{\scalebox{0.7}{\ $|100\rangle$  \  \ketbare{010}\ \ }}}{%
    \begin{pmatrix}
    \omega_{1} & g_{\mathrm{1c}} \\
    g_{\mathrm{1c}} & \omega_{\mathrm{c}} 
    \end{pmatrix}%
  },
  &H_{\mathrm{2}}^{\mathrm{CZ}} = \stackrel{\mbox{\scalebox{0.7}{ \ketbare{101}   \quad \qquad \ketbare{200}  \quad \qquad \ketbare{011}  }}}{%
    \begin{pmatrix}
    \omega_{1}+\omega_{2} & \sqrt{2} g_{12}  & g_{\mathrm{1c}}\\
    \sqrt{2} g_{12}& \omega_1+\omega_2 & 0 \\
    g_{\mathrm{1c}} & 0 & \omega_{\mathrm{c}} + \omega_2
    \end{pmatrix}%
  },       
\end{align}
where we have replaced $\omega_1+\eta_1$ by $\omega_2$, since $\omega_1+\eta_1=\omega_2$ is assumed here.
Note that this Hamiltonian truncation is only valid in the regime where $|\eta_{1}|, |\eta_{2}| \gg g_{1\mathrm{c}}, g_{2\mathrm{c}}$ (in our device, $|\eta_1|\approx|\eta_2|\approx 3 g_{\mathrm{1c}}\approx 3 g_{\mathrm{2c}}$). To analyze the leakage dynamics under general conditions, the leakage contribution from additional levels need to be considered and will be of interest in future research. 

To further simplify the three-level dynamics of $H_{\mathrm{2}}^{\mathrm{cz}}$, we introduce a partially hybridized basis: a bright state {$|\mathrm{B}\rangle \equiv \cos\Theta$\ketbare{011}$+\sin\Theta$\ketbare{200}} and a dark state {$|\mathrm{D}\rangle \equiv \cos\Theta$\ketbare{200}$-\sin\Theta$\ketbare{011}}, where ${\Theta\equiv\tan^{-1}(\sqrt{2}g_{12}/g_{1\mathrm{c}})}$~\cite{Lambropoulos2007}. To this end, we rewrite $H_{\mathrm{2}}^{\mathrm{CZ}}$ in the hybridized basis as follows:
\begin{align}
\label{suppeq:H2}
  &\overline{H}_{\mathrm{2}}^{\mathrm{CZ}} = \stackrel{\mbox{\scalebox{0.7}{ \quad \ketbare{101} \quad \ \ \ \ketbare{D}  \ \ \ \ketbare{B}  }}}{%
    \begin{pmatrix}
    \omega_1+\omega_2 & 0  & g_{\mathrm{B}}\\
    0 & \tilde{\omega}_{\mathrm{D}} & g_{\mathrm{r}} \\
    g_{\mathrm{B}} & g_{\mathrm{r}} & \tilde{\omega}_{\mathrm{B}}
    \end{pmatrix}%
  },      
\end{align}
where the eigenenergies of \ketbare{D} and \ketbare{B} are given as 
\begin{align}
    \tilde{\omega}_{\mathrm{D}}&=\cos^2\Theta(\omega_1+\omega_2)+ \sin^2\Theta(\omega_{\mathrm{c}}+\omega_2), \\
    \tilde{\omega}_{\mathrm{B}}&=  \cos^2\Theta(\omega_{\mathrm{c}}+\omega_2)+ \sin^2\Theta(\omega_1+\omega_2).
\end{align}
The coupling strength $g_{\mathrm{B}}$ between \ketbare{B} and \ketbare{101} is given as
\begin{align}
    g_{\mathrm{B}}=g_{\mathrm{1c}}\cos\Theta  +  \sqrt{2}g_{\mathrm{12}}\sin\Theta,
\end{align}
and the coupling strength $g_{\mathrm{r}}$ between \ketbare{B} and \ketbare{D} is given as 
\begin{align}
    g_{\mathrm{r}} = \cos\Theta \sin\Theta (\omega_{\mathrm{1}}-\omega_{\mathrm{c}}).
\end{align}
In the parameter regime, where $g_{\mathrm{1c}}\gg g_{12}$ ($\Theta\approx 0$), $g_{\mathrm{r}}$ becomes zero, and therefore  \ketbare{101} only interacts with the bright state \ketbare{B}; the dark state \ketbare{D} is decoupled from both of the states. This allows us to further reduce the three-level dynamics onto an effective two-level system, as described by Eq.~\eqref{eq:H_eff} in the main text. 
 As a result, the two effective Hamiltonians $H_{\mathrm{1}}^{\mathrm{CZ}}$ (\ketbare{100} and \ketbare{001} subspace) and $\overline{H}_{\mathrm{2}}^{\mathrm{CZ}}$ (\ketbare{101} and \ketbare{B} subspace) are equivalent to the following effective Hamiltonian $H_{\mathrm{eff}}^{\mathrm{CZ}}$ up to offset energies:
\begin{align}
    H_{\mathrm{eff}}^{\mathrm{CZ}}&= 
    \begin{pmatrix}
   0 & g_{\mathrm{1c}} \\
    g_{\mathrm{1c}} & \omega_{\mathrm{c}} -  \omega_{1}
    \end{pmatrix}.%
\end{align}
Optimal control techniques are well-studied for this class of effective Hamiltonians, which we will further discuss in Appendix~\ref{suppsec:slepian_approach}.

\begin{figure}[h!]
    \centering
    \includegraphics[width=8.9cm]{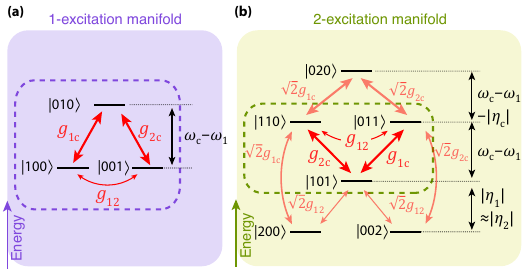}
    \caption{ 
    Energy level diagrams of the single-excitation manifold (\textbf{a}) and the double-excitation manifold (\textbf{b}) when performing the iSWAP gate. The dashed boxes indicate subspaces spanned by energy levels that are relevant to coherent leakage during the iSWAP gate. The red double-headed arrows denote exchange interactions between the energy levels.
    }
    \label{suppfig:effective_H_for_iSWAP}
\end{figure}

Next, we identify the leakage states for the iSWAP gate. When performing the iSWAP gate, we bring $|100\rangle$ (QB1) in resonance with $|001\rangle$ (QB2), and bias the coupler closer to the qubits to switch on the effective qubit-qubit coupling $\tilde{g}_{\mathrm{iSWAP}}$.
Unlike the CZ gate, the computational qubit states \ketbare{100} and \ketbare{001} are equally detuned from a leakage state \ketbare{010} in terms of energy. Therefore, we consider leakage from both \ketbare{001} and \ketbare{100} to \ketbare{010}. Accordingly, we determine the relevant subspace in the single-excitation manifold as shown in Fig.~\ref{suppfig:effective_H_for_iSWAP} (spanned by states in the purple dashed box). In the double-excitation manifold, we rule \ketbare{020} out as a leakage state, because it couples to the computational qubit state \ketbare{101} via a second-order process. We also rule out \ketbare{200} and \ketbare{002}, since they are detuned from \ketbare{101} by QB1 and QB2 anharmonicities, respectively, of which both are much greater than the QB1-QB2 direct coupling strength $\sqrt{2}g_{12}$  ($|\eta_{1}|,|\eta_{2}|\gg \sqrt{2}g_{12}$). Given that, we determine the relevant subspace in the double-excitation manifold as shown in Fig.~\ref{suppfig:effective_H_for_iSWAP}b (spanned by states in the green dashed box). 

We truncate the system Hamiltonian to the relevant subspaces for the iSWAP gate. Within the relevant subspaces, the effective Hamiltonians $H_{1}^{\mathrm{iSWAP}}$ and $H_{2}^{\mathrm{iSWAP}}$ in the single- and double-excitation manifolds, respectively, are given as follows.
\begin{align}
\label{suppeq:H_iswap_eff}
H_{\mathrm{1}}^{\mathrm{iSWAP}}&=  \stackrel{\mbox{\scalebox{0.7}{ \ \ketbare{010} \ \ketbare{100}  \  \ketbare{001} \ }}}{%
    \begin{pmatrix}
    \omega_{\mathrm{c}} & g_{\mathrm{1c}}  & g_{\mathrm{2c}}\\
    g_{\mathrm{1c}}& \omega_{\mathrm{1}} & g_{\mathrm{12}} \\
    g_{\mathrm{2c}} & g_{\mathrm{12}} & \omega_{\mathrm{1}}
    \end{pmatrix}%
   }
   ,
  \nonumber \\ 
  H_{\mathrm{2}}^{\mathrm{iSWAP}} &= \stackrel{\mbox{\scalebox{0.7}{  \quad \ketbare{101}    \qquad \ketbare{011} \ \ \ \qquad  \ketbare{110}  \qquad \ \ \ }}}{%
    \begin{pmatrix}
    2\omega_{1} & g_{\mathrm{1c}}  & g_{\mathrm{2c}}\\
     g_{\mathrm{1c}}& \omega_1+\omega_{\mathrm{c}} & g_{12} \\
    g_{\mathrm{2c}} & g_{12} &  \omega_1 + \omega_{\mathrm{c}}
    \end{pmatrix},%
  }
\end{align}
where we have replaced $\omega_2$ by $\omega_1$, since $\omega_1=\omega_2$ is assumed here.

To simplify the three-level dynamics of $H_1^{\mathrm{iSWAP}}$, we introduce a hybridized basis: a bright state $|B_1\rangle \equiv \cos\xi |001\rangle + \sin\xi |100\rangle$ and a dark state $|D_1\rangle \equiv \cos\xi |100\rangle - \sin\xi |001\rangle$, where $\xi\equiv\tan^{-1}(g_{\mathrm{1c}}/g_{\mathrm{2c}})$. Along the same line, we introduce a hybridized basis for $H_2^{\mathrm{iSWAP}}$ as follows: a bright state $|B_2\rangle \equiv \cos\xi |110\rangle + \sin\xi |011\rangle$ and a dark state $|D_2\rangle \equiv \cos\xi |011\rangle - \sin\xi |110\rangle$. Using these hybridization bases, we can rewrite the effective Hamiltonians as follows.
\begin{align}
\label{suppeq:H_iswap_eff_hybridized}
\overline{H}_{\mathrm{1}}^{\mathrm{iSWAP}}&=  \stackrel{\mbox{\scalebox{0.7}{ \ketbare{010} \ \ \ketbare{B_1} \  \ \ \ketbare{D_1} \ }}}{%
    \begin{pmatrix}
    \omega_{\mathrm{c}} & g_{\mathrm{B_1}}  & 0\\
    g_{\mathrm{B_1}} & \tilde{\omega}_{\mathrm{B_1}} & g_\mathrm{r1} \\
    0 & g_\mathrm{r1} & \omega_{\mathrm{D_1}}
    \end{pmatrix}%
   },
  \nonumber \\ 
  \overline{H}_{\mathrm{2}}^{\mathrm{iSWAP}} &= \stackrel{\mbox{\scalebox{0.7}{    \ketbare{101} \ \ \ketbare{B_2}   \ \ \ \ketbare{D_2}\  }}}{%
    \begin{pmatrix}
    2\omega_{1} &  g_{\mathrm{B_2}}  & 0\\
    g_{\mathrm{B_2}} & \tilde{\omega}_{\mathrm{B_2}} & g_{\mathrm{r2}} \\
    0 & g_{\mathrm{r2}} & \tilde{\omega}_{\mathrm{D_2}}
    \end{pmatrix}%
  },       
\end{align}
where the coupling strengths $g_{\mathrm{B_1}}$, $g_{\mathrm{B_2}}$, and $g_\mathrm{r}$ are given as
\begin{align}
    g_{\mathrm{B_1}} &= g_{\mathrm{B_2}} = g_{\mathrm{1c}} \sin\xi + g_{\mathrm{2c}} \cos\xi, \\
    g_{\mathrm{r1}} &= g_{\mathrm{r2}} =  g_{12}(\cos^2\xi -\sin^2\xi),
\end{align}
and the eigenenergies $\tilde{\omega}_{\mathrm{B_1}}$, $\tilde{\omega}_{\mathrm{D_1}}$, $\tilde{\omega}_{\mathrm{B_2}}$, and $\tilde{\omega}_{\mathrm{D_2}}$ are given as
\begin{align}
    \tilde{\omega}_{\mathrm{B_1}}&=\omega_1+2g_{12}\sin\xi\cos\xi,
    \\ 
     \tilde{\omega}_{\mathrm{D_1}}& =\omega_1-2g_{12}\sin\xi\cos\xi,
     \\
    \tilde{\omega}_{\mathrm{B_2}}&= \omega_1+\omega_{\mathrm{c}}+2g_{12}\sin\xi\cos\xi,
    \\
     \tilde{\omega}_{\mathrm{D_2}} &= \omega_1+\omega_{\mathrm{c}}-2g_{12}\sin\xi\cos\xi.
\end{align}

We assume $g_{\mathrm{1c}}=g_{\mathrm{2c}}\gg g_{12}$ ($\xi = \pi/4$), which is the case in our device and a practical parameter regime for tunable couplers~\cite{Yan2018}. 
In this regime, \ketbare{010} only interacts with \ketbare{B_1} and \ketbare{101} only interacts with \ketbare{B_2}. Assuming $\omega_{\mathrm{c}}-\omega_1 \gg g_{12}$, the corresponding two-level Hamiltonians  $\overline{H}_{\mathrm{2}}^{\mathrm{iSWAP}}$ (\ketbare{010} and \ketbare{B_1} subspace) and  $\overline{H}_{\mathrm{2}}^{\mathrm{iSWAP}}$ (\ketbare{101} and \ketbare{B_2} subspace) are approximately equal  to the following effective Hamiltonian $H_{\mathrm{eff}}^{\mathrm{iSWAP}}$ up to offset energies:
\begin{align}
    H_{\mathrm{eff}}^{\mathrm{iSWAP}}&= 
    \begin{pmatrix}
   0 & \sqrt{2} g_{\mathrm{1c}} \\
\sqrt{2} g_{\mathrm{1c}} & \omega_{\mathrm{c}} -  \omega_{1}
    \end{pmatrix}.%
\end{align}
Once again, we reduce the system description to this two-level Hamiltonian so that we can easily apply optimal control techniques for the gate.

\newpage

\section{Suppression of leakage using a Slepian-based optimal control}
\label{suppsec:slepian_approach}

As detailed in Appendix~\ref{suppsec:effective_H}, the effective Hamiltonians that describe the leakage dynamics of CZ and iSWAP gates are given as follows:
\begin{align}
    H_{\mathrm{eff}}^{\mathrm{CZ}}(t) & = 
    \begin{pmatrix}
   0 & g_{\mathrm{1c}} \\
    g_{\mathrm{1c}} & \omega_{\mathrm{c}}(t) -  \omega_{1}
    \end{pmatrix},
    \\ 
    H_{\mathrm{eff}}^{\mathrm{iSWAP}}(t) & = 
    \begin{pmatrix}
   0 & \sqrt{2} g_{\mathrm{1c}} \\
    \sqrt{2} g_{\mathrm{1c}} & \omega_{\mathrm{c}}(t) -  \omega_{1}
    \end{pmatrix}.
\end{align}
We optimize the control waveform $\omega_{\mathrm{c}}(t)$ for adiabatic behavior under these two-level systems. Note that these effective two-level systems address only predominant leakage channels, not all possible leakage channels during the two qubit gates. Specifically, $H_{\mathrm{eff}}^{\mathrm{CZ}}(t)$ addresses leakage from \ketbare{100} to \ketbare{010} (in the single-excitation manifold) and leakage from \ketbare{101} to \ketbare{011} (in the double-excitation manifold) during the CZ gate. In the case of the iSWAP gate, $H_{\mathrm{eff}}^{\mathrm{iSWAP}}(t)$ addresses leakage from \ketbare{100} and \ketbare{001} to \ketbare{010} (in the single-excitation manifold) and leakage from \ketbare{101} to \ketbare{110} and \ketbare{011} (in the double-excitation manifold).

\begin{figure}[h!]
    \centering
    \includegraphics[width=8.9cm]{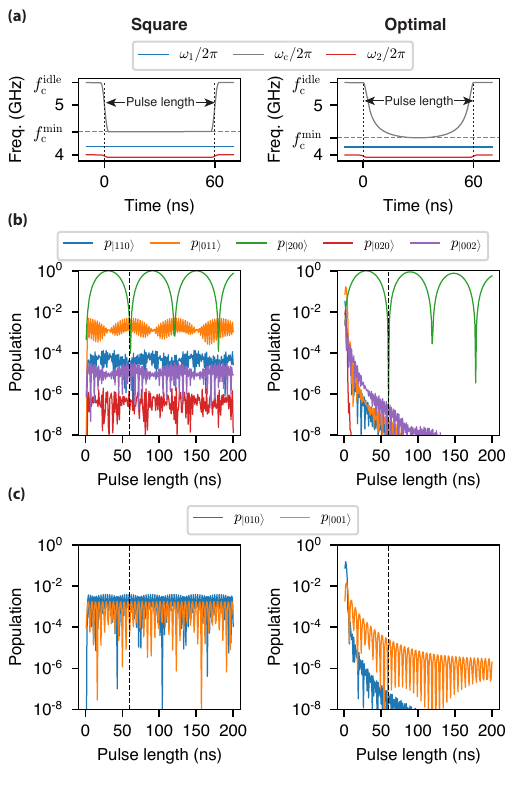}
    \caption{ 
    \textbf{Numerical simulation of coherent leakage of CZ gates.}
    \textbf{(a)} Square-shaped and optimal control waveforms for \SI{60}{\nano\second}-long CZ gates, respectively. 
    \textbf{(b)} Coherent leakage in the double-excitation manifold.
    We prepare \ketbare{101} and apply a control pulse, and then measure the state populations. 
    By using the optimal pulse shaping, we suppress population of the leakage state \ketbare{011} (orange curve) below $10^{-7}$ for pulses longer than \SI{60}{\nano\second}. 
    \textbf{(c)} Coherent leakage in the single-excitation manifold. We prepare \ketbare{100} and apply a control pulse, and then measure the state populations. 
    By using optimal pulse shaping, we suppress population of the leakage state \ketbare{010} (blue curve) below $10^{-7}$ for pulses longer than \SI{60}{\nano\second}. 
    The leakage to \ketbare{001} is not suppressed as much, since the optimal control relies on the effective Hamiltonian $H_{\mathrm{eff}}^{\mathrm{CZ}}(t)$ that only addresses leakage from \ketbare{100} to \ketbare{010} in the single-excitation manifold. The data points in \textbf{(b-c)} are obtained every \SI{1}{\nano\second}. 
    }
    \label{suppfig:optimal_vs_square_cz_simulation}
\end{figure}

\begin{figure}[h!]
    \centering
    \includegraphics[width=8.9cm]{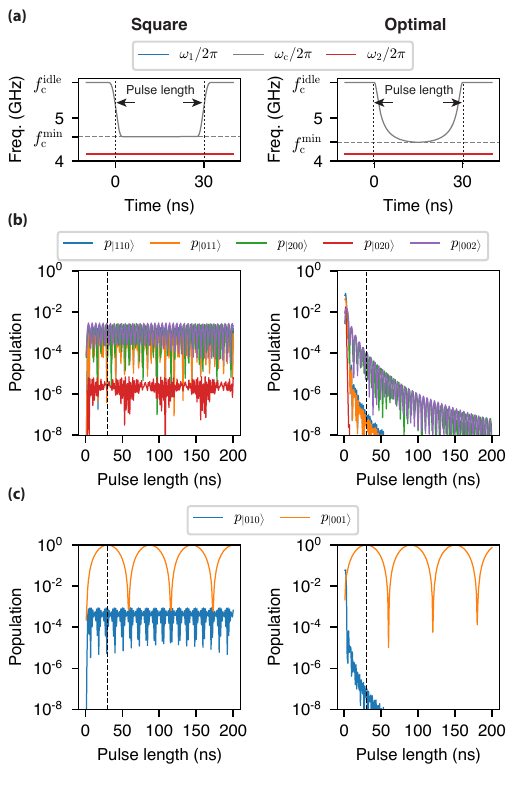}
    \caption{ 
    \textbf{Numerical simulation of coherent leakage of iSWAP gates.}
    \textbf{(a)} Square-shaped and optimal control waveforms for \SI{30}{\nano\second}-long iSWAP gates, respectively. 
    \textbf{(b)} Coherent leakage in the double-excitation manifold.
     We prepare \ketbare{101} and apply a control pulse, and then measure the state populations. 
     By using optimal pulse shaping, we suppress population of the leakage states \ketbare{110} and \ketbare{011} (blue and orange curves) below $10^{-7}$ for pulses longer than \SI{30}{\nano\second} (black dashed line). The leakage to \ketbare{200} and \ketbare{002} are not suppressed as much, since the optimal control relies on the effective Hamiltonian $H_{\mathrm{eff}}^{\mathrm{iSWAP}}(t)$ that only addresses the leakage from \ketbare{101} to \ketbare{110} and \ketbare{011} in the double-excitation manifold. 
    \textbf{(c)} Coherent leakage in the single-excitation manifold.    %
     We prepare \ketbare{101} and apply a control pulse, and then measure the state populations. 
     By using the optimal pulse shaping, we suppress population of the leakage state \ketbare{010} (blue curve) below $10^{-7}$ for pulses longer than \SI{30}{\nano\second} (black dashed line).
      The data points in \textbf{(b-c)} are obtained every \SI{1}{\nano\second}. 
    }
    \label{suppfig:optimal_vs_square_iswap_simulation}
\end{figure}

Following Ref.~\cite{Martinis2014}, we take a Slepian-based approach to implement an optimal control pulse that minimizes leakage errors for any pulse longer than the chosen pulse length.
For example, a Slepian control pulse for a \SI{60}{\nano\second}-long CZ gate minimizes the leakage error of CZ pulses which have the same pulse amplitude, but longer pulse lengths than \SI{60}{\nano\second}.

In Fig.~\ref{suppfig:optimal_vs_square_cz_simulation}, we numerically simulate coherent leakage of CZ gates (see Appendix~\ref{suppsec:numerical_simulation} for details about the simulation). We assess the performance of an optimized control pulse by comparing to a simple square pulse (Fig.~\ref{suppfig:optimal_vs_square_cz_simulation}a). Considering the bandwidth limitation of our AWGs, the square pulse is smoothed by applying a Hanning filter with a window width of 5 points (\SI{1}{\nano\second} intervals).
 The control pulse amplitudes, which are parameterized by the minimum point of CPLR frequency  $f_{\mathrm{c}}^{\mathrm{min}}$, are chosen such that \SI{60}{\nano\second}-long control pulses perform the CZ gate. In Fig.~\ref{suppfig:optimal_vs_square_cz_simulation}b, to characterize the leakage in the double-excitation manifold, we prepare \ketbare{101} and apply a control pulse, and then measure the population of leakage states \ketbare{110}, \ketbare{011}, \ketbare{200}, \ketbare{020}, and \ketbare{002} with varying the pulse length. We note that the square pulse shaping causes significant leakage, especially to \ketbare{011} (an orange curve in Fig.~\ref{suppfig:optimal_vs_square_cz_simulation}b). By using the optimal pulse, we suppress leakage populations $p_{|110\rangle}$, $p_{|011\rangle}$, $p_{|020\rangle}$, and $p_{|002\rangle}$ below $10^{-7}$ for pulses longer than the chosen gate length: \SI{60}{\nano\second}. 
Fig.~\ref{suppfig:optimal_vs_square_cz_simulation}c shows leakage in the single-excitation manifold. Here we characterize leakage from the computational qubit state \ketbare{100} after applying a CZ pulse. 
As in the case of double-excitation manifold, a square-shaped control pulse causes significant leakage, to both \ketbare{010} and \ketbare{001}. By using the optimal control, we suppress the leakage population $p_{|010\rangle}$ to \ketbare{010} below $10^{-7}$. However, we note that the leakage to \ketbare{001} is not suppressed as much, compared to \ketbare{010}. This is because our theoretical model $H_{\mathrm{eff}}^{\mathrm{CZ}}(t)$ only addresses leakage from \ketbare{100} to \ketbare{010} without taking \ketbare{001} into account.

In Fig.~\ref{suppfig:optimal_vs_square_iswap_simulation}, we simulate coherent leakage of iSWAP gates. We compare the performance of an optimal control pulse to a square pulse (Fig.~\ref{suppfig:optimal_vs_square_iswap_simulation}a). The control pulse amplitudes ($f_{\mathrm{c}}^{\mathrm{min}}$) are chosen such that \SI{30}{\nano\second}-long control pulses perform the iSWAP gate. In Fig.~\ref{suppfig:optimal_vs_square_cz_simulation}c and d, we characterize leakage in the double-excitation manifold as in the case of the CZ gates. The square control pulse causes significant leakage. By using the optimized pulse, we suppress the leakage population to \ketbare{110} (a blue curve) and \ketbare{110} (a orange curve) below $10^{-7}$ for pulses longer than the chosen gate length: \SI{30}{\nano\second}. 
Fig.~\ref{suppfig:optimal_vs_square_iswap_simulation}c shows the leakage in the single-excitation manifold. Here we characterize leakage from the computational qubit state \ketbare{100}. 
The square-shaped pulse causes significant leakage errors. By using the optimal control, we suppress the state population of a leakage state \ketbare{010} below $10^{-7}$. 

In this section, we demonstrated our Slepian-based optimal control by presenting numerical simulation results. 
We suppress population of the predominant leakage states below $10^{-7}$, by using the optimized control.
However, not every leakage channel is suppressed to the same level, since our theoretical model addresses only the predominant leakage channels. 
Developing a theoretical framework for addressing the full leakage channels will be the subject of future work.

\section{Advantages of small anharmonicity $\eta_{\mathrm{c}}$ for the coupler}
\label{suppsec:advantages_small_eta_c}
\begin{figure}[ht!]
    \centering
    \includegraphics[width=8.9cm]{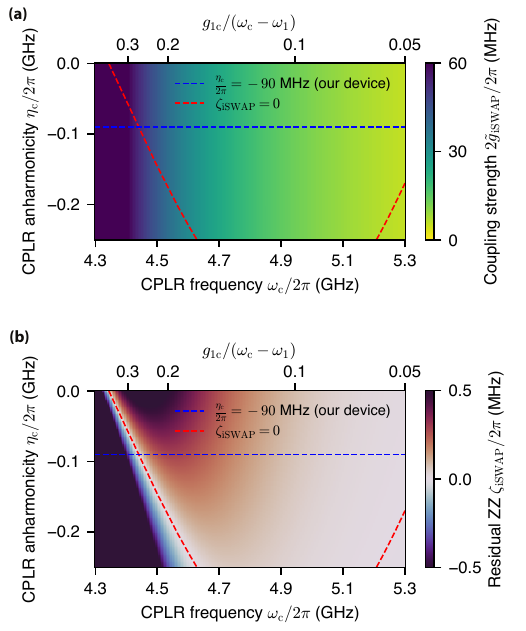}
    \caption{ 
 Numerical simulations of \textbf{(a)} coupling strength $\tilde{g}_{\mathrm{iSWAP}}$ and \textbf{(b)} residual $ZZ$ strength $\zeta_{\mathrm{iSWAP}}$ for an iSWAP interaction vs. the coupler anharmonicity $\eta_{\mathrm{c}}$ ($y$-axis) and the coupler frequency $\omega_{\mathrm{c}}$ ($x$-axis), where $\omega_1/2\pi=\omega_2/2\pi=\SI{4.16}{\giga\hertz}$. Note that smaller $\eta_{\mathrm{c}}$ enables $ZZ$-free iSWAP interaction ($\zeta_{\mathrm{iSWAP}}=0$, red dashed curves) with stronger $\tilde{g}_{\mathrm{iSWAP}}$, thereby enabling faster $ZZ$-free iSWAP gates.}
    \label{suppfig:engineered_anh_iswap}
\end{figure}

Smaller $\eta_{\mathrm{c}}$ enables the $ZZ$-free iSWAP interaction at a lower coupler frequency $\omega_{\mathrm{c}}$, which allows for a stronger $\tilde{g}_{\mathrm{iSWAP}}$ such that we can implement faster $ZZ$-free iSWAP gates. Figs.~\ref{suppfig:engineered_anh_iswap}(a) and (b) show numerical simulations of coupling strength (2$\tilde{g}_{\mathrm{iSWAP}}$) of the iSWAP interaction and its residual $ZZ$ strength ($\zeta_{\mathrm{iSWAP}}$) as a function of the coupler anharmonicity ($\eta_{\mathrm{c}}$). QB1 and QB2 frequencies are on resonance, which is the standard configuration for activating iSWAP gates. In each figure, the blue dashed curve represents our parameter choice for the coupler anharmonicity ($\eta_{\mathrm{c}}/2\pi=-\SI{90}{\mega\hertz}$) and red dashed curve represents a parameter regime, where residual $ZZ$ interaction becomes zero owing to the cancellation induced by the second excited state of the coupler. We note that as $\eta_{\mathrm{c}}$ decreases, the coupler frequency at which residual $ZZ$ interaction is cancelled also decreases (red dashed curve), while $\tilde{g}_{\mathrm{iSWAP}}$ remains constant approximately. Therefore, the resultant $\tilde{g}_{\mathrm{iSWAP}}$ for the $ZZ$-free iSWAP increases, thereby reducing the gate duration. Hence, owing to our relatively small coupler anharmonicity, we are able to make a short ($\SI{30}{\nano\second}$) $ZZ$-free iSWAP gate.

In addition, smaller $\eta_{\mathrm{c}}$ prevents $|020\rangle$ from being strongly hybridized with $|101\rangle$ and $|200\rangle$ during a CZ gate; this enables us to simplify multi-level leakage dynamics and use the Slepian-based optimal control. In the following paragraph, we explain how we numerically estimate the state overlap of $|020\rangle$ with $|101\rangle$ and $|200\rangle$. 

\begin{figure}[ht!]
    \centering
    \includegraphics[width=8.9cm]{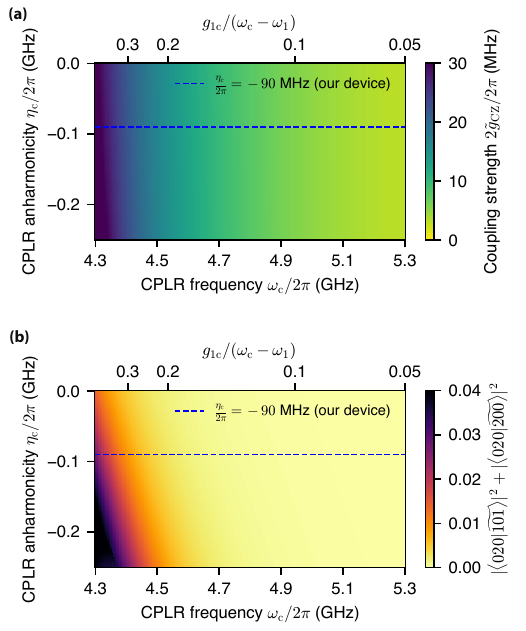}
    \caption{ 
 Numerical simulations of \textbf{(a)} coupling strength $\tilde{g}_{\mathrm{CZ}}$ for the CZ interaction and \textbf{(b)} the state overlap $|\langle 020|\widetilde{101}\rangle|^2+ \langle 020| \widetilde{200}\rangle |^2$ vs. CPLR anharmonicity $\eta_{\mathrm{c}}$ ($y$-axis) and CPLR frequency $\omega_{\mathrm{c}}$ ($x$-axis), where $(\omega_1+\eta_1)/2\pi=\omega_2/2\pi=\SI{3.94}{\giga\hertz}$. Note that smaller $\eta_{\mathrm{c}}$ prevents $|020\rangle$ from being strongly hybridized with $|101\rangle$ and $|200\rangle$ during the CZ interaction while keeping $\tilde{g}_{\mathrm{CZ}}$ constant; this enables us to simplify the multi-level leakage dynamics and use the Slepian-based optimal control.
 }
    \label{suppfig:engineered_anh_cz}
\end{figure}

The eigenstates (i.e., dressed states) $|\widetilde{\psi}\rangle$ of our system (see Eq.~\eqref{eq:H_lab} for the system Hamiltonian) can be expressed by a linear combination of basis states (i.e., bare states) $|\phi_{j}\rangle$ as follows.
\begin{align}
    |\widetilde{\psi}\rangle = \sum_{1}^{N}\langle \phi_j|\widetilde{\psi}\rangle|\phi_j\rangle,
\end{align}
where $N$ denotes the dimension of our Hilbert space. We estimate the complex state overlap coefficients $\langle 020|\widetilde{101}\rangle$ and $\langle 020 |\widetilde{200} \rangle$ by numerically diagonalizing the Hamiltonian based on the device parameters (see Appendix~\ref{suppsec:numerical_simulation} for the device parameters used for simulation). We define a sum of squares of these coefficients ($|\langle 020|\widetilde{101}\rangle|^2+ \langle 020| \widetilde{200}\rangle |^2$) as a metric that quantifies how strongly $|020\rangle$ hybridizes with $|101\rangle$ and $|200\rangle$.  
Figs.~\ref{suppfig:engineered_anh_cz}(a) and (b) show numerical simulation results of coupling strength ($2\tilde{g}_{\mathrm{CZ}}$) of the CZ interaction and the state overlap ($|\langle 020|\widetilde{101}\rangle|^2+ \langle 020| \widetilde{200}\rangle |^2$) as a function of CPLR anharmonicity ($\eta_{\mathrm{c}}$). QB1 and QB2 frequencies are set such that $\omega_1+\eta_1=\omega_2$, which is the standard configuration for activating CZ gates. We note that the as $\eta_{\mathrm{c}}$ decreases, the state overlap $|\langle 020|\widetilde{101}\rangle|^2+ \langle 020| \widetilde{200}\rangle |^2$ also decreases, while the coupling strength $\tilde{g}_{\mathrm{CZ}}$ remains constant approximately. Hence, by using a relatively small coupler anharmonicity, we are able to reduce the hybridization of $|020\rangle$ with $|101\rangle$ and $|200\rangle$ when performing CZ gates.

\section{Synchronization of XY axes for the iSWAP gate}
\label{suppsec:synchronization_xy_axes}
The computational qubit state is generally defined in a reference frame, rotating at the frequency of qubit driving tone (this frame is often called the logical frame).
Accordingly, in a multi-qubit system, we use multiple independently rotating frames to refer the computational state of each qubit.
Notably, performing iSWAP-like gates by tuning qubit frequencies into resonance~\cite{Barends2019} causes a non-trivial local phase shift in the logical frame due to the unmatched rotating frequencies. 
In this section, we explain how this phase shift occurs by presenting a simple example and discuss how it can be avoided.

We consider an uncoupled two-qubit system with Hamiltonian defined as follows in the laboratory frame ($\hbar\equiv1$)
\begin{align}
    H_{\mathrm{lab}} = \omega_1 (|1\rangle\langle1|) \otimes I + \omega_2 I \otimes (|1\rangle\langle1|),
\end{align}
where $\omega_1$ and $\omega_2$ denote the transition frequencies of each qubit. Consider an arbitrary state $\psi(t)$ evolving under the Hamiltonian $H_{\mathrm{lab}}$ as follows.
\begin{align}
    \psi (t) &=
    \begin{pmatrix}
    c_{00}(t) \\
    c_{01}(t) \\
    c_{10}(t) \\
    c_{11}(t)
    \end{pmatrix}
    =
    \begin{pmatrix}
    c_{00}(0) \\
    e^{i\omega_2t}c_{01}(0) \\
    e^{i\omega_1t}c_{10}(0) \\
    e^{i(\omega_1+\omega_2)t}c_{11}(0)
    \end{pmatrix},
\end{align}
where $c_m(t)$ denotes the probability amplitude of a basis state $|m\rangle\in \{|00\rangle, |01\rangle, |10\rangle, |11\rangle \}$ at time $t$. In the doubly rotating frame (\textit{i.e.} the logical frame), where each frame rotates at the corresponding qubit frequency, the logical state vector $\tilde{\psi}(t)$ is given by
\begin{align}
    \tilde{\psi}(t) &=  \begin{pmatrix}
    \tilde{c}_{00} \\
    \tilde{c}_{01} \\
    \tilde{c}_{10} \\
    \tilde{c}_{11}
    \end{pmatrix}=  \begin{pmatrix}
    c_{00}(0) \\
    \left(e^{-i\omega_2t}\right)e^{i\omega_2t}c_{01}(0) \\
    \left(e^{-i\omega_1t}\right)e^{i\omega_1t}c_{10}(0) \\
    \left(e^{-i(\omega_1+\omega_2)t}\right)e^{i(\omega_1+\omega_2)t}c_{11}(0)
    \end{pmatrix} \nonumber \\
    &=
    \begin{pmatrix}
    c_{00}(0) \\
    c_{01}(0) \\
    c_{10}(0) \\
    c_{11}(0)
    \end{pmatrix}.
\end{align}

Now, suppose that we apply an iSWAP gate at $t = \tau_1$ ($\tau_1>0$). In the lab frame, the iSWAP gate swaps the probability amplitudes $c_{01}(\tau_1)$ and $c_{10}(\tau_1)$ and adds a relative phase of $i$ as follows.
\begin{align}
    \psi (t)|_{t=\tau_1} & =
     \begin{pmatrix}
    1 & 0 & 0 & 0 \\
    0 & 0 & i & 0 \\
    0 & i & 0 & 0 \\
    0 & 0 & 0 & 1
    \end{pmatrix}
    \begin{pmatrix}
    c_{00}(\tau_1) \\
    c_{01}(\tau_1) \\
    c_{10}(\tau_1) \\
    c_{11}(\tau_1)
    \end{pmatrix}
    \nonumber \\
    & = \begin{pmatrix}
    c_{00}(\tau_1) \\
    ic_{10}(\tau_1) \\
    ic_{01}(\tau_1) \\
    c_{11}(\tau_1)
    \end{pmatrix}
    = \begin{pmatrix}
    c_{00}(0) \\
    ie^{i\omega_1\tau_1}c_{10}(0) \\
    ie^{i\omega_2\tau_1}c_{01}(0) \\
    e^{i(\omega_1+\omega_2)\tau_1}c_{11}(0)
    \end{pmatrix}
\end{align}
Subsequently, in the logical frame (the doubly rotating frame), the state vector $\tilde{\psi}(t)|_{t=\tau_1}$ can be written as
\begin{align}
    \tilde{\psi} (t)|_{t=\tau_1} 
    & = \begin{pmatrix}
    c_{00}(0) \\
    \left(e^{-i\omega_2\tau_1}\right)ie^{i\omega_1\tau_1}c_{10}(0) \\
    \left(e^{-i\omega_1\tau_1}\right)ie^{i\omega_2\tau_1}c_{01}(0) \\
    \left(e^{-i(\omega_1+\omega_2)\tau_1}\right)e^{i(\omega_1+\omega_2)\tau_1}c_{11}(0)
    \end{pmatrix} \nonumber \\
    & = \begin{pmatrix}
    \tilde{c}_{00} \\
    e^{i(\omega_1-\omega_2)\tau_1}(i\tilde{c}_{01}) \\
    e^{i(\omega_2-\omega_1)\tau_1}(i\tilde{c}_{10}) \\
    \tilde{c}_{11}
    \end{pmatrix}.
\end{align}
Note that the logical state vector has acquired additional local phase shifts $e^{i(\omega_1-\omega_2)\tau_1}$ and $e^{i(\omega_2-\omega_1)\tau_1}$ on the basis \ketbare{01} and \ketbare{10}, after the iSWAP gate. These phase shifts are artifacts of the frequency difference between the two rotating frames $|\omega_2-\omega_1|$. 
Notably, longitudinal entangling gates (e.g., the CZ gate) do not cause this phase shift, since they do not involve any energy exchange. 
Also, parametrically driven two-qubit gates~\cite{Mckay2016,Reagor2018}, which activate resonant exchange interactions in the logical frame (not the lab frame), do not result in this phase shift.

In this paper, we avoid this phase shift by putting the qubits in the same rotating frame; we drive the qubits using tones with the same frequency to synchronize their XY axes. 
However, driving one qubit, which is in resonance with other qubits, requires careful attention when implementing single qubit gates. 
Due to the microwave crosstalk, one microwave pulse can considerably drive multiple qubits at the same time. 
To resolve this issue, we actively cancel out the microwave crosstalk by applying cancellation tones simultaneously (see Appendix~\ref{suppsec:mw_crosstalk_calibration} for details).

\section{Microwave crosstalk cancellation}
\label{suppsec:mw_crosstalk_calibration}
We quantify the microwave crosstalk between the XY control lines and the qubits by measuring Rabi oscillations (Fig.~\ref{suppfig:mw_crosstalk_cancellation}). The normalized microwave crosstalk matrix $M_{\mathrm{mw}}$ is shown below, defined as $\Omega_{\mathrm{actual}}=M_{\mathrm{mw}}\Omega_{\mathrm{ideal}}$, where $|\Omega|$ is the Rabi frequency of each qubit and $\angle{\Omega}$ is the phase of the Rabi drive. 
\begin{align}
    M_{\mathrm{mw}} = \begin{pmatrix}
    1&0.1875 \angle{101.2^{\circ}}\\
    0.1505 \angle{-108.3^{\circ}}&1
    \end{pmatrix}
\end{align}
We apply cancellation drives to orthonormalize the XY control and find a remaining crosstalk of below {$3\times 10^{-5}$} (Fig.~\ref{suppfig:mw_crosstalk_cancellation}e). 

\begin{figure}[ht!]
    \centering
    \includegraphics[width=8.9cm]{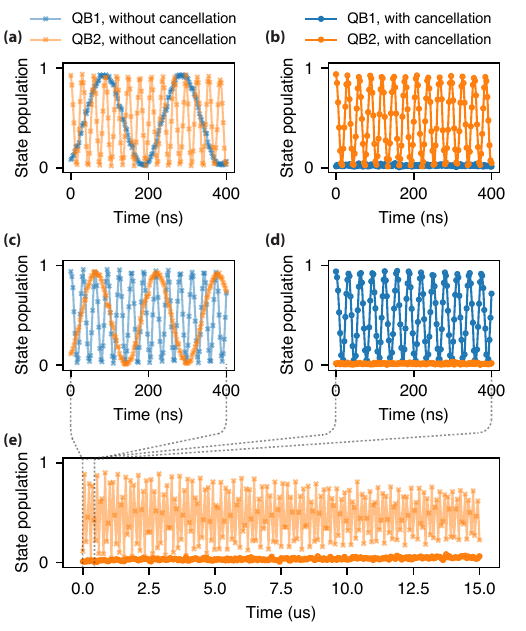}
    \caption{ 
    \textbf{Measurements and cancellation of a microwave crosstalk.}
    \textbf{(a,b)} Rabi oscillations of QB1 (blue) and QB2 (orange) when driving through the QB2 local drive line. 
    \textbf{(c,d)} Rabi oscillations of QB1 (blue) and QB2 (orange) when driving through the QB1 local drive line. 
    \textbf{(e)} Rabi oscillations of QB2 when driving through the QB1 local drive line.
    }
    \label{suppfig:mw_crosstalk_cancellation}
\end{figure}

\clearpage
\newpage

\section{Single-qubit gate fidelities}
\label{suppsec:1qb_rb}

\begin{figure}[H]
    \centering
    \includegraphics[width=8.9cm]{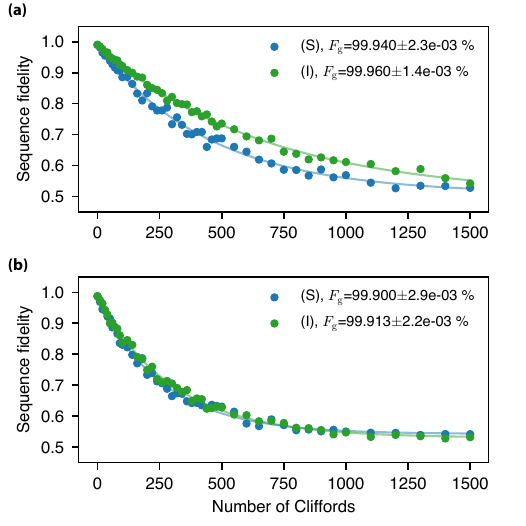}
    \caption{ 
    \textbf{Experimental results of single-qubit randomized benchmarking, when QB1 and QB2 are detuned by 150 MHz.}
    \textbf{(a)} Single-qubit RB measurement data for QB1,
    \textbf{(b)} Single-qubit RB measurement data for QB2.
    ``S" denotes the simultaneous application of single-qubit Cliffords, ``I" denotes the isolated application of single-qubit Cliffords. The pulse duration of X- and Y-rotation gates is \SI{30}{\nano\second}. QB1 and QB2 are biased at 4.16 and 4.00 GHz, respectively. The gate fidelities are degraded, when measured simultaneously, possibly due to microwave crosstalk. The data are averaged over 20 random sequences for each sequence length.
    }
    \label{suppfig:1qb_rb_cz}
\end{figure}

\begin{figure}[ht!]
    \centering
    \includegraphics[width=8.9cm]{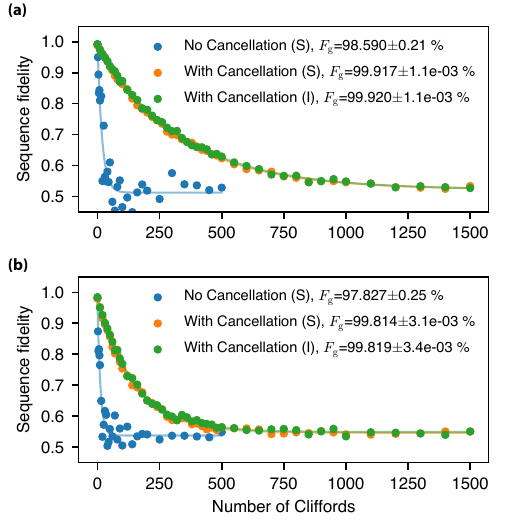}
    \caption{ 
    \textbf{Experimental results of single-qubit randomized benchmarking, when QB1 and QB2 are in resonance.}
    \textbf{(a)} Single-qubit RB measurement data for QB1,
    \textbf{(b)} Single-qubit RB measurement data for QB2.
    ``S" denotes the simultaneous application of single-qubit Cliffords, ``I" denotes the isolated application of single-qubit Cliffords. The pulse duration of X- and Y-rotation gates is \SI{70}{\nano\second}. Both QB1 and QB2 are biased at 4.16 GHz. We apply cancellation pulses to offset microwave crosstalk (orange and green circles) and reduce the gate errors by more than a factor of 10. The data are averaged over 20 random sequences for each sequence length.
    }
    \label{suppfig:1qb_rb_iswap}
\end{figure}

We measure the single qubit gate fidelities via Clifford-based randomized benchmarking~\cite{Corcoles2013,Megesan2012, Barends2014} in the following two configurations:
\begin{enumerate}
    \item QB1 and QB2 are detuned by $\approx$ 160 MHz ($\omega_1/2\pi=\SI{4.16}{\giga\hertz}$, $\omega_2/2\pi=\SI{4.00}{\giga\hertz}$) and CPLR is biased at \SI{5.45}{\giga\hertz}, where the static $ZZ$ coupling between the qubits is eliminated (see Fig.~\ref{suppfig:zz_vs_wc_idle}). This is the idling configuration when performing the CZ gate. We use a \SI{30}{\nano\second}-long microwave pulse for implementing X- and Y-rotation gates. Fig.~\ref{suppfig:1qb_rb_cz} shows the randomized benchmarking data.
    \item QB1 and QB2 are in resonance  ($\omega_1/2\pi=\omega_2/2\pi=\SI{4.16}{\giga\hertz}$) and CPLR is biased at \SI{5.8}{\giga\hertz}, where the effective QB-QB coupling $g_{\mathrm{iSWAP}}$ is switched off. This is the idling configuration when performing the iSWAP gate. For better cancellation of microwave crosstalk, we use a longer (\SI{70}{\nano\second}-long) microwave pulse for implementing X- and Y-rotation gates. Fig.~\ref{suppfig:1qb_rb_iswap} shows the randomized benchmarking data.
\end{enumerate}

\section{Two-qubit Clifford Randomized benchmarking for the iSWAP gate}
\label{suppsec:iswap_rb}
\begin{figure}[H]
    \centering
    \includegraphics[width=8.9cm]{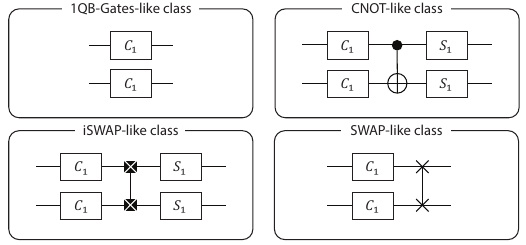}
    \caption{ 
    Two-qubit Clifford classes.
    }
    \label{suppfig:2qb_clifford_classes}
\end{figure}

Following Refs.~\cite{Corcoles2013, Barends2014}, we construct the two-qubit Clifford group, which has four distinct classes as shown in Fig.~\ref{suppfig:2qb_clifford_classes}. The single-qubit Clifford group $C_1$ is the group of 24 single-qubit rotations, which can be written in terms of the X- and Y- rotations~\cite{Barends2014}. One of three-element single-qubit rotation groups $S_1$ is shown in Table.~\ref{supptab:singleQ_rotation_groups}.

\begin{figure}[H]
    \centering
    \includegraphics[width=8.9cm]{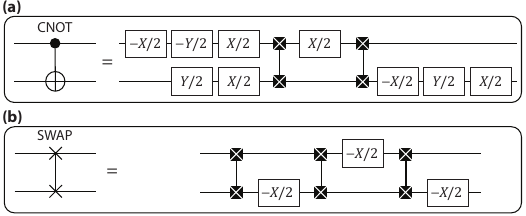}
    \caption{ 
    \textbf{(a)} Decomposition of the CNOT gate into the iSWAP gates.
    \textbf{(b)} Decomposition of the SWAP gate into the iSWAP gates.
    }
    \label{suppfig:iswap_decomposition}
\end{figure}

\begin{figure}[H]
    \centering
    \includegraphics[width=8.9cm]{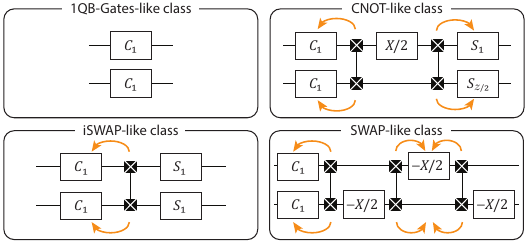}
    \caption{ 
    Two-qubit Clifford classes written in terms of the iSWAP gate and single-qubit gates. Since our native iSWAP gate accompanies additional unwanted single-qubit Z-rotations, we incorporate compensating Z-rotations into the existing single qubit gates that are either preceded or followed by the iSWAP gate to undo the unwanted Z-rotations. The orange colored arrows denote which single-qubit gates are subject to be updated to undo the Z-rotations of the iSWAP. 
    }
    \label{suppfig:2qb_cliffords_in_iswap}
\end{figure}

By rewriting the CNOT and the SWAP in terms of the iSWAP (Fig.~\ref{suppfig:iswap_decomposition}), we generate the two-qubit Cliffords in terms of the iSWAP and single-qubit XY gates as shown in Fig.~\ref{suppfig:2qb_cliffords_in_iswap}. Our native iSWAP gate accompanies single-qubit Z rotations since the qubit frequencies are dynamically tuned during the gate. We undo these unwanted Z-rotations by incorporating compensating Z-rotations into the existing single qubit gates that are either preceded or followed by the iSWAP gate. For example, in the case of ``the iSWAP-like Cliffords'' (Fig.~\ref{suppfig:2qb_cliffords_in_iswap}), we update the single-qubit gates that are preceded by an iSWAP gate such that they undo the Z-rotations of the iSWAP gate. Specifically, we replace the corresponding single-qubit Clifford gate ($C_1$) by three rotation gates along $x-y-x$ axes, which can implement an arbitrary single-qubit rotation according to Euler's rotation theorem (see also Appendix~\ref{suppsec:euler_z}). 

We now calculate the average gate composition for the two-qubit Cliffords. The single-qubit class has 576 elements and contains 90/24 single qubit gates per element, on average. The CNOT-like class has 5184 elements and contains 2 iSWAP gates and 13 single qubit gates per element, on average. The iSWAP-like class has 5184 elements and contains 1 iSWAP gate and 70/3 single qubit gates per element, on average. The SWAP-like class has 576 elements and contains 3 iSWAP gates and 14 single qubit gates per element, on average. Given these, we find that the two-qubit Cliffords are composed of 1.5 iSWAP gates and 10.9375 single qubit gates, on average. Therefore, the average error rate of two-qubit Cliffords can be calculated as follows: $r_{\mathrm{Clifford}}=10.9375\times r_{\mathrm{1qb}}+1.5\times r_{\mathrm{iSWAP}}$, where $r_{\mathrm{1qb}}$ and $r_{\mathrm{iSWAP}}$ are the average error rates of single-qubit gates and an iSWAP gate, respectively.

\begin{figure}[H]
    \centering
    \includegraphics[width=8.9cm]{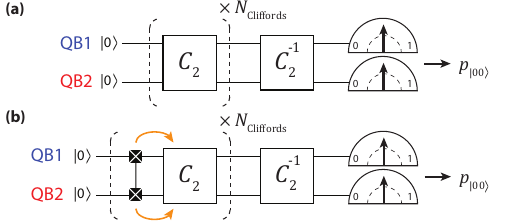}
    \caption{ 
   \textbf{(a)} A diagram of the standard (or reference) two-qubit RB sequence.
   \textbf{(b)} A diagram of the two-qubit RB sequence interleaved by the iSWAP gate. The additional unwanted Z-rotations of the interleaved iSWAP gate are canceled out by the subsequent two-qubit Clifford (orange arrows).
    }
    \label{suppfig:iswap_RB}
\end{figure}

To characterize the two-qubit interaction fidelity of the iSWAP gate, we perform interleaved randomized benchmarking~\cite{Corcoles2013, Megesan2012, Barends2014}. The benchmarking sequences are illustrated in Fig.~\ref{suppfig:iswap_RB}. Note that when interleaving the iSWAP gate, we incorporate compensating Z-rotations into the subsequent reference Clifford. For example, in the case of ``1QB-gates-like'' reference Clifford, the constituent single-qubit Clifford $C_1$ on each qubit is replaced by three single-qubit gates along $x-y-x$ axes. Other two-qubit Clifford classes already have three single-qubit gates along $x-y-x$ axes at their front ends (see Fig.~\ref{suppfig:2qb_cliffords_in_iswap}), so they do not require additional XY gates when interleaving the iSWAP gate. 
As a consequence, the iSWAP-interleaved sequence acquires additional 0.1125 XY gates on average. 
The error contribution of additional XY gates is accounted for when estimating the two-qubit interaction fidelity of the iSWAP gate.

\vfill

\begin{table}[H]
       \begin{ruledtabular}
       \begin{tabular}
       {>{\centering} p{1.5cm}|l l l}
            $S_1$ & \hspace*{1.5pt} I & Y/2, X/2  &  -X/2, -Y/2 \\
            $S_{z/2}$ & -X/2, Y/2, X/2 & Y/2, X & -X/2 
        \end{tabular}
    \caption{The three-element single-qubit rotation groups $S_1$ and $S_{z/2}$ written in terms of X- and Y-rotation gates in time.}
    \label{supptab:singleQ_rotation_groups}
    \end{ruledtabular}
\end{table}

\newpage

\section{Two-qubit gate tune-up procedures}
\label{suppsec:cz_tune-up}

\begin{figure}[htb!]
    \centering
    \includegraphics[width=8.9cm]{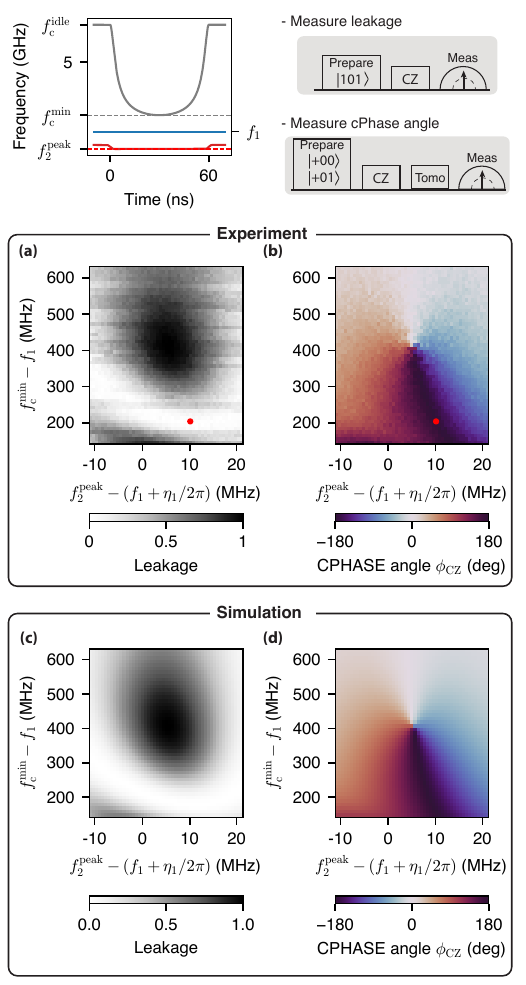}
    \caption{ 
    \textbf{Tune-up measurements for the CZ gate.}
    \textbf{(a,b)} Experimental data of tune-up measurements for a 60 ns-long CZ gate. We measure leakage from \ketidle{101} and conditional phase (CPHASE) angle $\phi_{\mathrm{CZ}}$ as functions of QB2 Z-pulse amplitude ($x$ axis) and CPLR Z-pulse amplitude ($y$ axis). The control pulse and sequences to measure leakage and CPHASE angle are illustrated at the top. We find an optimal parameter set that minimizes both the leakage and CPHASE angle error (red circles). \textbf{(c,d)} Numeric simulation reproducing the experimental data of tune-up measurements.
    }
    \label{suppfig:cz_tuneup}
\end{figure}

\begin{figure}[htbp!]
    \centering
    \includegraphics[width=8.9cm]{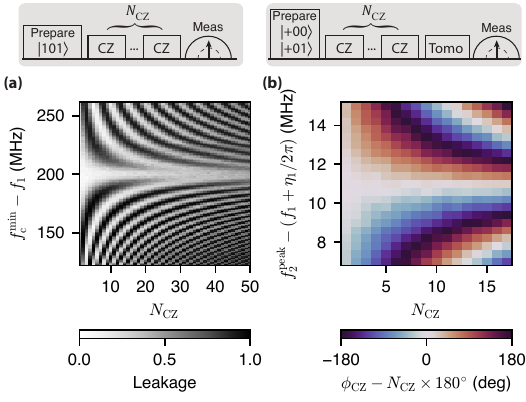}
    \caption{ 
    \textbf{Experimental data of fine-tuning measurements for the CZ gate.}
    \textbf{(a)} Measuring the leakage of multiple CZ pulses to finely adjust the CPLR Z-amplitude ($f_{\mathrm{c}}^{\mathrm{min}}$). \textbf{(b)} Measuring the CPHASE angle error ($\phi_{\mathrm{CZ}}- N_{\mathrm{CZ}}\times180^{\circ}$) of multiple CZ pulses to finely adjust the QB2 Z-amplitude ($f_{2}^{\mathrm{peak}}$).
    }
    \label{suppfig:cz_tuneup_finer}
\end{figure}

We calibrate the CZ gate by adjusting the Z control amplitudes for a fixed gate length ($\SI{60}{\nano\second}$) and measuring the
leakage from \ketbare{101} and the conditional phase (CPHASE) angle $\phi_{\mathrm{CZ}}$. 
A control pulse for the CZ gate and pulse sequences for these measurements are illustrated at the top of Fig.~\ref{suppfig:cz_tuneup}.
To measure the leakage from \ketbare{101}, we prepare \ketbare{101} by applying $\pi$ pulses to both QB1 and QB2 and measure the state population of \ketbare{101} after a CZ gate (Fig.~\ref{suppfig:cz_tuneup}a). 
To measure the CPHASE angle, we perform a cross-Ramsey type experiment, which measures the conditional phase accumulation of QB1, while initializing QB2 in its ground state or excited state (Fig.~\ref{suppfig:cz_tuneup}b).

We find the optimal spot (red circles in Figs.~\ref{suppfig:cz_tuneup}a and b) in the parameter space for the CZ gate, which minimizes both the leakage and the CPHASE angle error ($\equiv\phi_{\mathrm{CZ}}- 180^{\circ}$).
Notably, the measured data has a slight tilt (the leakage and the CPHASE angle data are not symmetric about their $x$-axes: the qubit-qubit detuning) due to the level repulsion induced by qubit-coupler interactions.
These tune-up measurements are qualitatively reproduced by time-dependent Hamiltonian simulations for three-interacting qutrits (Figs.~\ref{suppfig:cz_tuneup}c and d). See Appendix~\ref{suppsec:numerical_simulation} for details about the simulations.

Near the optimal spot, we note that the leakage is predominantly controlled by the CPLR Z-pulse amplitude (the $y$ axes of the plots), while the CPHASE angle is controlled by the QB2-Z pulse amplitude (the $x$ axes of the plots). Keeping this in mind, we individually adjust the CPLR-Z pulse amplitude and the QB2-Z pulse amplitude by measuring the leakage and the CPHASE angle error, respectively. For fine-adjustment of the amplitudes, we measure multiple CZ pulses to amplify the effects of small amplitude errors (Fig.~\ref{suppfig:cz_tuneup_finer}). The measurement data exhibits symmetric chevron patterns that allow us to easily find optimal values for the pulse amplitudes to minimize the leakage and the CPHASE angle error ($\equiv\phi_{\mathrm{CZ}}- N_{\mathrm{CZ}}\times180^{\circ}$). We repeat this class of fine-tuning measurements 2--3 times within narrower amplitude ranges so that we can make the most precise adjustments possible (ultimately limited by the amplitude resolution limit of our AWGs).

\begin{figure}[htbp!]
    \centering
    \includegraphics[width=8.9cm]{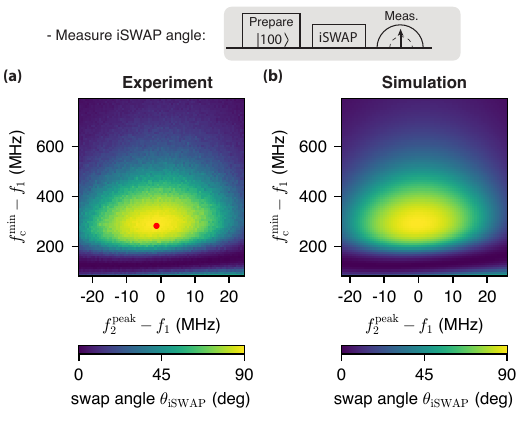}
    \caption{ 
    \textbf{Tune-up measurement for the iSWAP gate.} \textbf{(a)} Experimental data of tune-up measurements for a 30 ns-long iSWAP gate. We measure the iSWAP angle $\theta_{\mathrm{iSWAP}}$ as functions of QB2 Z-pulse amplitude ($x$ axis) and CPLR Z-pulse amplitude ($y$ axis). The control sequence to measure iSWAP angle is illustrated at the top. A red circle denotes an optimal parameter set that maximizes $\theta_{\mathrm{iSWAP}}$.  \textbf{(b)} Numeric simulation reproducing the experimental data of tune-up measurements.
    }
    \label{suppfig:iswap_tuneup}
\end{figure}

Finally, to offset single-qubit phase accumulation that accompanies the CZ gate, we subsequently apply virtual Z gates~\cite{Mckay2017}. To calibrate these virtual Z gates, we perform Ramsey experiments on QB1 and QB2 and measure the single-qubit phase accumulation of each qubit due to the CZ gate. Fine-tuning the angles of the virtual Z gates is done by a numerical optimization method (the Nelder-Mead algorithm) with the fidelity of two-qubit randomized benchmarking sequences as an objective function~\cite{Kelly2014}.

Along the same line, we calibrate the iSWAP gate by adjusting the Z control pulse amplitudes for a fixed gate length ($\SI{30}{\nano\second}$) and measure the swap angle (Fig.~\ref{suppfig:iswap_tuneup}). The swap angle $\theta_{\mathrm{iSWAP}}$ quantifies how much the population of QB1 has been transferred to QB2 and vice versa. Accordingly, to measure $\theta_{\mathrm{iSWAP}}$, we prepare \ketbare{100} and measure how much the population of \ketbare{100} has transferred to the population of \ketbare{001} by an iSWAP gate ($\theta_{\mathrm{iSWAP}} \equiv \tan^{-1}(p_{|001\rangle}/p_{|100\rangle})$, where $p_{|001\rangle}$ and $p_{|100\rangle}$ are the measured populations of \ketbare{001} and \ketbare{100}, respectively, at the end). We find an optimal spot for the iSWAP gate (red circle in Fig.~\ref{suppfig:iswap_tuneup}a) which maximizes $\theta_{\mathrm{iSWAP}}$ ($0^{\circ} \leq \theta_{\mathrm{iSWAP}}\leq 90^{\circ}$). In Fig.~\ref{suppfig:iswap_tuneup}b, we numerically simulate this tune-up measurement and show good qualitative agreement with the experimental result.

\begin{figure}[H]
    \centering
    \includegraphics[width=8.9cm]{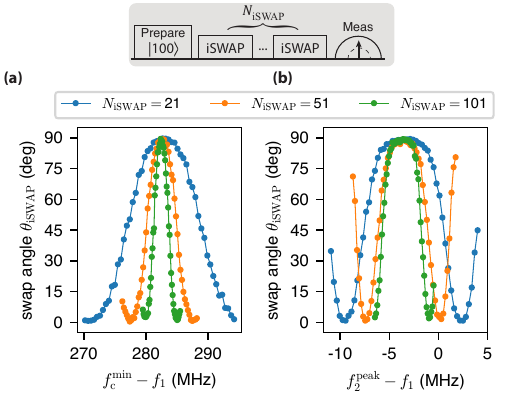}
    \caption{ 
    \textbf{Experimental data of fine-tuning measurements for the iSWAP gate.}
    \textbf{(a)}  Measurements of the swap angle $\theta_{\mathrm{iSWAP}}$ for multiple iswap pulses to finely adjust the CPLR Z-amplitude ($f_{\mathrm{c}}^{\mathrm{min}}$). \textbf{(b)} Measurements of the swap angle $\theta_{\mathrm{iSWAP}}$ for multiple iswap pulses to finely adjust the QB2 Z-amplitude ($f_{\mathrm{2}}^{\mathrm{peak}}$).
    }
    \label{suppfig:iswap_tuneup_finer}
\end{figure}

To finely adjust the CPLR-Z pulse amplitude and the QB2-Z pulse amplitude, we measure multiple iSWAP pulses (Fig.~\ref{suppfig:iswap_tuneup_finer}). Since the swap angle is controlled by both the CPLR-Z and QB2-Z amplitudes, we adjust the both amplitudes in an alternating manner---that is, adjusting the amplitudes of QB2-Z, CPLR-Z, QB2-Z, CPLR-Z, $\cdots$---with varying the number of iSWAP pulses ($N_{\mathrm{iSWAP}} \in \{21, 51, 101\}$). 

Finally, to offset single-qubit phase accumulation that is accompanied when performing the iSWAP gate, we apply actual Z gates using XY control (see Appendix~\ref{suppsec:euler_z} for details). To calibrate these Z gates, we perform Ramsey experiments on QB1 and QB2 and measure the single-qubit phase accumulation of each qubit due to the iSWAP gate. As in the case of the CZ gate, we numerically search the optimal angles of the Z gates that maximize the sequence fidelity of two-qubit RB sequences.

\section{Residual leakage of the CZ gate}
\label{suppsec:cz_residual_leakage}
\begin{figure}[H]
    \centering
    \includegraphics[width=8.9cm]{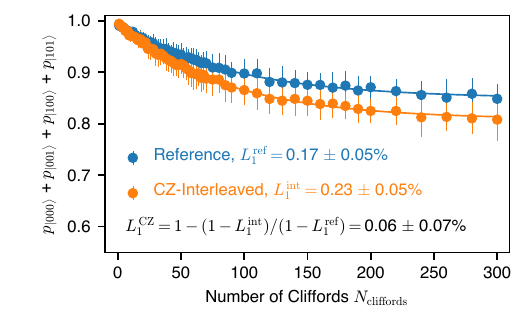}
    \caption{ 
    \textbf{Characterization of the leakage rate of the CZ gate with an optimized pulse shape.} Population in the computational subspace $p_{\mathcal{X}_1}=p_{|000\rangle}+p_{|001\rangle}+p_{|100\rangle}+p_{|101\rangle}$ for the reference and interleaved two-qubit randomized benchmarking sequences.
    }
    \label{suppfig:cz_optimal_leakage_rb}
\end{figure}

Following Refs.~\cite{Wood2018, Rol2019}, we estimate the average leakage rate of our \SI{60}{\nano\second}-long CZ gate with an optimized pulse shape from our interleaved randomized benchmarking measurement  (Fig.~\ref{fig:Fig4}(h)). To estimate the leakage rate, we fit the population in the computational subspace $p_{\mathcal{X}_1}\equiv p_{|000\rangle}+p_{|001\rangle}+p_{|100\rangle}+p_{|101\rangle}$ to an exponential model (for both reference and interleaved RB curves, see Fig.~\ref{suppfig:cz_optimal_leakage_rb}):
\begin{align}
    p_{\mathcal{X}_1,\mathrm{ref}} &= A_{\mathrm{ref}} + B_{\mathrm{ref}}(\lambda_{1,\mathrm{ref}})^{N_{\mathrm{Cliffords}}}, \\
    p_{\mathcal{X}_1,\mathrm{int}} &= A_{\mathrm{int}} + B_{\mathrm{int}}(\lambda_{1,\mathrm{int}})^{N_{\mathrm{Cliffords}}}.
\end{align}
To ensure accurate uncertainties of the leakage rates, we perform a weighted least-squares fit using the inverse of variance as the weights. Then we estimate the average leakage rates $L_{1,\mathrm{ref}}$, $L_{1,\mathrm{int}}$ per Clifford as follows.
\begin{align}
    L_{1,\mathrm{ref}} = (1-A_{\mathrm{ref}})(1-\lambda_{1,\mathrm{ref}}), \\
    L_{1,\mathrm{int}} = (1-A_{\mathrm{int}})(1-\lambda_{1,\mathrm{int}}).
\end{align}
The average leakage rate $L_1^{\mathrm{CZ}}$ per CZ gate is subsequently obtained by the following equation,
\begin{align}
    L_1^{\mathrm{CZ}} = 1- \frac{1-L_{1,\mathrm{int}}}{1-L_{1,\mathrm{ref}}}.
\end{align}
The leakage rate $L_{1}^{\mathrm{CZ}}$ per CZ gate is estimated of $0.06\pm0.07\%$.
We find that the most of the residual leakage is introduced into the second excited state of QB1 ($1-p_{\mathcal{X}_1}\approx p_{|200\rangle}+p_{|201\rangle}$), which indicates the residual leakage may be due to residual pulse distortion in Z-control pulses of QB2 and CPLR.

\section{Z-corrections for the two-qubit gates}
\label{suppsec:euler_z}
Two-qubit gates are accompanied by local phase shifts (single-qubit Z-rotations), since the qubit frequencies are dynamically tuned during the gates. To undo these phase shifts, we apply additional single Z-rotations either before or after the two-qubit gates. In the case of the CZ gate, we utilize virtual Z gates~\cite{Mckay2017} which are simply implemented by shifting phase offsets of  microwave pulses. In contrast, in the case of the iSWAP gate, we implement actual Z gates  since the iSWAP gate that we consider in this work is not compatible with virtual Z gates in general~\cite{Mckay2017}.

\begin{figure}[H]
    \centering
    \includegraphics[width=8.9cm]{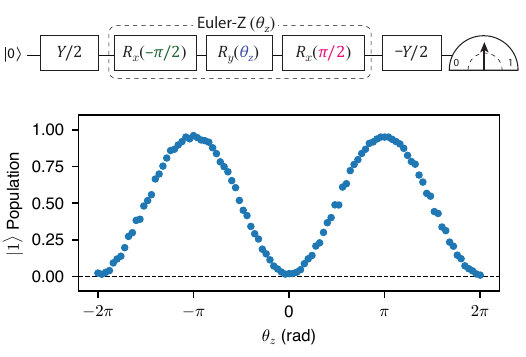}
    \caption{ 
    \textbf{Implementation of arbitrary Z gates by combining X- and Y-rotations.} Experimental data of the Ramsey-type experiment to validate the Euler-Z gate. We vary the angle $\theta_{z}$ of Y rotation, which effectively adjusts the rotation angle of the Euler-Z gate. 
    }
    \label{suppfig:test_euler_z}
\end{figure}

We implement actual Z rotations by combining X and Y rotations. According to Euler’s rotation theorem, any rotation matrix can be described by the multiplication of three rotation matrices along $x-y-x$ axes. Subsequently, arbitrary Z gates (we call the Euler-Z gate) with rotation angle $\theta_z$ can be implemented by a series of X- and Y-rotations: $R_x(-\pi/2)-R_y(\theta_{z})-R_x(\pi/2)$, where $R_x, R_y$ are single-qubit rotations along the $x$ and $y$ axes, respectively (Fig.~\ref{suppfig:test_euler_z}). 

\newpage
\section{Numerical simulation of the dynamics}
\label{suppsec:numerical_simulation}
We numerically simulate the three-body dynamics (Eq.~\eqref{eq:H_lab}) presented in this work by treating our system as three interacting qutrits (for both time-dependent and time-independent Hamiltonian simulations).
Given the control waveforms $\omega_{1}(t)$, $\omega_{2}(t)$, and $\omega_{\mathrm{c}}(t)$, we modulate the $b_i^{\dagger}b_i$ ($i\in\{1,2,\mathrm{c}\}$) terms in the system Hamiltonian.
The coupling strengths $g_{\mathrm{1c}}(t)$, $g_{\mathrm{2c}}(t)$, and $g_{12}(t)$ are subsequently modulated, as they are determined by $\omega_1(t)$, $\omega_{2}(t)$, $\omega_{\mathrm{c}}(t)$ and the capacitance matrix of the superconducting circuit (see Appendix A of Ref.~\cite{Yan2018} for details). 
The capacitances that are used to model the circuit are summarized in Table.~\ref{supptab:capacitances}.
The anharmonicities of the qubits and the coupler are assumed to be fixed and set as follows: $\eta_1/2\pi = \SI{-220}{\mega\hertz}$, $\eta_2/2\pi = \SI{-210}{\mega\hertz}$, and $\eta_{\mathrm{c}}/2\pi = \SI{-90}{\mega\hertz}$.
\begin{table}[H]
    \begin{ruledtabular}
        \begin{tabular}
                {cccccc}
                $C_1$ & $C_{\mathrm{c}}$ &  $C_2$ &$C_{\mathrm{1c}}$ &  $C_{\mathrm{2c}}$ & $C_{\mathrm{12}}$  \\ \hline  \\[-0.3cm] \SI{95}{\femto\farad} & \SI{228}{\femto\farad} &
                \SI{98}{\femto\farad} & \SI{5.36}{\femto\farad} & \SI{5.36}{\femto\farad} & \SI{0.125}{\femto\farad}
        \end{tabular}
        \caption{The capacitances used for the numerical simulations (see Ref.~\cite{Yan2018} for the notations).}
        \label{supptab:capacitances}
    \end{ruledtabular}
\end{table}

\newpage
\section{$T_1$ contribution to gate errors}
\label{suppsec:T1_contribution_to_gate_errors}
We perform numerical simulations to estimate the contributions of (both qubits' and the coupler's) energy relaxations to the errors of the CZ and iSWAP gates. 
\begin{figure*}[ht!]
    \begin{center}
    \includegraphics[width=18.3cm]{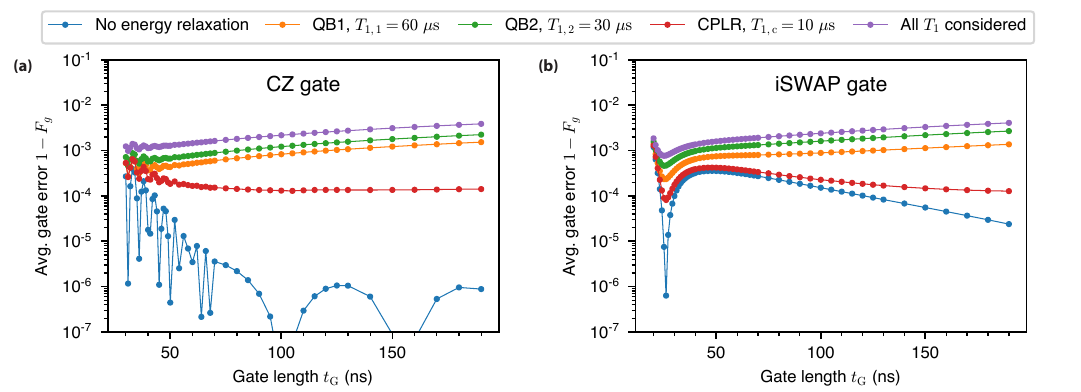}
    \end{center}
    \caption{
    \textbf{Numerical simulation results for the average gate errors of the CZ and the iSWAP gates.}
    \textbf{(a)} The average gate infidelity ($1-F_{g}$) of the iSWAP gate as a function of its gate length $t_{\mathrm{G}}$. For each gate length, the control pulse shape is optimized as detailed in Appendices~\ref{suppsec:effective_H} and \ref{suppsec:slepian_approach}. Each data point is obtained numerically simulating quantum process tomography and computing the corresponding gate infidelity. 
    \textbf{(b)} The average gate infidelity ($1-F_{g}$) of the iSWAP gate as a function of its gate length $t_{\mathrm{G}}$. The lowest gate error is achieved when $t_{\mathrm{G}}\approx \SI{25}{\nano\second}$. At this point, the residual $ZZ$ interaction of the iSWAP is minimized, owing to the cancellation induced by the higher level of CPLR.
    }
    \label{suppfig:Fg_vs_tg}
\end{figure*}
The time evolution of the system
is calculated by solving the Lindblad master equation in QuTiP~\cite{Johansson2013}:
\begin{align}
    \dot{\rho}(t)=-\frac{i}{\hbar}[H(t),\rho(t)] + \sum_{j=1,2,\mathrm{c}} \Gamma_{1,j}\mathcal{L}[b_{j}]\left(\rho(t)\right),
\end{align}
where $\rho(t)$ is the density matrix of the system, $H(t)$ is the system Hamiltonian (Eq.~\eqref{eq:H_lab}), and $\Gamma_{1,j}\equiv 1/T_{1,j}$ ($j$=1,2, and c) are the relaxation rates of QB1, QB2, and CPLR respectively. The Lindblad superoperator acting on a density matrix $\rho$, for a generic operator $C$, is defined by
\begin{align}
\mathcal{L}[C]\left(\rho \right)\equiv  C\rho C^{\dagger}-\{ \rho, C^{\dagger} C \}/2.
\end{align}

Following Ref.~\cite{Yan2018}, we compute the average gate fidelity $F_{g}$ by numerically simulating quantum process tomography. We prepare 16 different input states $\{( |0\rangle, |1\rangle, |+\rangle, |-\rangle)\}\otimes\{( |0\rangle, |1\rangle, |+\rangle, |-\rangle)\}$ and reconstruct the process matrix $\chi$ from the resulting density matrices. 
By comparing it to the ideal process matrix $\chi_{\mathrm{ideal}}$, we compute the process fidelity $F_p=\mathrm{Tr}[\chi_{\mathrm{ideal}}\chi]$.
The ideal process matrix include additional unwanted single-qubit Z-rotations that accompany by our native two-qubit gates.
The average gate fidelity $F_g$ is subsequently obtained from the simple relationship between $F_p$ and $F_g$~\cite{Horodecki1999, Nielsen2002}.

Fig.~\ref{suppfig:Fg_vs_tg}a shows the average gate infidelity ($1-F_{g}$) of the CZ gate as a function of its gate length $t_{\mathrm{G}}$.
For each gate length, the optimized control pulse is used and calibrated in a manner similar to that described in Appendix~\ref{suppsec:cz_tune-up}.
We run simulations in the absence of energy relaxation (blue circles), in the presence of only QB1's relaxation (orange circles), only QB2's relaxation (green circles), only CPLR's relaxation (red circles), and all possible relaxations (purple circles). 
Experimental values of $\Gamma_{1,j}$ are used for the simulations.
We find that gate errors due to parasitic interactions (blue circles) diminish drastically, when $t_{\mathrm{G}} \geq \SI{60}{\nano\second}$. 
We extract  dissipation-induced ($T_1$-induced) gate errors by taking the difference between the fidelities $F_g$ in the presence and the absence of energy relaxations. 
Table~\ref{supptab:T1_contribution_to_Fg} summarizes $T_1$ contributions to the average gate error of a  \SI{60}{\nano\second}-long CZ gate (which is realized in our experiments).

\begin{table}[H]
\begin{ruledtabular}
    \begin{tabular}{c | c c}
        & \SI{60}{\nano\second}-long CZ & \SI{30}{\nano\second}-long iSWAP \\ \hline \\[-0.3cm] QB1, $T_{1,1}=\SI{60}{\micro\second}$ & 5.2E-4 & 2.6E-4 \\ 
        QB2, $T_{1,2}=\SI{30}{\micro\second}$ & 7.8E-4 & 5.2E-4 \\ 
        CPLR, $T_{1,\mathrm{c}}=\SI{10}{\micro\second}$ & 1.6E-4 & 7.6E-5 \\ 
        Total $T_1$-induced error & 1.5E-3 & 8.6E-4 \\ 
    \end{tabular}
    \caption{\textbf{$\boldsymbol{T_1}$ contributions to the average gate errors of a {60}\hspace{1pt}{ns}-long CZ and a {30}\hspace{1pt}{ns}-long iSWAP gate.} 
    Each $T_1$ contribution is computed by taking the difference between the gate errors in the presence and the absence of corresponding energy relaxation.
    We find that the sum of individual $T_1$ contributions is approximately equal to the total $T_1$-induced error computed from a separate simulation that takes all possible relaxations into account (purple circles in Fig.~\ref{suppfig:Fg_vs_tg}).}
    \label{supptab:T1_contribution_to_Fg}
\end{ruledtabular}
\end{table}

Fig.~\ref{suppfig:Fg_vs_tg}b shows the average gate errors of the iSWAP gate in the absence and the presence of energy relaxations.
The lowest gate error is achieved when $t_{\mathrm{G}}\approx \SI{25}{\nano\second}$; this is the point where the residual $ZZ$ interaction of the iSWAP is minimized, owing to the cancellation induced by the higher level of CPLR.
The $T_1$ contributions of QB1, QB2, and CPLR to the \SI{30}{\nano\second}-long iSWAP gate error are summarized in Table.~\ref{supptab:T1_contribution_to_Fg}.

\newpage
\section{$1/f^{\alpha}$ flux noise contribution to gate errors}
\label{suppsec:Flux_noise_contribution_to_gate_errors}
We simulate the error contribution of $1/f^{\alpha}$ flux noise, which predominantly limits the dephasing times $T_2^*, T_2^{\mathrm{echo}}$ of QB2 and CPLR. 
While there are other noise sources affecting the qubits and coupler such as charge noise and photon-shot noise, we assume that their impacts are negligible.
Specifically, our transmon qubits and coupler have large $E_{\mathrm{J}}/E_{\mathrm{c}}$ $(> 60)$, which makes their charge dispersions smaller than $\approx \SI{1}{\kilo \hertz}$ such that they are insensitive to charge noise. Notably, owing to large $E_{\mathrm{J}}/E_{\mathrm{c}}$ and negligible photon shot noise, QB1 (which has a fixed-frequency) exhibits nearly $T_1$-limited dephasing times $T_2^{*}\approx T_1, T_2^{\mathrm{echo}}\approx2T_1$.

We extract the power spectral density $S_{\Phi}(f)\approx \frac{(1.43 \upmu \Phi_0)^2}{f}+\frac{(1.77 \upmu \Phi_0 )^2\times(\mathrm{Hz})^{0.33}}{f^{(1.33)}}$ of flux noise from the repeated Ramsey and echo experiments of QB2 (see Appendix~\ref{suppsec:flux_noise_in_the_device} for details). We simulate contribution of $1/f^{\alpha}$ flux noise to gate error by performing Monte Carlo simulations with 1,000 random flux noise realizations; flux noise samples are generated based on the power spectral density $S_{\Phi}(f)$. We assume that QB2 and CPLR experience the same flux noise power (but we independently sample noise waveforms for QB2 and CPLR). We set the low-cutoff frequency of the noise PSD at $10^{-4}$ Hz, which is close to $1/(2\times$ the total length of RB experiment). We set the high-cutoff frequency at $5\times 10^{9}$ Hz, which is close to the $|0\rangle \rightarrow |1\rangle$ transition frequencies of the qubits and coupler; following Ref.~\cite{Martinis2003}, we assume that only the noise below the qubit frequency results in the phase noise.

\begin{figure}[H]
    \centering
    \includegraphics[width=8.9cm]{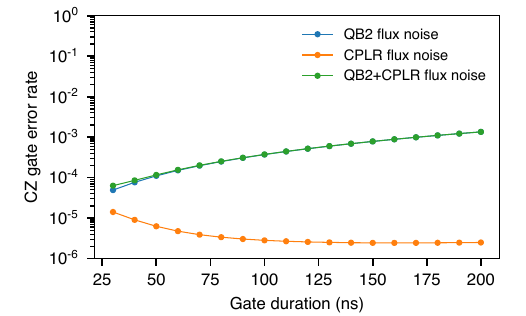}
    \caption{ 
    Contribution of $1/f^{\alpha}$ flux noise to error rates of CZ gates. Notably, the error contribution from CPLR flux noise increases as the gate duration decreases (orange circles), possibly due to stronger qubit-coupler hybridization.
    }
    \label{suppfig:contribution_flux_noise_CZ_fidelity}
\end{figure}

\begin{figure}[ht!]
    \centering
    \includegraphics[width=8.9cm]{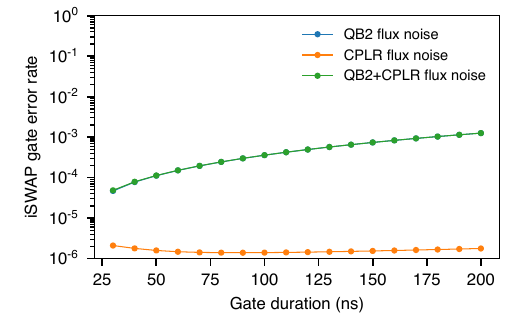}
    \caption{ 
    Contribution of $1/f^{\alpha}$ flux noise to error rates of iSWAP gates. Notably, the error contribution from CPLR flux noise increases as the gate duration decreases (orange circles), possibly due to stronger qubit-coupler hybridization.
    }
    \label{suppfig:contribution_flux_noise_iSWAP_fidelity}
\end{figure}

\begin{table}[H]
\begin{ruledtabular}
    \begin{tabular}{c | c c}
        & \SI{60}{\nano\second}-long CZ & \SI{30}{\nano\second}-long iSWAP \\ \hline \\[-0.3cm] 
        QB2 flux noise & 1.5E-4 & 4.7E-5 \\ 
        CPLR flux noise & 4.8E-6 & 2.1E-6 \\ 
        QB2+CPLR flux noise & 1.6E-4 & 4.9E-5 \\ 
    \end{tabular}
    \caption{\textbf{$1/f^{\alpha}$ flux noise contributions to the average gate errors of a {60}\hspace{1pt}{ns}-long CZ and a {30}\hspace{1pt}{ns}-long iSWAP gate.} 
    Each flux noise contribution is computed by taking the difference between the gate errors in the presence and the absence of flux noise.}
    \label{supptab:flux_noise_contribution_to_Fg}
\end{ruledtabular}
\end{table}

Figs.~\ref{suppfig:contribution_flux_noise_CZ_fidelity} and \ref{suppfig:contribution_flux_noise_iSWAP_fidelity} show the estimated contribution of the flux noise to gate error rates for CZ and iSWAP gates, respectively. We extract the error contribution of $1/f^{\alpha}$ flux noise by subtracting the error rate in the presence of flux noise by the error rate in the absence of flux noise. We compare the flux noise induced error rates and T1-induced error rates for \SI{60}{\nano\second}-long CZ and \SI{30}{\nano\second}-long iSWAP gates in Table~\ref{supptab:flux_noise_contribution_to_Fg}. We find that flux noise contribution is quite small. This is because our gate lengths are short (30--\SI{60}{\nano\second}) such that the impact of long-time correlated noise, i.e., $1/f$ noise, is significantly suppressed. This result is consistent with Ref.~\cite{O'Malley2015}. It also explains how flux-tunable qubits can achieve high-fidelity gates (both single- and two-qubit gates) in general, even though they operate at flux sensitive points at the idling configuration.
\clearpage
\section{Supplementary experimental data for Fig.~\ref{fig:Fig4} in the main text}
\begin{figure}[H]
    \centering
    \includegraphics[width=8.9cm]{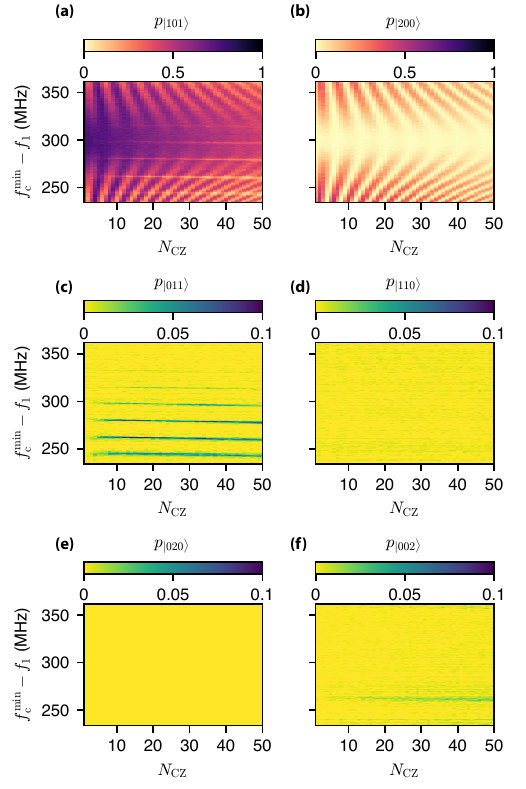}
    \caption{ 
    State population of \textbf{(a)} \ketbare{101}, \textbf{(b)} \ketbare{200}, \textbf{(c)} \ketbare{011}, \textbf{(d)} \ketbare{110}, \textbf{(e)} \ketbare{020}, and \textbf{(f)} \ketbare{002} for the repeated square CZ pulses. 
    }
    \label{suppfig:cz_square_leakage_experiment}
\end{figure}

\begin{figure}[H]
    \centering
    \includegraphics[width=8.9cm]{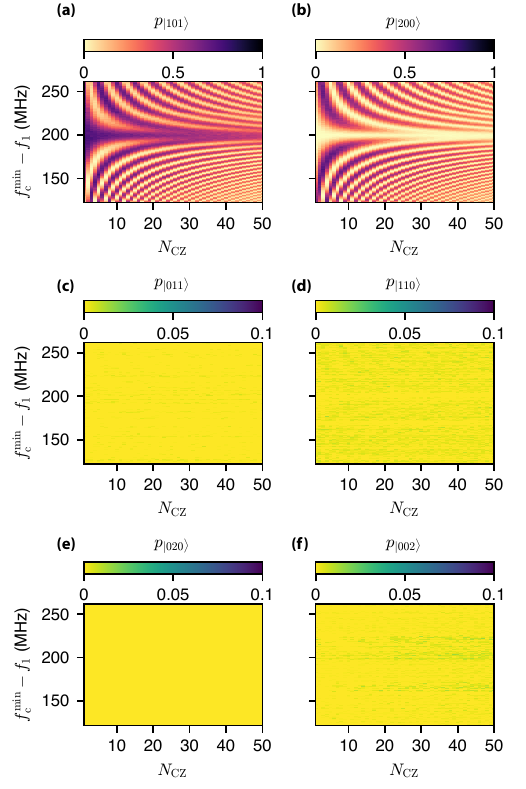}
    \caption{ 
    State population of \textbf{(a)} \ketbare{101}, \textbf{(b)} \ketbare{200}, \textbf{(c)} \ketbare{011}, \textbf{(d)} \ketbare{110}, \textbf{(e)} \ketbare{020}, and \textbf{(f)} \ketbare{002} for the repeated optimal CZ pulses. 
    }
    \label{suppfig:cz_optimal_leakage_experiment}
\end{figure}

In Figs.~\ref{suppfig:cz_square_leakage_experiment} and \ref{suppfig:cz_optimal_leakage_experiment}, we present state population of \ketbare{101}, \ketbare{200}, \ketbare{011}, \ketbare{110}, \ketbare{020} and \ketbare{002} as supplementary experimental data for Fig.~\ref{fig:Fig3} in the main text.

\label{suppsec:supp_exp_cz_leakage}

\bibliography{ref.bib}

\end{document}